\documentclass[11pt,a4paper]{article}
\usepackage{jheppub,amsmath, upgreek, amssymb,slashed,url,bm}
\usepackage{graphicx}
\usepackage{epstopdf}

\input xyv2
\catcode `\@=11
\catcode `\@=12
\usepackage{amsmath}
\usepackage{amsfonts}
\usepackage{amssymb}
\usepackage[mathscr]{eucal}
\usepackage{xr}

\input epsf
\def\Bbb{\mathbb}
\def\h{\widehat}
\def\bb{b}
\def\cF{{\sf F}}
\def\frak{\mathfrak}
\def\underarrow#1{\vbox{\ialign{##\crcr$\hfil\displaystyle
 {#1}\hfil$\crcr\noalign{\kern1pt\nointerlineskip}$\longrightarrow$\crcr}}}
 \def\Tr{{\rm Tr}}

\def\N{{\cal N}}
\def\Pic{{\mathrm{Pic}}}

\def\pp{{\sf p}}
\def\qq{{\sf q}}
  
\def\hat{\widehat}
\font\teneurm=eurm10 \font\seveneurm=eurm7  \font\fiveeurm=eurm5
\newfam\eurmfam
\textfont\eurmfam=\teneurm \scriptfont\eurmfam=\seveneurm
\scriptscriptfont\eurmfam=\fiveeurm
\def\eurm#1{{\fam\eurmfam\relax#1}}
\font\teneusm=eusm10 \font\seveneusm=eusm7 \font\fiveeusm=eusm5
\newfam\eusmfam
\textfont\eusmfam=\teneusm \scriptfont\eusmfam=\seveneusm
\scriptscriptfont\eusmfam=\fiveeusm
\def\eusm#1{{\fam\eusmfam\relax#1}}
\font\tencmmib=cmmib10 \skewchar\tencmmib='177
\font\sevencmmib=cmmib7 \skewchar\sevencmmib='177
\font\fivecmmib=cmmib5 \skewchar\fivecmmib='177
\newfam\cmmibfam
\textfont\cmmibfam=\tencmmib \scriptfont\cmmibfam=\sevencmmib
\scriptscriptfont\cmmibfam=\fivecmmib

\def\FF{{\mathscr F}}
\def\G{{\Gamma}}
\def\eG{{\eurm G}}
\def\neg{\negthinspace}
\def\MM{\mathscr{M}}
\def\M{{\MM}}
\def\CP{{\Bbb{CP}}}
\def\L{{\mathcal L}}
\def\U{{\mathcal U}}

\def\Y{{\mathcal Y}}
\def\V{{\mathcal V}}
\def\ade{{\mathrm{ad}(E)}}
\def\H{{\mathcal H}}
\def\W{{\mathcal W}}
\def\MH{{{\MM}_H}}
\def\ppi{{\boldsymbol{\mathscr{\pi}}}}
\def\tilde{\widetilde}
\def\bar{\overline}
\def\Z{{\Bbb Z}}
\def\AA{{\eurm A}}

\def\B{{\mathcal B}}
\def\T{{\bf T}}
\def\t{\widetilde }

\def\O{{\mathcal O}}
\def\F{{\mathcal F}}
\def\FF{{\eusm F}}
\def\A{{\mathcal A}}
\def\CW{{\mathcal W}}

\def\R{{\Bbb R}}

\font\zfont = cmss10 %scaled \magstep1
 
\def\bigone{\hbox{1\kern -.23em {\rm l}}}
\def\ZZ{\hbox{\zfont Z\kern-.4emZ}}

\def\C{{\Bbb C}}
\def\G{{\mathcal G}}

\def\Im{{\rm Im ~}}

\def\CA{{\cal A}}
\def\CF{{\cal F}}
\def\ad{{\rm ad}}
\def\EE{{\mathcal E}}
\def\PP{\large{\mathscr P}}
\def\DD{\large{\mathscr D}}

\def\RR{\mathcal{R}}

\def\D{{\mathcal D}}
\def\rr{{\sf r}}

\title{More On Gauge Theory And Geometric Langlands}

 \author{Edward Witten }
\affiliation{School of Natural Sciences, Institute for Advanced Study,\\ 1 Einstein Drive, Princeton, NJ 08540 USA}
\abstract{  The geometric Langlands correspondence was described some years ago in terms of $S$-duality of $\N=4$ super Yang-Mills theory.   Some additional
matters relevant to this story are described here.   The main goal is to explain
directly why an $A$-brane of a certain simple kind can be an eigenbrane for the action of 't Hooft operators.  To set the stage, we review some
facts about Higgs bundles and the Hitchin fibration.   We consider only the simplest examples, in which
many technical questions can be avoided.}
\begin{document} \maketitle

\section{Introduction}\label{intro}

Some years ago, it was shown  by A. Kapustin and the author \cite{KW} that the geometric Langlands correspondence (see for example
\cite{frenkel} for an introduction) can be formulated in terms of
$S$-duality of $\N=4$ super Yang-Mills theory in four dimensions.  The basic idea was to consider, in the context of compactification to two dimensions,
 boundary conditions in this theory, 
the action of $S$-duality on those boundary conditions, and the behavior of supersymmetric line operators near a boundary.   
A number of generalizations have been developed subsequently.
The story has been 
extended to encompass tame \cite{GW} and wild \cite{Wild} ramification, by including in the analysis surface operators and line
operators that live on them.  The simplest issues involving geometric Langlands duality for  branes supported at singularities  were explored in 
\cite{FW}.  The study in  \cite{GaiW,GaiW2} of half-BPS boundary conditions and domain walls and their behavior under $S$-duality was
partly motivated by potential applications to geometric Langlands.
 Some of the ingredients that enter in the gauge theory approach to
geometric Langlands are also relevant for understanding Khovanov homology via gauge theory \cite{WittenKhov}.    For informal explanations
of some aspects of gauge theory and geometric Langlands, see \cite{EW1,EW2,EW3}.

The purpose of the present paper is to present some additional topics relevant to this subject.  We consider only the unramified case, so we do not
include surface operators.  Our main goal is to explain concretely (without assuming electric-magnetic duality) why certain simple $A$-branes on the moduli
space $\MH$ of Higgs bundles are ``magnetic eigenbranes,'' as predicted by duality.   As described in \cite{KW}, the branes in question are branes of rank 1 supported
on a fiber of the Hitchin fibration and dual to zero-branes.  In order to show directly that these branes are magnetic eigenbranes, we first
need to review some facts about Higgs bundles and the Hitchin fibration.  

Though we repeat some details to make this paper more nearly self-contained, the reader will probably need to be familiar with parts of the previous paper \cite{KW}.   
Our notation largely follows that previous paper, with some
minor modifications.  In particular, as in \cite{KW}, we write $^L\neg G$ for the Langlands or GNO dual of a compact Lie group $G$ (in some
of the above-cited papers, the dual group is denoted as $G^\vee$).

As a preliminary, we review in sections \ref{compacthitchin} and \ref{dualtori} some basic facts about Higgs bundles and the Hitchin fibration.
None of this material is new, but some  may be relatively inaccessible. (In these sections, we explain a few useful details that are not strictly needed
for our application.) In section \ref{thooftheckeop},  following \cite{KW}, we describe
the concept of an ``eigenbrane,'' describe the electric eigenbranes on $\MH(^L\neg G)$, and describe the predictions of $S$-duality for magnetic
eigenbranes on $\MH(G)$.  The rest of the paper is devoted to showing directly that the branes in question really are magnetic eigenbranes.
We do this only in the simplest cases, that is for those gauge groups and 't Hooft operators for which this is most straightforward.

\section{Compactification And Hitchin's Moduli Space}
\label{compacthitchin} 

\subsection{Preliminaries}\label{prelims}

The basic reason that it is natural to derive the geometric Langlands corresponence
from gauge theory in four dimensions is that $2+2=4$.  As usually formulated mathematically,
the geometric Langlands correspondence relates categories associated respectively to representations of the fundamental group
or to holomorphic vector bundles on a Riemann surface $C$.  From a physical point of view, 
a representation of the fundamental group or a holomorphic vector bundle is described by a {\it gauge field} on $C$.  So we should
expect to do some sort of gauge theory.
However, we must take into account the fact that the geometric Langlands correspondence is
a statement not about numbers, or vector spaces, associated to $C$, but  about {\it categories}.

A $d$-dimensional quantum field theory associates a number -- the partition function or the value
of the path integral -- to a $d$-manifold $M_d$.  To a $d-1$-manifold $M_{d-1}$, it associates
a vector space, the space $\H_d$ of quantum states obtained in quantization of the theory on $M_{d-1}$.   The next step in this hierarchy is slightly less familiar (see for example \cite{Freed,BD,
Kapustin}).  A $d$-dimensional quantum field theory associates a {\it category} to a manifold $M_{d-2}$
of dimension $d-2$.  For example, a $d=2$ field theory associates a category to a 0-manifold, that is, to a point.  Since any two points are isomorphic,
this just means that a two-dimensional field theory determines a category -- the category of boundary conditions.  (The most familiar examples
are probably the categories of $A$-branes and $B$-branes in topologically twisted $\sigma$-models in two dimensions.)

If, therefore, we want a quantum field theory to associate a category to an arbitrary  two-manifold $C$,
we have to start with a quantum field theory in $2+2=4$ dimensions.  If, moreover, the category
is supposed to be associated to gauge fields on $C$, then we should start with gauge fields in 
four dimensions.  
Finally, if we are hoping to find a duality between a category associated in some way to a
gauge group $G$ and a category associated in some way to the Langlands or GNO dual group $^L
\neg G$, then we should start with a theory in four dimensions that has a duality 
that exchanges these two groups.  The gauge theory with the right properties is $\N=4$ super Yang-Mills
theory in four dimensions.

For the application to geometric Langlands, the $\N=4$ theory is topologically twisted so as to produce, formally, a topological field theory in
four dimensions.\footnote{\label{partial} This is very likely only a partial topological field theory.  It is not clear
that its partition function is well-defined in general; to define it, one would have to grapple with
integration over noncompact spaces of zero-modes.  (On a four-manifold $M$, the space of classical minima of the action is the generically noncompact space
of homomorphisms $\pi_1(M)\to G_\C$, up to conjugation.  The noncompactness of this space will cause difficulty in a proof of topological invariance and of cutting and gluing properties
expected in a topological field theory.)  However, observables for which the noncompactness
of the field space do not come into play are well-defined.  In particular, there are well-defined categories of branes, and other observables relevant to geometric Langlands are also
well-defined.  Technically, because $\MH$ is not compact, different conditions on the behavior of a brane at infinity are possible and a correct choice is needed to agree in detail
with standard mathematical formulations of geometric Langlands.}  
For details of this twisting, and now it produces a pair of topological
field theories, the $A$-model and the $B$-model, exchanged by $S$-duality, see \cite{KW}. 

A Riemann surface 
$C$ is introduced simply by considering compactification on $C$ from four to two dimensions.
Thus we consider the four-dimensional $\N=4$ theory on the four-manifold $\R^2\times C$, or
more generally on $\Sigma\times C$, where $\Sigma$ is a two-manifold that is assumed to be much
larger than $C$.  The (partial, as in the footnote) topological field theory on $\Sigma\times C$ reduces
after compactification on $C$ to a topological field theory on $\Sigma$.    

In studying this theory
on $\Sigma$, we can always assume that $\Sigma$ is much larger than $C$ and work at large
distances on $\Sigma$.  This means, for many purposes, that we can replace the full four-dimensional
theory on $\Sigma\times C$ by a two-dimensional theory on $\Sigma$ \cite{BJSV,HMS}.   This theory is a supersymmetric $\sigma$-model whose target space is the space of classical supersymmetric
vacua that arise in compactification on $C$.  Away from singularities\footnote{In this statement,
we assume that either the genus $g_C$ of $C$ is at least 2, or  ramification is included in genus 0 or 1.
In genus 0 or 1, in the absence of ramification, the solutions of Hitchin's equations have abelian
structure group and a more careful formulation is needed.  Also, we assume that $G$ is simple or semi-simple.
For $G=U(1)$ or $U(N)$, matters are different as there is always an unbroken $U(1)$ subgroup of $G$.  See section
\ref{elmag} for a discussion of this point.} at which some of the gauge symmetry is restored, this space is the Hitchin hyper-Kahler moduli space of ``Higgs bundles'' on $C$ with structure
group $G$.   We call this space $\MH$, or $\MH(G,C)$
if we wish to specify the gauge group $G$ and the Riemann surface $C$ on which we have compactified.  

Although a description by a $\sigma$-model with target $\MH$ is very useful, this description is
really only generically valid. That is because of the singularities of  $\MH$ at which
some of the gauge symmetry becomes restored and additional degrees of freedom must be included
to give a good description.   Exactly how many additional degrees of freedom must be included for a good description depends on which question one asks.  The universal description, valid for all purposes
without any approximation, is the underlying four-dimensional gauge theory on $\Sigma\times C$,
with no attempt to reduce to an effective two-dimensional theory.   If one wants to describe this complete four-dimensional theory 
as a two-dimensional $\sigma$-model, one should call it a two-dimensional supersymmetric
$\sigma$-model
with target space the space $\Y$ of all gauge fields on $C$ (or more precisely, as we discuss later, the corresponding cotangent bundle $\CW=T^*\Y$ to
include the Higgs field), and gauge group the infinite-dimensional
group $\G$ of all maps
from $C$ to the finite-dimensional group $G$.   If one topologically twists this $\sigma$-model to make
either an $A$-model or a $B$-model, then $\G$ is effectively replaced by its complexification, the
group $\G_\C$ of maps from $C$ to the complexification $G_\C$ of $G$.

To make contact between this picture and standard statements in geometric Langlands, we should
specialize to either the $A$-model or the $B$-model side of the duality.  For brevity, we consider
here the $A$-model  side; analogous statements hold for the $B$-model.  
Algebraic geometers formulate the $A$-model in terms of the ``stack'' of all holomorphic $G$ bundles
over $C$, not necessarily stable.\footnote{Thus according to \cite{BeD}, on one side of the duality, one must consider $\mathcal D$-modules (modules
for the sheaf of differential operators)
on the stack of $G$-bundles (and not just on the moduli space of stable $G$-bundles).  The relation of the $A$-model to $\mathcal D$-modules depends on the
fact that the moduli space of Higgs bundles can be approximated by a cotangent bundle 
and is explained in \cite{KW}.}  However, according to Atiyah and Bott \cite{AB}, a concrete model of this stack is the
infinite-dimensional space $\Y$ acted on by the infinite-dimensional group $\G_\C$.  Thus a physicist's interpretation of
the statement that a complete description involves the full ``stack'' is just to say that a full description involves a supersymmetric $\sigma$-model
with target the full infinite-dimensional space $\Y$ (acted on by $\G_\C$) or in other words the full four-dimensional gauge theory.

\subsection{Hitchin's Equations}\label{hitcheq}

In this paper, however, we largely consider  questions for which it is adequate to consider a $\sigma$-model with target $\MH$.
$\MH$ is defined by a familiar system of equations known as Hitchin's equations.  These are equations for a pair $A,\phi$,
where $A$ is a connection on a $G$-bundle $E$ over the two-dimensional surface $C$, and $\phi$ is a one-form on $C$ with
values in the adjoint representation, that is, in the adjoint bundle $\ade$ associated to $E$.  In writing these equations, $C$ is
assumed to be a Riemann surface, with a chosen complex structure. Introducing a local complex
coordinate $z$ on $C$, the equations can be written 
\begin{align}
\label{secondway} F_{z\bar z}-[\phi_z,\phi_{\bar z}]&=0\nonumber
\\    D_{\bar z}\phi_z=D_z\phi_{\bar z}&=0. \end{align}
Here $D_{\bar z}=\partial_{\bar z}+[A_{\bar z},\cdot],$ $D_z=\partial_z+[A_z,\cdot]$.  

 Alternatively, we can combine $A$ and $\phi$ to the complex-valued connection
 $\CA=A+i\phi$.   We view this as a connection on a $G_\C$-bundle $E_\C\to C$ that
 is obtained by complexifying $E\to C$.  It has structure group $G_\C$ and curvature
 $\F=d \A+\A\wedge \A$.
 The real and imaginary parts of $\F$ are  ${\rm 
Re}\,\CF=F-\phi\wedge\phi$, ${\rm Im}\,\CF=D\phi$. 
Hitchin's equations are equivalent to 
\begin{align}\label{firstview}
 \F & = 0 \nonumber \\
                     D\star \phi&=0. \end{align}
Here $D\star\phi=D_{\bar z}\phi_z+D_z\phi_{\bar z}$.            
                     
These two ways to write Hitchin's equations lead to two different complex structures on the moduli space $\MH$ of solutions of
those equations.      Starting from eqn. (\ref{firstview}), we observe that the equation $\F=0$ is invariant under complex-valued gauge transformations of $\A$,
and that modulo such a gauge transformation, the space of solutions is the space of equivalence classes of homomorphisms $\pi_1(C)\to G_\C$, up to
conjugation.  In particular,
the equation $\F=0$ is topologically-invariant; it does not really depend on the complex structure on $C$.
A classic theorem \cite{Corlette} says that given a mild  and generically valid condition of semi-stability,\footnote{This condition says that
if the monodromies of the flat connection $\A$ can be simultaneously conjugated to a block-triangular form $\begin{pmatrix} * & * \cr 0 & * \end{pmatrix}$,
then in fact they can be conjugated to the block-diagonal form   $\begin{pmatrix} * & 0 \cr 0 & * \end{pmatrix}$.  This condition is generically satisfied
since (for $C$ of genus $\geq 2$) generically the monodromies of a complex flat connection on $C$ cannot be conjugated to a block-triangular form.} it is equivalent
to classify complex flat connections on $C$  up to $G_\C$-valued gauge transformations, or to impose the ``moment map'' equation $D\star\phi=0$ and
divide only by $G$-valued gauge transformations. 

Thus, in one perspective, $\MH$ can be interpreted as the moduli space of semi-stable homomorphisms $\pi_1(C)\to G_\C$.  Since $G_\C$ is itself
a complex manifold, this interpretation endows $\MH$ with a complex structure.  This complex structure has been called $J$ by Hitchin \cite{Hitchin}.     
                     
In another perspective, we view $A_{\bar z}$ and $\phi_z$ as holomorphic variables (and their complex conjugates $A_z$ and $\phi_{\bar z}$ as antiholomorphic
variables).  Hitchin's equations (\ref{secondway}) thus split up as a holomorphic equation $D_{\bar z}\phi_z=0$ (which implies the complex conjugate equation
$D_z\phi_{\bar z}=0$) and the ``moment map'' equation $F-\phi\wedge\phi=0$.   To interpret the holomorphic equation, we observe first that for any $G$-valued
connection $A$ on a bundle $E\to C$, once a complex structure is picked on $C$,
the corresponding $\bar\partial$ operator $D_{\bar z}$ endows $E$ with a holomorphic structure.  The equation $D_{\bar z}\phi_z=0$ says that in this
complex structure, $\phi_z$ can be viewed as a holomorphic section of $K\otimes \ade$, where $K$ is the canonical bundle of $E$.  A Higgs bundle,
in Hitchin's terminology, is a pair consisting of a holomorphic $G_\C$-bundle $E\to C$ and a holomorphic ``Higgs field'' $\varphi=\phi_z d z\in H^0(C,
K\otimes \ade)$, or in other words a solution of $D_{\bar z}\phi_z=0$. We will call such a pair $(E,\varphi)$ a Hitchin pair.

$G_\C$-valued gauge transformations act naturally on the space of pairs $A_{\bar z}$, $\phi_z$ that obey $D_{\bar z}\phi_z=0$.  However, Hitchin proves
that -- given again a mild condition of semistability\footnote{\label{semicon} For simplicity, we will state this condition only for $G=SU(N)$.  In this case, the holomorphic
$G_\C$-bundle $E_\C\to C$ is semi-stable if any subbundle has non-positive first Chern class. A Higgs bundle $(E,\varphi)$ is semi-stable if 
any $\varphi$-invariant subbundle has non-positive first Chern class. (The concept of $\varphi$-invariance is as follows.  As $\varphi$ is a section of
$K\otimes \ade$, it can be understood naturally as a map $E\to K\otimes E$.  Picking a local trivialization of $K$, this gives a map $E\to E$, and
a subbundle of $E$ is $\varphi$-invariant if it is invariant under this map.  This criterion does not depend on the choice of local trivialization of $K$.)
 In particular, if $E$ is semi-stable, then $(E,\varphi)$ is semi-stable for any
$\varphi\in H^0(C,K\otimes\ade)$.  In these statements, to replace ``semi-stable'' by ``stable,'' one merely has to replace ``non-positive'' by ``negative.''
As a matter of terminology, we should note that what is usually called the moduli space of stable objects of any given kind is usually defined to
parametrize objects of the given type that are stable or semi-stable.  This is why one calls $\MH$ the moduli space of stable Higgs bundles, even
though in general some of the objects that it parametrizes are only semi-stable.} -- 
it is equivalent to divide the space of such pairs by $G_\C$-valued gauge transformations or to impose
the ``moment map'' equation  $D\star\phi=0$ and divide only by $G$-valued gauge transformations.

Thus, from this point of view, $\MH$ acquires another interpretation as the moduli space of stable Higgs bundles.  This description makes manifest
another complex structure
that Hitchin calls $I$.  Hitchin shows, moreover, that the complex structures $I$ and $J$ form part of a hyper-Kahler structure on $\MH$, with in particular
an action on the tangent space to $\MH$ at any smooth point of the quaternion units $I$, $J$, and $IJ=K$.             

Many aspects of this picture that are relevant to geometric Langlands have been described in \cite{KW} and will not be repeated here.  
About the statement that $\MH$ is hyper-Kahler, we remark only that it can be constructed as a hyper-Kahler quotient.  The space $\CW$ of
pairs $A,\phi$ can be regarded as an infinite-dimensional hyper-Kahler manifold with flat hyper-Kahler metric
\begin{equation}
\label{igob}
ds^2=-{1\over
2\pi}\int_C|d^2z|\,\Tr\bigl(\delta A_z\otimes \delta A_{\bar
z}+\delta \phi_z\otimes
\delta \phi_{\bar z}\bigr),~~~|d^2z|=i\,dz\,d\bar z.
\end{equation}
The group $\G$ of $G$-valued gauge transformations acts on $\CW$, preserving its hyper-Kahler structure.  Hitchin's equations assert the vanishing
of the hyper-Kahler moment map for the action of $\G$ on $\CW$.  The space $\MH$ of solutions of Hitchin's equations, modulo $\G$, can thus be interpreted
as the hyper-Kahler quotient $\CW///\G$.  As such it naturally carries a hyper-Kahler structure.

We will focus here on aspects of the picture in complex structure $I$ that were not explained in \cite{KW} and
that are important in a deeper understanding of geometric
Langlands.  For example, we will use some of this understanding later in discussing magnetic eigenbranes. 

Since we will concentrate  on just one complex structure,
namely $I$, we will not see the full hyper-Kahler structure of $\Sigma$.  However, there is an important part of the structure that is quite visible in complex
structure $I$.  In general, a hyper-Kahler manifold $X$ has a three-dimensional space of real symplectic structures.  In any one of the complex structures on
$X$, two of the three real symplectic structures can be combined as the real and imaginary parts of a holomorphic symplectic structure.  In the case of the
complex structure $I$ on $\MH$, the holomorphic symplectic structure is
\begin{equation}\label{zolk}\Omega_I=\frac{1}{\pi}\int |d ^2z|\,\Tr\,\delta\phi_z\wedge\delta A_{\bar z}. \end{equation}
Here $-\Tr$ is an invariant quadratic form on the Lie algebra $\frak g$ of $G$.  (We will use a standard normalization in which short roots
have length squared 2.) 
In complex structure $I$, the hyper-Kahler
quotient $\CW///\G=\MH$ can
be understood as a complex symplectic quotient, that is a symplectic quotient with respect to the action of
$\G_\C$ on $\CW$, viewed as a complex symplectic manifold with symplectic form $\Omega_I$.   (Such a symplectic quotient is defined by setting to zero the complex moment
map, which in the present case is $\nu_I=D_{\bar z}\phi_z$, and dividing by the complex symmetry group, here $\G_\C$.  In the present instance, this is the
operation that produces $\MH$.)

We can write $\Omega_I=\omega_J+i\omega_K$, where $\omega_J$ and $\omega_K$ are real symplectic forms
that are Kahler forms for complex structure $J$ and $K$ respectively.  In geometric Langlands, one is primarily interested
in the $A$-model with symplectic form $\omega_K$; we call this more briefly the $A$-model of type $K$.

\subsection{$\MH$ and the Cotangent Bundle}
\label{hitchfib} 

To begin our more detailed study of the geometry of $\MH$ in
complex structure $I$, recall first that a Hitchin pair $(E,\varphi)$, where $E\to C$ is a holomorphic $G$ bundle and $\varphi\in
H^0(C,K\otimes\ade)$,  is stable or semi-stable if the underlying bundle $E$ is stable or semi-stable.  (The converse is not true.)
So in particular, if $E$ is a stable bundle, the Hitchin pair $(E,0)$ is always stable.  This gives a natural embedding
 of $\M$, the moduli
space of stable $G$-bundles on $C$, into the Hitchin moduli space
$\MH$.  $\M$ is a holomorphic submanifold of $\MH$ in complex
structure $I$, since it is defined by the equation $\varphi=0$,
which is holomorphic in complex structure $I$.  (In complex
structures $J$ and $K$, $\M$ is not holomorphic.  But it is Lagrangian for $\Omega_I$ and hence
also for $\omega_J$ and $\omega_K$.)

If $E$ is stable, then the Hitchin pair $(E,\varphi)$ is stable
for every $\varphi\in H^0(C,K\otimes\ad(E))$.  The tangent space
to $\M$ at the point represented by $E$ is $H^1(C,\ad(E))$, and by
Serre duality, the dual of this, or in other words the cotangent
space to $\M$ at the point $E$, is $H^0(C,K\otimes\ad(E))$.
Since $\varphi$ takes values in this  space, it follows that  the
space of all pairs $(E,\varphi)$ with stable $E$ is the cotangent
bundle $T^*\M$. We thus actually get an embedding of $T^*\M$ in
$\MH$.    The holomorphic symplectic form $\Omega_I$ of $\MH$ in complex
structure $I$ restricts on $T^*\M$ to its natural symplectic
structure as a holomorphic cotangent bundle.  

The image of $T^*\M$ in $\MH$ is not all of $\MH$ because a
Hitchin pair $(E,\varphi)$ may be stable even if $E$ is
unstable.\footnote{For $G=SU(2)$, this happens if $E$ has a
holomorphic line sub-bundle ${\cal L}$ of positive first Chern
class (so $E$ is not stable), but ${\cal L}$ is not
$\varphi$-invariant (so $(E,\varphi)$ is stable). We give examples
in section \ref{distinguished}.} However, the stable Hitchin pairs
$(E,\varphi)$ for which $E$ is unstable are of sufficiently high
codimension to be unimportant for many applications. Upon throwing
away this set, $\MH$ becomes isomorphic to $T^*\M$, and has a
natural map to $\M$ by forgetting $\varphi$.  (This map is used in \cite{KW}
in understanding the relation of $A$-branes of type $K$ on $\MH(G,C)$ to $\mathcal D$-modules on $\M(G,C)$.)

Instead of making a  projection from $\MH$ to $\M$, which is only
generically defined, we can consider the foliation of  $\MH$ by
the holomorphic type of the bundle $E$ that underlies a stable
Hitchin pair $(E,\varphi)$.  This foliation is defined throughout
the smooth part of $\MH$.  It has middle-dimensional leaves, which
are Lagrangian with respect to the complex symplectic structure
$\Omega_I$.

\subsection{The Hitchin Fibration}
\label{hitchfibr}

\def\calP{{\mathcal P}}
What is usually called the Hitchin fibration is not the map that sends $\varphi$ to 0
but another map that sends a Hitchin pair $(E,\varphi)$ to the characteristic polynomial
of $\varphi$.  
 For $G=SU(2)$, this characteristic polynomial is simply the quadratic differential  $w=\Tr\,\varphi^2$.  $w$ is
 holomorphic since $\varphi$ is, so it takes values in  ${\cal
V}=H^0(C,K^2)\cong \Bbb{C}^{3g-3}$. The Hitchin fibration is the map $\ppi:\MH\to {\cal V}$
that sends $(E,\varphi)$ to $w=\Tr\,\varphi^2$.

For any $G$, the Hitchin fibration is defined similarly, incorporating all of the independent
Casimirs of $G$, and not just the quadratic Casimir. For
example, for $G=SU(N)$, we  define $w_n=\Tr\,\varphi^n\in H^0(C,K^n)$,
$n=2,\dots,N$,  and let ${\cal V}=\oplus_{n=2}^NH^0(C,K^n)$. The
Hitchin fibration is then defined to take  $(E,\varphi)$ to $(w_2,w_3,\dots,w_n)\in {\cal V}$. For any 
Lie group $G$ of rank $r$, the ring of invariant polynomials on the Lie algebra
$\frak g$ is a polynomial algebra with $r$ generators
 ${\cal P}_i$. The degrees $d_i$ of these polynomials obey
 \begin{equation}
\label{yty}
\sum_i(2d_i-1)={\rm dim}(G).
\end{equation}
For example, for $G=SU(N)$,  for the ${\cal P}_i$ we can take
the polynomials $\Tr\,\varphi^n$, $n=2,\dots, N$, of degree $n$, so that the identity (\ref{yty})
becomes $\sum_{n=2}^N(2n-1)=N^2-1={\rm
dim}(G)$. For any $G$, the   Hitchin fibration is defined to take $(E,\varphi)$ to the collection of invariant polynomials $\calP_i(\varphi)\in H^0(C,K^{d_i})$. So the base of the Hitchin fibration is ${\cal V}=\oplus_i
H^0(C,K^{d_i})$.

Since $\dim \,H^0(C,K^d)=(2d-1)(g-1)$, it follows from
(\ref{yty}) that the complex dimension of ${\cal V}$ is $(g-1){\rm
dim}(G)$, which equals the dimension of $\M$, and  one-half of the
dimension of $\MH$. The Hitchin fibration $\ppi:\MH\to {\cal V}$
is surjective, as we will discuss momentarily.  
A generic fiber $\FF$ of the Hitchin fibration therefore also
has half the dimension of $\MH$:
\begin{equation}
\label{yrof}
\dim\,\FF=\dim\,{\cal V}={1\over
2}\dim\,\MH=(g-1)\dim\,G.
\end{equation}

In section \ref{distinguished}, we will construct explicitly, for each point $w\in {\cal
V}$, a stable Hitchin pair $(E,\varphi)$ that projects to $w$
under the Hitchin fibration. In the meantime, we give a more
qualitative argument that the Hitchin fibration is surjective.
 For example, take $G=SU(2)$. Pick a stable $SU(2)$ bundle $E$, and look for
 a Hitchin pair $(E,\varphi)$ that maps to a given point in $\V$, defined by a quadratic
 differential $w$.  For this we need to find $\varphi\in H^0(C,K\otimes \ad(E))$ with
 $\Tr\,\varphi^2=w$.  That is a system of $3g-3$ equations for $3g-3$ unknowns
 so the generic number of solutions is $2^{3g-3}$.  A similar counting is possible for other
 $G$.

\subsection{Complete Integrability}
\label{complete}

One of Hitchin's main results \cite{HitchinDuke} is the statement that $\MH$ is a completely integrable Hamiltonian
system in the complex structure $I$.  In fact, we can find
${1\over 2}{\rm dim}\MH$ functions $H_a$ on $\MH$ that are
holomorphic in complex structure $I$, are algebraically
independent, and commute in the Poisson brackets
obtained from the holomorphic symplectic form
$\Omega_I$.\footnote{By taking a real slice of $\MH$, one can extract from this construction
a conventional integrable system with a real phase space and real commuting Hamiltonians.  See \cite{DonagiMarkman} for some examples and references.}

In fact, we can take the $H_a$ to be linear functions on ${\cal
V}$, since the dimension of ${\cal V}$ is the same as the desired
number of functions.   For $G=SU(2)$, we simply begin by picking  a basis $\alpha_a$,
$a=1,\dots,3g-3$ of the $(3g-3)$-dimensional space $ H^1(C,T)$,
which is dual to $H^0(C, K^2)\cong {\cal V}$. (Here $T$ is the
holomorphic tangent bundle to $C$.) We represent $\alpha_a$ by
$(0,1)$-forms valued in $T$, which we also denote as $\alpha_a$,
and we define
\begin{equation} \label{miro} H_a=\int_C\alpha_a \wedge
\Tr\,\varphi^2.
\end{equation}
 We
claim that these functions are Poisson-commuting with respect to
the holomorphic symplectic form $\Omega_I$.

A natural proof uses the fact that the definition of the $H_a$
makes sense on the infinite-dimensional space $\CW$, before taking
the hyper-Kahler quotient.  Using the symplectic structure
$\Omega_I$ on $\CW$ to define Poisson brackets, the $H_a$ are
obviously Poisson-commuting. For in these Poisson brackets, given
the form (\ref{zolk}) of $\Omega_I$, $\varphi_z$ has vanishing
Poisson brackets with itself (its Poisson brackets with $A_{\bar
z}$ are of course nonzero). But the $H_a$ are functions of
$\varphi_z$ only, not $A_{\bar z}$.

The $H_a$ can be restricted to the locus with $0=D_{\bar z}\phi_z$, and then,
because they are invariant under the $G_{\Bbb{C}}$-valued gauge
transformations, they descend to holomorphic functions on $\MH$. A
general property of symplectic reduction (in which one sets to
zero a moment map, in this case $\nu_I=D_{\bar z}\phi_z$, and then divides by the
corresponding group, in this case the group of
$G_{\Bbb{C}}$-valued gauge transformations) is that it maps
Poisson-commuting functions to Poisson-commuting functions.  So
the $H_a$ are Poisson-commuting as functions on $\MH$.  There are
enough of them to establish the complete integrability of $\MH$.

The generalization from $G=SU(2)$ to arbitrary $G$ is made by
simply using all the independent gauge-invariant polynomials
${\cal P}_i$, not just $\Tr\,\varphi^2$, to define the
Hamiltonians. The commuting Hamiltonians are now
$H_{a,i}=\int_C\alpha_{a,i}{\cal P}_i(\varphi)$, where now
$\alpha_{a,i}\in H^1(C,K^{1-d_i})$. These commuting Hamiltonians
are the full set of linear functions on the base ${\cal V}$ of the
Hitchin fibration, and equal in number to one-half the dimension
of ${\cal M}_H$.

In this construction, we started with a 
particular choice of $(0,1)$-forms $\alpha_{a,i}$ that are used to
construct the commuting Hamiltonians.  But after restricting and descending to $\MH$, the functions we get on $\MH$
depend only on the cohomology classes of the $\alpha_a$. In fact,
once we have $\bar D\varphi=0$ and hence $\bar\partial {\cal
P}_i(\varphi)=0$, a simple integration by parts shows that the
$H_{a,i}$ are invariant under
$\alpha_{a,i}\to\alpha_{a,i}+\bar\partial\epsilon_{a,i}$.

The Poisson-commuting functions $H_a$ generate commuting flows on
$\MH$ that are holomorphic in complex structure $I$. Moreover,
these flows commute with the Hitchin fibration -- since the
commuting Hamiltonians are precisely the functions on the base of
this fibration.  So the flows act on the fibers of the Hitchin
fibration, and in particular those fibers admit a maximal set of
commuting flows.

Complex tori admit such a maximal set of commuting flows, and one
might surmise that the orbits generated by the $H_a$ are complex
tori at least generically. This follows from general arguments
given the ``properness'' of the Hitchin fibration (the compactness
of the fibers), but we will demonstrate it more directly,
following \cite{HitchinDuke}, by using the theory of the spectral curve.

One easy and important consequence of complete integrability is
that the fibers of the Hitchin fibration are Lagrangian
submanifolds in the holomorphic symplectic structure $\Omega_I$.  Indeed, a fiber of this fibration is defined by
equations $H_{a,i}-h_{a,i}=0$, where $H_{a,i}$ are the commuting
Hamiltonians and $h_{a,i}$ are complex constants.  In general, a middle-dimensional
submanifold defined by the vanishing of a
 collection of Poisson-commuting
functions, such as $H_{a,i}-h_{a,i}$ in the present case, is
Lagrangian.

\subsection{The Spectral Curve}
\label{spectral}

\subsubsection{Basics}

To describe the idea of the spectral curve,\footnote{In understanding the following material, I was greatly assisted by explanations by R. Donagi
in the period 2004-6.}  let us consider the
case $G=SU(N)$. We think of $E$ as a rank $N$ complex vector
bundle. Because $\varphi$ takes values in the adjoint
representation, we can think of it locally as an $N\times N$
matrix of holomorphic one-forms -- which we can take to act on the
fiber of $E$. Then, fixing a point $\pp\in C$, and denoting as $\uppsi$ an element of
the fiber of $E$ at $\pp$, we can consider the
``eigenvalue problem''
\begin{equation}
\label{udu}
\varphi(\pp)\uppsi=y\uppsi.
\end{equation}
 Since the 
matrix elements of   $\varphi(\pp)$ take values in $K|_\pp$ -- the
fiber at $\pp$ of the canonical bundle $K$ -- we cannot interpret
$y$ as a number. But the eigenvalue problem makes sense if we
interpret $y$ as an element of $K|_\pp$.

We know how to find the eigenvalues of an $N\times N$ matrix. They
are the zeroes of the characteristic polynomial.  Thus, they obey
the equation
\begin{equation}
\label{bludu}
\det(y-\varphi)=0.
\end{equation}
 For generic $\varphi$
and $\pp$, the equation (\ref{bludu}) has $N$ distinct roots.  For $N=2$,
the equation simplifies. Since the $2\times 2$ matrix $\varphi$ is
everywhere traceless, the equation reduces to
\begin{equation}
\label{tudlu}
y^2-{1\over 2}\Tr\,\varphi^2=0.
\end{equation}

So far we have presented this at a single point $\pp\in C$, but
obviously we can consider the equation for all $\pp$.  As $\pp$
varies, the $N$ roots of (\ref{bludu}) sweep out an algebraic
curve $D$ which is an $N$-fold cover of $C$. $D$ maps to $C$ by
``forgetting'' $y$.  We let $W$ be the algebraic surface which is
the total space of the line bundle $K$ over $C$.  The curve $D$
naturally lies in the surface $W$.  We can think of $y$ as
parametrizing the fiber of the fibration $W\to C$. The equation
$\det(y-\varphi)=0$ singles out $N$ points in each fiber, making
up the spectral curve $D$, with its $N$-fold covering map
$\psi:D\to C$.

Now let us consider the problem of describing the fiber $\FF$ of the
Hitchin fibration  $\ppi:\MH\to {\cal V}$.  Choosing a particular
fiber $\FF$ means making a particular choice of $\Tr\,\varphi^2$
(for $SU(2)$) or of the characteristic polynomial
$\det(y-\varphi)$ of $\varphi$ (for any unitary group). Hence the choice
of a fiber determines a particular spectral curve $D$; every fiber
is associated with its own $D$. If we pick a sufficiently generic
fiber, the curve $D$ is smooth and irreducible. For example, for
$SU(2)$, this is so precisely if the zeroes of the quadratic
differential $\Tr\,\varphi^2$ are distinct.

These zeroes are the branch points of the cover $D\to C$. A
quadratic differential on $C$ has $4g_C-4$ zeroes.  The double
cover of a curve of genus $g_C$ branched over $4g_C-4$ points has
genus $4g_C-3$. To verify this, recall that a curve of genus $g_C$
has Euler characteristic $2-2g_C$. If $C$ is a curve of genus
$g_C$, and $D$ is a double cover of $C$ with $4g_C-4$ branch
points, then $D$ has Euler characteristic $2(2-2g_C)-(4g_C-4)$.
This is the same as $2-2g_D$ with $g_D=4g_C-3$, so that is the genus
of $D$.

For $SU(N)$, let $\PP(y)=\det(y-\varphi)$. The ``discriminant'' of
a polynomial $\PP$ with roots $\lambda_i$ is
$\prod_{i<j}(\lambda_i-\lambda_j)^2$. In the present context, the
discriminant is a gauge-invariant polynomial $\Delta(\varphi)$
homogeneous of degree $N(N-1)$, and thus an element of
$H^0(C,K^{N(N-1)})$. $D$ is smooth and irreducible if
$\Delta(\varphi)$ has only simple zeroes (and more generally if
and only if $\varphi$ always has only a single Jordan block for
each eigenvalue).  A zero of the discriminant is the same as a
point on $C$ over which there is a  point on $D$ at which
$\PP(y)=\PP'(y)=0$. The equation $\PP(y)=0$ defines $D$, and the
points on $D$ with $\PP'(y)=0$ are exactly the ramification points
of the cover $D\to C$. The zero set of $\PP'(y)$ thus defines a
divisor ${\cal R}$ on $D$ that we will call the ramification
divisor. It consists generically of $2N(N-1)(g_C-1)$ distinct points (one for each zero
of $\Delta(\varphi)\in H^0(C,K^{N(N-1)})$), and this implies
that the genus of $D$ is $g_D=g_C+(N^2-1)(g_C-1)=g_C+\dim({\cal
V})=g_C+\dim(\FF)$.   If the ramification points are distinct, this suffices to ensure
that $D$ is smooth.

Now we will explain a key fact: the eigenvectors of $\varphi$
furnish a line bundle over $D$. Suppose that a Hitchin pair
$(E,\varphi)$ is given. For some $\pp\in C$, we find a root $y$ of
the characteristic equation (\ref{bludu}). This corresponds to a
point $\qq\in D$ -- one of the $N$ points $\qq_1,\dots, \qq_N\in D$ that
lie above $\pp$.  For each such $y$, the eigenvector problem
(\ref{udu}) has a solution for some nonzero $\uppsi$. To be more precise, assuming the roots of (\ref{bludu}) are distinct, the eigenvector problem
for given $y$ and thus given $q\in D$ has a one-dimensional space of solutions that we will call ${\cal
N}_\qq$. As $\qq$ varies, ${\cal N}_\qq$ varies as the fiber of a
complex line bundle ${\cal N}$ over $D$.

Does the definition of the line bundle ${\cal N}$ break down when
the roots fail to be distinct?  When the curve $D$ is smooth, it
does not.  Let us explain this just for $G=SU(2)$, the general
case being similar. The roots fail to be distinct precisely where
$\Tr\,\varphi^2=0$. We pick a local complex coordinate $z$ on $C$
and assume that $\Tr\,\varphi^2=0$ at $z=0$. As we require
$\Tr\,\varphi^2$ to have simple zeroes, the behavior of $\varphi$
near $z=0$ is, up to conjugacy,
\begin{equation}
\label{ygor}
\varphi\sim
\begin{pmatrix}0 & z\\ 1 & 0\end{pmatrix}.
\end{equation}
 The spectral curve is
given near $z=0$ by
\begin{equation}
\label{jygor}
0=\det(y-\varphi)=y^2-z.
\end{equation}
 The
solutions of this equation are given, for given $z$, by $y=\pm
\sqrt{z}$.  However, a better way to describe the spectral curve
near $z=0$ is to regard $y$ as a local parameter, with $z=y^2$.
Plugging $z=y^2$ into the local formula (\ref{ygor}) for $\varphi$, the
eigenvalue equation (\ref{udu}) becomes
\begin{equation}
\label{zygor}\begin{pmatrix}0 &
y^2\\ 1 & 0\end{pmatrix}\begin{pmatrix}\psi_1
\\ \psi_2\end{pmatrix}=y\begin{pmatrix}\psi_1 \\ \psi_2\end{pmatrix}.
\end{equation}
This has, for every $y$, a one-dimensional space of solutions,
generated by
\begin{equation}
\label{calleds}
s=\begin{pmatrix}\psi_1
\\ \psi_2\end{pmatrix}=\begin{pmatrix}y \\ 1\end{pmatrix},
\end{equation}
 showing
that the definition of the line bundle ${\cal N}$ works perfectly
well at simple branch points of the spectral curve $D$.

In effect, we have defined ${\cal N}$ as the kernel of
$\varphi-y$.  We can define a related line bundle ${\cal L}$ as
the {\it cokernel} of $\varphi-y$.  As long as the eigenvalues of
$\varphi$ are distinct, the linear transformation $\varphi-y$,
regarded as a map from sections of $E\otimes K^{-1}$ to sections
of $E$, has a one-dimensional kernel and therefore an
$(N-1)$-dimensional image $E'=\Im(\varphi-y)$.  The quotient
${\cal L}=E/E'$ is therefore one-dimensional away from the
ramification points, so this gives us another holomorphic  line
bundle away from those points.

We can use the above local model to verify that ${\cal L}$
naturally extends as a line bundle over the ramification points.
In the local model,
\begin{equation}
\label{tygor}
\varphi-y=\begin{pmatrix}-y & z \\
1 & -y\end{pmatrix},
\end{equation}
 so the image $E'$ of $\varphi-y$ is generated by
\begin{equation}
\label{defu}
u=\begin{pmatrix}-y \\ 1\end{pmatrix}.
\end{equation}
  Hence the
quotient ${\cal L}=E/E'$ is generated by
\begin{equation}
\label{timely}
t=\begin{pmatrix}1\\ 0\end{pmatrix},
\end{equation}
 that is, $u$ and
$t$ give a basis for every $y$.  This exhibits a natural extension
of ${\cal L}$ across the point $z=0$.

There is also an important relationship between ${\cal L}$ and
${\cal N}$.  We defined ${\cal N}$ as the subbundle of $E$ (or
more precisely, of $\psi^*(E)$, the pullback of $E$ to $D$)
annihilated by $\varphi-y$, so there is a natural embedding
$i:{\cal N}\to \psi^*(E)$.  Also, ${\cal L}$ is a quotient of
$\psi^*(E)$ given by reducing mod $E'=\Im(\varphi-y)$, so there is
a natural map $r:\psi^*(E)\to {\cal L}$. The composition
\begin{equation}
\label{huty}
{\cal N}\underarrow{i} \psi^*E\underarrow{r} {\cal L}
\end{equation}
gives a holomorphic map $\theta=ri:{\cal N}\to {\cal L}$.

 Away from ramification points,
$\varphi-y$ can be block-diagonalized with a 1-dimensional kernel
and an invertible block of codimension 1:
\begin{equation}\label{zelbo}\varphi-y=\begin{pmatrix} 0 & 0\cr 0&*\end{pmatrix}.\end{equation}
So the kernel and cokernel are generated by the upper element
(the zero-mode) and the map $\theta:\cal N\to \cal L$ is an isomorphism.

To see what happens at a ramification point, we again use the
local model for $N=2$.  ${\cal N}$ is generated by the vector called $s$ in
(\ref{calleds}).  To evaluate $\theta(s)$, we just reduce mod $u$,
defined in (\ref{defu}), and express the result as a multiple of the
generator $t$ of ${\cal L}$, found in (\ref{timely}).  We have $s=u+2yt$,
or
\begin{equation}
\label{burble}
s = 2yt ~{\rm mod}~ u.
\end{equation}
 So a generator $s$ of
${\cal N}$ maps to $2y$ times a generator of ${\cal L}$. As $y$
has a simple zero at the ramification divisor ${\cal R}$, the
result is that
\begin{equation}
\label{troygo}
{\cal L}\cong {\cal N}({\cal R}),
\end{equation}
 in
other words a section of ${\cal L}$ is equivalent to a section of
${\cal N}$ that may have a pole at the ramification divisor.

The ramification divisor, as noted above, is precisely the zero
set on $D$ of the polynomial $\PP'(y)$, which is a section of
$\psi^*(K^{N-1})$ (since $y$ is a section of $\psi^*(K)$, and
$\PP'(y)$ is of degree $N-1$ in $y$). So we can alternatively
write
\begin{equation}
\label{roygo} {\cal L}\cong {\cal N}\otimes \psi^*(K^{N-1}).
\end{equation}

\subsubsection{Abelianization}

The key insight is now that one can reconstruct the bundle $E\to
C$ (and the Higgs field $\varphi$) from the line bundle ${\cal
L}\to D$. This represents progress, because $E$ is a $G$-bundle
where $G$ is {\it nonabelian}, while a line bundle is the analog
of a $G$-bundle with $G$ replaced by the abelian group $U(1)$. The
price we pay for this abelianization is that instead of working on
the curve $C$, we have to work on its spectral cover $D$.

Suppose first that a point $\pp\in C$ is not a branch point of the
fibration $\psi:D\to C$.  Then the fiber of $E$ at $\pp$, which we
denote as $E_\pp$, can be decomposed as a sum of the $N$
one-dimensional eigenspaces of $\varphi$.  These eigenspaces are
the fibers  of what we have called ${\cal N}$ at the points $\qq_i$
that lie above $\pp$.  Thus, as long as $\pp$ is not a branch point,
we get
\begin{equation}
\label{ytot}
E_\pp=\oplus_{i=1}^N{\cal N}_{\qq_i}.
\end{equation}
  This
certainly shows that we can recover $E$ from ${\cal N}$ away from
the branch points.  Since ${\cal N}$ and ${\cal L}$ are naturally
isomorphic away from the branch points, we can equally well write
\begin{equation}
\label{bytot}
E_\pp=\oplus_{i=1}^N{\cal L}_{\qq_i}
\end{equation}
 away from branch
points.  This description turns out to extend more simply over the
branch points.

The extension of this formula over the branch points involves
another notion in algebraic geometry, the ``push-forward.'' This
is defined for any map $\psi:D\to C$ and any sheaf on $D$. Our
sheaves will be sheaves of sections of a line bundle or vector
bundle, and we will not distinguish in the notation between a
bundle and the corresponding sheaf of sections. If ${\cal L}$ is a
line bundle on $D$, the push-forward $\psi_*({\cal L})$ is the
sheaf on $C$ defined by saying that sections of $\psi_*({\cal L})$
over a sufficiently small open set $U\subset C$ are the same as
sections of ${\cal L}$ over $\psi^{-1}(U)$.

The claim is that $E$ can be reconstructed from ${\cal L}$ as
\begin{equation}
\label{hytot}E=\psi_*({\cal L}).
\end{equation}
  Away from the ramification
points, this is just a fancy restatement of (\ref{bytot}). Let us see
what happens at ramification points, again using the local model.
We do this just for $N=2$, though the result is general.

A section of ${\cal L}$ is the same thing as a section of
$\psi^*E$ except that we must reduce mod $u$, the generator of
${\rm Im}(\varphi-y)$. A section of $\psi_*(\psi^*E)$ is,
informally, the same as a section of $E$ except that it is allowed
to depend on $y$.  So a section of $\psi_*({\cal L})$ is a section
of $E$ which {\it (i)} may depend on $y$; {\it (ii)} is considered
trivial if it is a multiple of $u$.

A section of $E$ takes the form
\begin{equation}
\label{tyro}
\begin{pmatrix}a(z)\\
b(z)\end{pmatrix},
\end{equation}
 with holomorphic functions $a(z)$, $b(z)$. A
section of $\psi_*({\cal L})$ can be written
\begin{equation}
\label{wytot}
\begin{pmatrix}A(z)+yC(z)\\ B(z)+yD(z)\end{pmatrix}~{\rm
mod}~ \left(G(z)+yH(z)\right)\begin{pmatrix}-y \\ 1\end{pmatrix},
\end{equation}
where we allow the $y$ dependence and reduce mod $u$. For $G=-C$,
$H=D$, we have
\begin{equation}
\label{qytot}
(G(z)+yH(z))\begin{pmatrix}-y \\
1\end{pmatrix} =y\begin{pmatrix}C \\ D\end{pmatrix}
+\begin{pmatrix}-zD\\ -C\end{pmatrix},
\end{equation}
 showing that the equivalence
relation in (\ref{wytot}) suffices to set $C=D=0$ in a unique  fashion,
 thus rendering (\ref{wytot}) equivalent to (\ref{tyro}) and
 showing that the statement $E=\psi_*({\cal L})$ remains
true at the ramification points.

One similarly can recover the Higgs field $\varphi$ by
\begin{equation}
\label{vyro} \varphi=\psi_*(y).
\end{equation}
 This formula means that if $f$ is a
section of ${\cal L}\to D$ and $s=\psi_*(f)$ is the corresponding
section of $E\to C$, then $\varphi(s)=\psi_*(yf)$, a condition
that suffices to determine $\varphi$.  To see that this is true,
note that a section $f$ of ${\cal L}$ is the same as a section
$\hat f$ of $\psi^*(E)$ modulo the equivalence relation of setting
to zero $(\varphi-y)\chi$ for any $\chi$.  Modulo this equivalence
relation, $y\hat f=\varphi\hat f$.
%(or more fastidiously, $y\hat f=\psi^*(\varphi)\hat f$).
Since this is true for all $\hat f$, it pushes down to the
relation (\ref{vyro}) on $C$.

\subsubsection{Which Line Bundles Appear?}
\label{whichline}

So every Hitchin pair $(E,\varphi)$ comes from a line bundle
${\cal L}$ on the spectral cover $D$.  But which line bundles
appear this way?  We will first consider the degree of ${\cal L}$
and then its moduli.

We start by computing the push-forward $\psi_*({\cal O})$, where
${\cal O}$ is a trivial line bundle over $D$.  A local holomorphic
section $s$ of ${\cal O}\to D$ can be expanded in powers of $y$:
\begin{equation}
\label{rofo}
s=\alpha_0+y\alpha_1+y^2\alpha_2+\dots+y^{N-1}\alpha_{N-1}.
\end{equation}
 The series stops at $y^{N-1}$ because of the equation
$\det(y-\varphi)=0$.  As $y$ is a section of $\psi^*(K)$, each
$\alpha_i$ is a section of the line bundle $K^{-i}$ over $C$. So
\begin{equation}
\label{jofo}\psi_*({\cal O})={\cal O}\oplus K^{-1}\oplus
K^{-2}\oplus\dots\oplus K^{-(N-1)}.
\end{equation}
 For future reference, we can
also easily calculate $\psi_*({\cal L})$ if ${\cal L}=\psi^*({\cal
L}_0)$ is the pullback of a line bundle ${\cal L}_0$ over $C$.  In
this case, a local section $s$ of  ${\cal L}$ can be expanded just
as in (\ref{rofo}), but $\alpha_i$ is now a section of ${\cal
L}_0\otimes K^{-i}$.  So
\begin{equation}
\label{xofo} \psi_*(\psi^*({\cal L}_0))={\cal
L}_0\otimes\left({\cal O}\oplus K^{-1}\oplus K^{-2}\oplus
\dots\oplus K^{-(N-1)}\right).
\end{equation}

As $K^{-i}$ has degree $-2i(g_C-1)$, (\ref{jofo}) implies that
the first Chern class of $\psi_*({\cal O})$ is $d=-N(N-1)(g_C-1)$.
If instead ${\cal L}\to D$ has degree $c$, then $\psi_*({\cal L})$
will be a vector bundle on $C$ of first Chern class
$c+d$.\footnote{One way to see this is to use the Riemann-Roch
formula and the fact that the holomorphic Euler characteristic of
${\cal L}\to D$ equals that of $\psi_*({\cal L})\to C$.  If the
first Chern class of ${\cal L}\to D$ is increased by 1, the
holomorphic Euler characteristic increases by 1, and hence the
first Chern class of $\psi_*({\cal L})\to C$ must increase by 1.}
To get an $SU(N)$ bundle, the first Chern class should vanish, so
the degree of ${\cal L}$ must be
\begin{equation}
\label{ywo} c_0=N(N-1)(g_C-1).
\end{equation}

\def\Jac{{\rm Jac}}
Therefore, a line bundle ${\cal L}$ associated with an $SU(N)$
Hitchin pair is of this degree and represents a point in ${\rm
Pic}_{c_0}(D)$, the component  of the Picard group of $D$ that
parametrizes line bundles of degree $c_0$.

The dimension of each component of the Picard group is equal to
the genus $g_D$ of $D$, which is $g_D=g_C+\dim(\FF)$. So the number
of continuous parameters upon which the line bundle ${\cal L}$
depends exceeds by $g_C$ the dimension of the fiber $\FF$ of the
Hitchin fibration. To parametrize that fiber, we must impose
$g_C$ conditions on ${\cal L}$.

It is clear what those conditions must be.  The condition
(\ref{ywo}) on the degree of ${\cal L}$ ensures that the
determinant line bundle $\det(E)$ is of degree zero and so is
topologically trivial.  To get an $SU(N)$ Hitchin pair
$(E,\varphi)$, $\det(E)$ must also be trivial holomorphically.
{\it A priori}, $\det(E)$ takes values in $\Jac(C)$, the Jacobian
of $C$, which is $g_C$-dimensional.  So asking for $\det(E)$ to be
holomorphically trivial imposes $g_C$ conditions, reducing to the
correct dimension. If we let $\Lambda$ be the map that takes a
line bundle ${\cal L}$ over $D$ to the line bundle $\Lambda({\cal
L})=\det\psi_*({\cal L})$ over $C $, then the Hitchin fiber for
$SU(N)$ consists of line bundles such that $\Lambda({\cal L})\cong
{\cal O}$.  In other words, this fiber is
\begin{equation}
\label{yudo}
\FF_{SU(N)}=\Lambda^{-1}({\cal O}).
\end{equation}

Away from the branch points, we can express the condition on
${\cal L}$ more simply.  If $\pp$ is not a branch point, then $E_\pp$
has the decomposition (\ref{bytot}), by virtue of which
$\det(E)_\pp=\otimes_{i=1}^N{\cal L}_{\qq_i}$. Thus away from the
branch points, there is a holomorphically varying isomorphism
\begin{equation}
\label{milto}
\otimes_{i=1}^N{\cal L}_{\qq_i}\cong \Bbb{C}.
\end{equation}

In general, the kernel of a holomorphic map between complex tori,
in this case $\Lambda:{\rm Pic}_{c_0}(D)\to {\rm Jac}(C)$, is a
complex torus or a union of tori.  We show in section
\ref{topview} that $\FF_{SU(N)}$ is connected, so it is actually a
complex torus. Is it in a natural way an abelian variety? To
exhibit it as one, we must pick in a natural fashion a point in
$\FF_{SU(N)}$, that is, we must pick a line bundle ${\cal L}\to D$
of the appropriate degree that is in the kernel of $\Lambda$. For
this, pick a square root $K^{1/2}$ of the canonical bundle of
$C$ and let ${\cal L}_0=\psi^*(K^{(N-1)/2})$. This does have the
appropriate degree ($K^{(N-1)/2}$ has degree $(g_C-1)(N-1)$, and
because ${\cal L}_0$ is the pullback by a map of degree $N$, its
degree is $N$ times greater).  It is also true that ${\cal L}_0$
is in the kernel of the map $\Lambda$. Indeed, from (\ref{xofo}),
we have in this case $\psi_*({\cal L})=K^{(N-1)/2}\otimes
\left({\cal O}\oplus K^{-1}\oplus K^{-2}\oplus \dots\oplus
K^{-(N-1)}\right)$, from which it follows that
$\det(\psi_*({\cal L}_0))\cong {\cal O}$.

If $N$ is odd, $(N-1)/2$ is an integer, and this construction did
not really depend on choosing a square root of $K$.  So for odd
$N$, we have exhibited $\FF_{SU(N)}$ as an abelian variety, with a
distinguished representative  $(E,\varphi)$ of each fiber of the
Hitchin fibration. For even $N$, this is not quite the case, since
the choice of $K^{1/2}$ does matter. For even $N$, the best we
can do is to construct a pair $(E,\varphi)$ that is natural up to
the possibility of tensoring by a flat line bundle of order 2.
This distinction was pointed out and generalized to other $G$ in
\cite{donga}.  See also our
discussion below at the end of section \ref{distinguished}.

\subsubsection{Relation To K-Theory}\label{dolfo}

We obtained the degree $c_0$ of the line bundle $\L\to D$ in a rather technical fashion, but
actually the answer
(eqn. (\ref{ywo})) has a nice interpretation, which was essentially described in section 4.3
of \cite{FW}.

Let us view $W=T^*C$ as a real symplectic manifold, with the symplectic form being the imaginary (or real)
part of the holomorphic symplectic form $\Omega_I$.  We consider the $A$-model of $W$ with this symplectic form.
We view $C\subset W$ as a Lagrangian submanifold and consider $A$-branes supported on $C$.

Naively, a rank 1 $A$-brane supported on $C$ is endowed with a flat Chan-Paton line bundle $\U\to C$.  But actually,
the $K$-theory interpretation of $D$-branes means that there is a twist that involves the square root of the normal bundle
to $C$ in $W$.  The upshot is that $\U$, rather than a flat line bundle, should be a flat ${\mathrm{Spin}_c}$ structure on $C$.
In terms of complex geometry, this means that the degree of $\U$ should be not zero but the degree of a square root of the canonical bundle $K\to C$,
that is, it should be $g_C-1$.

Similarly, the Chan-Paton bundle $E$ of a rank $N$ $A$-brane supported on $C$ should be $N(g_C-1)$.  This means that if we want the line
bundle $\L\to D$ to have the property that $E=\psi_*(\L)$ is the Chan-Paton bundle of an $A$-brane supported on $C$, then its degree must be
not $c_0$ as defined in eqn. (\ref{ywo}) but
\begin{equation}\label{tywo}c_0'=c_0+N(g_C-1)=N^2(g_C-1)=g_D-1. \end{equation}
But  $D$ is also a Lagrangian submanifold of $W$, and this value of the degree of $\L\to D$ simply means that $D$, endowed with the Chan-Paton
 bundle $\L$, can be interpreted as an $A$-brane.  

Thus we can think of the Higgs bundle and spectral cover construction as giving a correspondence between an $A$-brane of rank 1 supported on $D$ and
an $A$-brane of rank $N$ supported on $C$.

\subsubsection{The Unitary Group}
\label{unitary}

In a similar fashion, we can analyze the fiber of the Hitchin
fibration for the related groups $U(N)$ and
$PSU(N)=SU(N)/\Bbb{Z}_N$.

First we consider $G=U(N)$.  For this example, we should modify
the construction at the beginning by not requiring $\varphi$ to be
traceless.   Accordingly, in defining ${\cal V}$, we include
$\Tr\,\varphi$ along with the higher traces, with the result that
the dimension of ${\cal V}$ is $(g_C-1)N^2+1$. Furthermore, for
$U(N)$, we would not want to restrict the determinant of $E$ to be
trivial, so $(g_C-1)N^2+1$ is the appropriate dimension of the
fiber $\FF$.  These are equal to each other and to 
$\frac{1}{2}\dim\,\MH$, which is $N^2(g_C-1)+1$ for $G=U(N)$.  Since the
determinant of $E$ is not constrained to be trivial, $\psi_*({\cal
L})$ can be anything, and roughly speaking, ${\cal L}$ can be any
line bundle on $D$.  Thus, the fiber of the Hitchin fibration for
$U(N)$ is simply the Picard group ${\rm Pic}(D)$ that
parametrizes line bundles on $D$.

To be more precise, we need to observe that there are two related
versions of the question.   For $U(N)$, a Higgs bundle
$(E,\varphi)$ has an integer topological invariant, which is the
first Chern class of $E$, denoted $c_1(E)$. A stable Higgs bundle
can have any first Chern class.  The Hitchin fibration makes sense
for any value of this invariant.  The fiber of the Hitchin
fibration for $c_1(E)=m$ is the moduli space ${\rm
Pic}_{c_0+m}(D)$ parametrizing line bundles on $D$ of degree
$c_0+m$, where $c_0$ was given in (\ref{ywo}).  This follows by
the same analysis as before.

However, the equivalence of stable Higgs bundles with solutions of
Hitchin's equations holds only when the first Chern class
vanishes.  One of Hitchin's equations, $F-\phi\wedge \phi=0$
reduces upon taking the trace to $\Tr\,F=0$, implying that the
first Chern class must vanish.  So the moduli space $\MH(U(N),C)$
of solutions of Hitchin's equations (or classical vacua of our
field theory) actually corresponds to the component of the moduli
space of stable Higgs bundles with $c=0$. The fiber of the Hitchin
fibration of $\MH(U(N),C)$ is therefore
\begin{equation}\label{picun}\FF_{U(N)}={\rm
Pic}_{c_0}(D).\end{equation} Precisely as we discussed above for
$SU(N)$, this fiber is canonically an abelian variety if $N$ is
odd, but it is not quite an abelian variety for even $N$. Rather,
it becomes one once a spin structure $K^{1/2}$ is chosen on $C$.

For $U(N)$, we are mainly interested in Higgs bundles
$(E,\varphi)$ of degree zero, because they represent points in the
moduli space $\MH(U(N),C)$ of classical vacua of our $\sigma$-model.
One can ask nevertheless whether Higgs bundles $(E,\varphi)$ of
other degrees are relevant to geometric Langlands.  The answer is that they can be relevant
if one asks questions for which a reduction from the four-dimensional gauge theory
to the two-dimensional $\sigma$-model with target
$\MH$ is not adequate. See section \ref{elmag} for more on this.

\subsubsection{The Group $PSU(N)$}
\label{psun}

Finally, we will determine the fiber of the Hitchin fibration for
the group $PSU(N)=SU(N)/\Bbb{Z}_N$, where here $\Bbb{Z}_N$ is the
center of $SU(N)$.

A  $PSU(N)$ bundle over $C$ can be topologically non-trivial. Let
us first consider the topologically trivial case. A topologically
trivial bundle $E'$ over $C$ with structure group $PSU(N)$ can be
lifted to an $SU(N)$ bundle $E$ over $C$, but not quite uniquely.
The failure of uniqueness is that $E$ can be twisted by a line
bundle ${\cal S}$ that is of order $N$, or alternatively by a flat
line bundle whose monodromies lie in the center of $SU(N)$. Such a
flat bundle of order $N$ disappears when one projects to $PSU(N)$.
Let $\Gamma_N$ be the group of line bundles on $C$ of order $N$.
Then for topologically trivial $PSU(N)$ bundles, the fiber of the
Hitchin fibration is
\begin{equation}
\label{mormot}
\FF^{(0)}_{PSU(N)}=\FF_{SU(N)}/\Gamma_N=\Lambda^{-1}({\cal
O})/\Gamma_N.
\end{equation}
(An element ${\cal S}\in \Gamma_N$ acts on ${\cal L}\in
\Lambda^{-1}({\cal O})$ by ${\cal L}\to {\cal L}\otimes
\psi^*({\cal S})$.)   $\FF^{(0)}_{PSU(N)}$ is always an abelian
variety in a canonical way. For odd $N$, $\FF_{SU(N)}$ is already an
abelian variety, even before dividing by $\Gamma_N$. For even $N$,
to make $\FF_{SU(N)}$ into an abelian variety, we must pick a spin
structure on $C$. But dividing by $\Gamma_N$ precisely cancels the
dependence on the spin structure.  Indeed, changing the spin
structure amounts to twisting by a line bundle of order 2, but for
even $N$ such a line bundle is an element of $\Gamma_N$.

We find another useful way to describe $\FF^{(0)}_{PSU(N)}$  if we
observe that  $PSU(N)$ coincides with $PU(N)=U(N)/U(1)$.
 For ${\cal S}$ a line bundle on $C$ of degree zero, $\psi^*({\cal
S})$ is a line bundle on $D$ of degree zero. So for ${\cal L}\in
\FF_{U(N)}= {\rm Pic}_{c_0}(D)$, the tensor product ${\cal L}\otimes
\psi^*({\cal S})$ is also in ${\rm Pic}_{c_0}(D)$.  This gives a
group action of ${\rm Jac}(C)$ on ${\rm Pic}_{c_0}(D)$, and we
write ${\rm Pic}_{c_0}(D)/{\rm Jac}(C)$ for the quotient.  Then
\begin{equation}\label{publy} \FF^{(0)}_{PSU(N)}={\rm
Pic}_{c_0}(D)/{\rm Jac}(C). \end{equation} The basis for this
assertion is that, since $\psi_*({\cal L}\otimes \psi^*{\cal
S}))=\psi_*({\cal L})\otimes {\cal S}$ (according to
(\ref{xofo})), and $\psi_*({\cal L})$ is of rank $N$, we have
$\det(\psi_*({\cal L}\otimes \psi^*({\cal S})))=\psi_*({\cal
L})\otimes {\cal S}^N$. Hence, ${\cal S}$ can be chosen to set
$\det(\psi_*({\cal L}\otimes \psi^*({\cal S})))={\cal O}$, so that
${\cal L}\otimes \psi^*({\cal S})$ is an element of $\FF_{SU(N)}$.
This condition does not quite uniquely fix ${\cal S}$.  We are
still free to twist ${\cal S}$ by an element of  $\Gamma_N$. After
allowing for this, we see that (\ref{mormot}) and (\ref{publy})
are equivalent.

This second description for the topologically trivial case is a
useful starting point for analyzing the Hitchin fibration for
topologically non-trivial $PSU(N)$ bundles. Topologically, a
$PSU(N)$ bundle $E$ is  classified by a characteristic class $\xi$
that takes values in $H^2(C,\Bbb{Z}_N)\cong \Bbb{Z}_N$. (For $N=2$, $\xi$ is the second
Stieffel-Whitney class $w_2(E)$.)   This
characteristic class measures the obstruction to ``lifting'' a
$PSU(N)$ bundle to $SU(N)$. Let $\MH^{(d)}$ be the component of
$\MH(PSU(N),C)$ that parametrizes Higgs bundles with
characteristic class $\xi=d$.

Although a topologically non-trivial $PSU(N)$ bundle cannot be
lifted to $SU(N)$, it can be lifted non-uniquely to $U(N)$.  In
fact, if $E$ is any $U(N)$ bundle, then the associated adjoint
bundle ${\rm ad}(E)$ is a $PSU(N)$ bundle, and any $PSU(N)$ bundle
can be obtained this way for some $E$. If the first Chern class of
$E$ is $d$, or more generally if it is congruent to $d$ modulo
$N$, then the characteristic class of the $PSU(N)$ bundle ${\rm
ad}(E)$ is $\xi=d$.

If $E_1$ and $E_2$ are two different $U(N)$ bundles, then the
corresponding $PSU(N)$ bundles ${\rm ad}(E_1)$ and ${\rm ad}(E_2)$
are isomorphic if and only if $E_1\cong E_2\otimes{\cal S}$, for
some line bundle ${\cal S}\to C$.  (Note that in this operation,
the first Chern classes of $E_1$ and $E_2$ are equal modulo $N$,
since $c_1(E_1)=c_1(E_2)+Nc_1({\cal S})$.  They reduce mod $N$ to
$d=\xi({\rm ad}(E))$.) This leads to an analog of (\ref{publy}).
The Hitchin fiber $\FF^{(d)}_{PSU(N)}$ for $PSU(N)$ bundles of
$\xi=d$ is
\begin{equation}\label{blonko} \FF^{(d)}_{PSU(N)}={\rm
Pic}_{c_0+d}(D)/{\rm Jac}(C).\end{equation} The idea here is that
a point in ${\rm Pic}_{c+d}(D)$ determines a $U(N)$ bundle $E$
over $C$ with $c_1(E)=d$; acting by ${\cal S}\in {\rm Jac}(C)$ has
the effect $E\to E\otimes {\cal S}$, and does not affect the
associated $PSU(N)$ bundle ${\rm ad}(E)$.

There is also a useful, though less canonical-looking, analog of
(\ref{mormot}).  Given a $PSU(N)$ bundle that we will call ${\rm
ad}(E)$, its lift $E$ to $U(N)$ is far from unique, because of the
freedom $E\to E\otimes {\cal S}$.  The action on $\det(E)$ is
$\det(E)\to \det(E)\otimes {\cal S}^N$.  So, restricting ourselves
to ${\cal S}$ of degree zero, we can adjust $\det(E)$ in an
arbitrary fashion within its component of ${\rm Pic}(C)$, and then
we are still free to twist $E$ by a line bundle  of
order $N$, that is, by an element of $\Gamma_N$.  We may pick any
convenient line bundle ${\cal S}_0$ over $C$ of degree $d=\xi({\rm
ad}(E))$, and require $\det(E)={\cal S}_0$.  For example, we may
pick a point $\pp\in C$ and choose ${\cal S}_0={\cal O}(\pp)^d$.
So we get the analog of (\ref{mormot}):
\begin{equation}
\label{normat} \FF^{(d)}_{PSU(N)}=\Lambda^{-1}({\cal S}_0)/\Gamma_N,
\end{equation}
expressing the fact that by the action of $J(C)$, we can fix the
determinant of $E$, and we are then still free to act by
$\Gamma_N$.

One last comment is that what we have determined in (\ref{mormot})
and (\ref{normat}) is precisely the fiber of the Hitchin fibration
of $\MH(PSU(N),C)$, which is defined by dividing the space of all
solutions of Hitchin's equations by the group of all
$PSU(N)$-valued gauge transformations.  It is also sometimes
convenient to work on the
universal cover of $\MH(PSU(N),C)$, which one achieves by dividing
only by the subgroup of gauge transformations that can be
continuously deformed to the identity. On the fiber of the Hitchin
fibration, this has the effect precisely of not dividing by
$\Gamma_N$ in the above constructions.  So if $\hat
\MH^{(d)}(PSU(N),C)$ is the cover of $\MH^{(d)}(PSU(N),C)$
obtained by dividing only by connected gauge transformations, then
the corresponding fiber of the Hitchin fibration is
\begin{equation}\label{palooko} \hat \FF^{(d)}_{PSU(N)}=\Lambda^{-1}({\cal
L}_0).\end{equation} If we set $d=0$, $\hat\MH^{(d)}(PSU(N),C)$
reduces to $\MH(SU(N),C)$.  Indeed, for $d=0$, we can take ${\cal
L}_0={\cal O}$, and $\hat \FF^{(d)}_{PSU(N)}$ reduces to
$\FF_{SU(N)}$.

One might wonder what is the dual of these topologically
non-trivial components of $\MH(PSU(N))$.  
Here the relevant duality is the $S$-duality or mirror symmetry between $\MH(PSU(N))$ and $\MH(SU(N))$ that underlies
geometric Langlands duality for these gauge groups; see \cite{KW} for a full explanation.
As explained from a
mathematical point of view in 
\cite{HauselThaddeus},
 the dual of $\MH^{(d)}(PSU(N))$ is our friend
$\MH(SU(N))$, but endowed with a certain non-trivial gerbe.
We elaborate on the relevant concept of a gerbe in section \ref{gerbes} below.
See also section 7 of \cite{KW}, where the gerbe in question is interpreted in terms of a discrete
$B$-field of the $\sigma$-model and is related to a discrete version of electric-magnetic duality.
And see  \cite{donga} for an analysis of these issues for arbitrary $G$.

\subsubsection{Spectral Covers For Other Gauge Groups}\label{spectrother}

We have described the theory of the spectral cover only for $G=SU(N)$ and for the closely related groups $U(N)$ and $PSU(N)$.
What happens for other $G$?

If $G$ has a convenient ``small'' representation, one can develop a somewhat similar story to what has been described above.
This is practical for $SO(N)$, $Sp(N)$, and even $G_2$  \cite{HitchinDuke, KP, Hthree}.

In general -- for example for $G=E_8$ -- there is no conveniently ``small'' representation and more abstract
methods are necessary.  See \cite{donga}.

\subsection{The Distinguished Section}
\label{distinguished}
\subsubsection{The Case Of $SU(N)$}

For practice with these ideas, and because it turns out to be useful, we will look more closely at a certain family of
distinguished Hitchin pairs $(E,\varphi)$. 

We begin with $G=SU(2)$ and we take $E$ to be the rank two complex
vector bundle $E=K^{1/2}\oplus K^{-1/2}$.  This bundle is
unstable; it contains the line subbundle $K^{1/2}$, which is of
positive degree. Nonetheless, there are stable Hitchin pairs
$(E,\varphi)$, and we will describe them, following \cite{Hitchin}.

One reflection of the fact that $E$ is unstable is that, unlike a
stable bundle, it has non-constant automorphisms.  If we write a
section of $E$ as $\begin{pmatrix}s\\ t\end{pmatrix}$, where $s$
is a section of $K^{1/2}$ and $t$ of $K^{-1/2}$, then $SL(2,\C)$ automorphisms of $E$  act by
\begin{equation}
\label{uti}
\begin{pmatrix}\lambda & \tau\\
0&\lambda^{-1}\end{pmatrix}.
\end{equation}
 Here $\lambda$ is a nonzero complex
number, and $\tau$ an element of $H^0(C,K)$. We want to classify
stable Hitchin pairs $(E,\varphi)$ up to the action of these
automorphisms on $\varphi$.

A general holomorphic section $\varphi$ of
$H^0(C,K\otimes\ad(E))$ can be written
\begin{equation}
\label{buti}
\begin{pmatrix} v &\; w/2 \\ u &\;
-v\end{pmatrix},
\end{equation}
 where $u$ is a complex number, $v$ an element of
$H^0(C,K)$, and $w$ an element of $H^0(C,K^2)$.  (The factor of 1/2 multiplying $w$ is for convenience.) $u$ must be
nonzero in order for the pair $(E,\varphi)$ to be stable, since if
$u=0$, the line bundle $K^{1/2}\subset E$, whose degree is
positive, is $\varphi$-invariant.  If $u\not=0$, we can 
choose $\lambda$ and $\tau$ to set $u=1$, $v=0$ (in a way that is unique modulo the center $\{\pm 1\}$ of $SL(2,\C)$), in which case
\begin{equation}
\label{hgy}
\varphi = \begin{pmatrix} 0 & w/2 \\ 1 & 0\end{pmatrix}.
\end{equation}
As shown in \cite{Hitchin}, any such pair
$(E,\varphi)$ is stable (and $E$ is the most unstable bundle for
which there are such stable pairs). Clearly $\Tr\,\varphi^2=w$, so
with this particular $E$, there is up to automorphism a unique
stable pair $(E,\varphi)$ with given $\Tr\,\varphi^2$.  So the family
$V_E$ of stable Hitchin pairs with this $E$ intersects each fiber
$\FF$ of the Hitchin fibration in precisely one point. The family $V_E$ thus gives,
modulo the choice of $K^{1/2}$, a distinguished section 
\begin{equation}\label{zelo}\zeta:\V\to\MH\end{equation}
 of the
Hitchin fibration. As we explained at the end of section
\ref{hitchfibr}, for generic stable $E$, the number of
intersections of $V_E$ with $\FF$  is $2^{3g-3}$.

The spectral curve $D$ is defined by the equation
$\det(y-\varphi)=0$ or $y^2=w/2$.  The line bundle ${\cal N}\to D$
is found by solving the eigenvector equation $\varphi\uppsi=y\uppsi$
or
\begin{equation}\label{ivox}
\begin{pmatrix} 0 & w/2 \\ 1 & 0\end{pmatrix}\uppsi = y
\uppsi.
\end{equation}
 Locally this can be handily solved by picking
\begin{equation}
\label{yupo}
\uppsi=\begin{pmatrix} y\alpha \\ \alpha\end{pmatrix},
\end{equation}
where $\alpha$ is any local holomorphic section of $K^{-1/2}$.
This gives a natural map from sections of $K^{-1/2}$ to sections
of ${\cal N}$, implying that in this example, ${\cal N}$ coincides
with $\psi^*(K^{-1/2})$.

The ``cokernel'' of $\varphi-y$ defines another line bundle ${\cal
L}$, which according to (\ref{roygo}) is ${\cal L}={\cal N}\otimes
\psi^*(K)=\psi^*(K^{1/2})$.  The bundle $E$ is supposed to be
$\psi_*({\cal L})$, which according to (\ref{xofo}) is
$K^{1/2}\oplus K^{-1/2}$.  This is indeed isomorphic to the
bundle $E$ with which we began.

The  analog of this for $G=SU(N)$ is to take
$E=K^{(N-1)/2}\oplus K^{(N-1)/2-1}\oplus\dots\oplus
K^{-(N-1)/2}$.  We express a section of $E$ as a column vector
\begin{equation}
\label{plook}
\begin{pmatrix}s_1\\
s_2\\ \vdots\\ s_N\end{pmatrix},
\end{equation}
 where $s_i$ is a section of
$K^{(N+1)/2-i}$. An  analysis similar to that above shows that
stable Hitchin pairs $(E,\varphi)$ with this $E$ can be placed in
the form
\begin{equation}
\label{plyo}
\varphi=\begin{pmatrix}0 & w_2/2 & w_3/3 & \dots
&w_{N-1}/(N-1)& w_N/N\\ 1 & 0 & 0 & \dots &0& 0 \\ 0 & 1 & 0 &
\dots & 0&0\\ \vdots & \vdots &\vdots & \vdots& \ddots &\vdots\\
0&0&0&\dots &1&0\end{pmatrix},
\end{equation}
 with 1's just below the main
diagonal, and $w_n$, $n=2,\dots,N$, taking values in
$H^0(C,K^n)$. We have $\Tr\,\varphi^n=w_n$. So for this $E$,
there is again a unique stable Hitchin pair on each fiber $\FF$ of
the Hitchin fibration, giving a natural section of the Hitchin fibration.
 Similarly to what we found for $SU(2)$, by
solving the equation for an eigenvector, one finds that ${\cal
N}\cong \psi^*(K^{-(N-1)/2})$, and therefore, using
(\ref{roygo}), ${\cal L}\cong \psi^*(K^{(N-1)/2})$.  We expect
to recover $E$ as $\psi_*({\cal L})$. Indeed, from (\ref{xofo}),
we do find that $\psi_*({\cal L})=K^{(N-1)/2}\oplus
K^{(N-1)/2-1}\oplus \dots\oplus K^{-(N-1)/2}=E$.

\subsubsection{Section Of The Hitchin Fibration For Any $G$}

All of this has an analog for any simple Lie group $G$, though we will not aim
to give a complete explanation.

The starting point in general is to embed the $SU(2)$
bundle $K^{1/2}\oplus K^{-1/2}$ in $G$, using Kostant's
principal $SU(2)$ embedding. This gives a special unstable $G$-bundle $E$.
For this $E$, there is a unique stable Hitchin pair $(E,\varphi)$ on any
given fiber of the Hitchin fibration. This is shown, in a slightly
different formulation, in \cite{BeD}
using the theory of ``opers.''  Thus this construction gives for any $G$ a
nearly canonical section of the Hitchin fibration. We call this
section nearly canonical since in general it may depend on the
choice of $K^{1/2}$.

In fact, the Kostant embedding is for any $G$ an embedding of the
Lie algebra  of $SU(2)$ in that of $G$.  At the group level, what
is embedded in $G$ may be either $SU(2)$ or $PSU(2)=SO(3)$. In the
latter case, the dependence on a choice of $K^{1/2}$ disappears
when we make the Kostant embedding, and we get a truly canonical
section of the Hitchin fibration.  In the former case, we get a
section of the Hitchin fibration for each choice of $K^{1/2}$.

For example, let $\frak{su}(N)$ denote  the Lie algebra of
$SU(N)$. The Kostant embedding of $\frak{su}(2)$ in $\frak{su}(N)$
is the one in which the $N$-dimensional representation of
$\frak{su}(N)$ is irreducible with respect to $\frak{su}(2)$ (and
hence transforms with spin $(N-1)/2$). At the group level, the
image of the Kostant map for $SU(N)$ is $SO(3)$  for odd $N$, but
is $SU(2)$  for even $N$.  More generally, if
$G=SU(N)/\Gamma$, where $\Gamma$ is a subgroup of the center of
$SU(N)$, then the image of the Kostant map is $SO(3)$ except when
$N$ is even and the element $-1$ of $G$ is not contained in
$\Gamma$.  For example, for $G=PSU(N)$, the image of the Kostant
map is always $SO(3)$.
These results are of course  consistent with what we found in
section \ref{whichline}.

 In general, if $G$ is of adjoint type, the image of the Kostant map is always $SO(3)$.
So for example, for $G=E_8$, the fiber of the Hitchin fibration is canonically an abelian variety.

\section{Dual Tori And Hitchin Fibrations}
\label{dualtori}
\markright{\thesection. Dual Tori And Hitchin Fibrations}

As we have reviewed in section \ref{compacthitchin},
for any simple Lie group,  the fiber $\FF$ of the Hitchin fibration
is generically a complex torus, because of the complete integrability
of the space $\MH(G,C)$.  As has been argued
physically \cite{BJSV,HMS},  
$S$-duality in four
dimensions acts by $T$-duality on the fiber $\FF$ of the Hitchin
fibration, along with a geometrical symmetry\footnote{For the unitary groups that we give as examples, the
geometrical symmetry of the base is a simple rescaling that does
not play an important role.} of the base ${\cal
V}$.   As was originally explained in \cite{HauselThaddeus}, the mirror symmetry between
$\MH(^L\neg G,C)$ and $\MH(G,C)$ is a rare example in which the the SYZ approach
to mirror symmetry \cite{SYZ} can be implemented in some detail, because the hyper-Kahler structure
makes it possible to construct rather explicitly a fibration by Lagrangian tori.  This interpretation of the duality
between $\MH(^L\neg G,C)$ and $\MH(G,C)$ was important in \cite{KW}.
  Here
we will give a more thorough explanation of some aspects than was provided there.

To be precise, the claim is that 
 if $G$ and $^LG$ are two dual
groups, the Hitchin fibers $\FF$ and $^L\neg \FF$ over corresponding points
in the base are dual complex tori.  
We recall that this means that $\FF$ parametrizes flat line bundles
on $^L\neg \FF$, and vice-versa.  A duality between complex tori $\cF$ and
$\cF'$ can be described most symmetrically by presenting a unitary
line bundle ${\cal T}$ with connection over the product $\cF\times
{}\cF'$, such that ${\cal T}$ is flat when restricted to $f\times
{}\cF'$ for any $f\in \cF$, or to $\cF\times {}f'$ for any $f'\in \cF'$.
(If $\cF$ and $\cF'$ have complex or algebraic structures, one wants
${\cal T}$ to be a holomorphic or algebraic line bundle.) Thus,
letting $f$ vary, we get a family ${\cal Y}'$ of flat line bundles
over $\cF'$, and letting $f'$ vary, we get a family ${\cal Y}$ of
flat line bundles over $\cF$. 
 If ${\cal Y}'$ is the moduli space of flat bundles over $\cF'$ and
${\cal Y}$ is the moduli space of flat bundles over $\cF$, then
$\cF$ and $\cF'$ are called dual tori, and
${\cal T}$ is called a Poincar\'e line bundle for the pair.

An immediate consequence of the definition is that if $\cF$ and $\cF'$
are dual tori, then there is always a unique distinguished point
$f\in \cF$ such that ${\cal T}$ is trivial when restricted to
$f\times {}\cF'$; conversely, there is always a unique distinguished
point in ${}f'\in{} \cF'$ such that ${\cal T}$ is trivial when
restricted to $\cF\times {}f'$.  As a result, if $\cF$ and $\cF'$ are
complex algebraic  tori that are dual in the above sense, they always are
abelian varieties in a canonical way.

This section is devoted to discussing these issues in the context of Hitchin fibrations.
On a number of points, we go into more detail  than is needed in the rest
of the paper.

\subsection{Examples}
\label{examplo}

For an elementary example of dual tori, let $S$ be a circle
parametrized by an angular variable $x$, $0\leq x\leq 2\pi$, and
$S'$ another circle parametrized by an angular variable $y$,
$0\leq y\leq 2\pi$. Of course $S\times S'\cong
\Bbb{R}^2/\Bbb{Z}^2$ where ${\Bbb{Z}}^2$ acts by $(x,y)\to (x+2\pi
n,y+2\pi m)$. Let $\omega= dx\wedge dy/2\pi$. A line bundle ${\cal
T}\to S\times S'$ with curvature $F=-i\omega$ can serve as a
Poincar\'e line bundle. We can define such a ${\cal T}$ by picking
the connection form $B=- ix\,dy/2\pi$ on a trival line bundle
${\cal O}$ over $\Bbb{R}^2$, and then descending to $S\times S'$
by dividing by unit translations in the $x$ and $y$ directions.
(We take $(x,y)\to (x+2\pi n,y+2\pi m)$ to act on ${\cal O}$ as
multiplication by $e^{ iny}$.) The holonomy of ${\cal T}$ around
the circle $x\times S'$ is $e^{ ix}$ and the holonomy around the
circle $S\times y$ is $e^{iy}$.  Each possible holonomy of a flat
line bundle appears once in each family, so ${\cal T}$ is a
Poincar\'e line bundle that establishes a duality between $S$ and
$S'$.

Instead of picking an explicit connection form $B$ to construct
the line bundle ${\cal T}$, we could accomplish the same by simply
asking that ${\cal T}$ have curvature $F=-i\omega$ and trivial
holonomy when restricted to $x=0$ or to $y=0$. These conditions
completely characterize ${\cal T}$ up to isomorphism.

This elementary example has the following generalization.  Let
$\Gamma$ and $\Gamma^*$ be any two dual lattices, with a
unimodular bilinear form $\langle ~,~\rangle:\Gamma\times
\Gamma^*\to \Bbb{Z}$. Define the  vector spaces
$V=\Gamma{\otimes}_{\Bbb{Z}}\Bbb{R}$, $V^*=\Gamma^*{\otimes}
_{\Bbb{Z}}\Bbb{R}$ and the tori $\cF=V/\Gamma$, $\cF^*=V^*/\Gamma^*$.
$V$ and $V^*$ are dual vector spaces, as the form
$\langle~,~\rangle$ extends naturally to a bilinear pairing
$V\otimes V^*\to \Bbb{R}$. Moreover, the tori $\cF$ and $\cF^*$ are
always dual tori in a natural way.  One simply makes the same
construction that we have just seen, beginning with the symplectic
form $\omega=\langle dx,dy\rangle/2\pi$, where we write $x$ or $y$
for a lift of a point in $\cF$ or $\cF^*$ to $V$ or $V^*$.

For an example closer to our subject, let $D$ be a compact
two-dimensional oriented closed manifold -- for the moment endowed
with no complex structure -- and let ${\cal J}$ be the moduli
space of flat unitary complex line bundles on $D$.  Then ${\cal
J}$ is a self-dual torus, because it is
$(\Gamma\otimes_{\Bbb{Z}}\Bbb{R})/\Gamma$, where $\Gamma$ is the
self-dual lattice $\Gamma= H_1(D,\Bbb{Z})$.  Consequently, the
remarks of the last paragraph apply to generate a canonical
self-duality of ${\cal J}$.

Alternatively, it will be helpful in our later work to understand
the self-duality of ${\cal J}$ in the following way, using gauge
theory. We define a line bundle ${\cal T}\to {\cal J}\times {\cal
J}$ as follows. Let ${\cal L}$ be a flat line bundle over $D$,
with connection $A$, and let ${\cal M}$ be a second flat line
bundle over $D$, with connection $B$. Being flat, the connections
obey
\begin{equation}
\label{olpp}
0=F_A=dA,~~0=F_B=dB.
\end{equation}
 The pair
${\cal L}$, ${\cal M}$ determines a point  $P\times P'\subset
{\cal J}\times {\cal J}$. We define a two-form over ${\cal
J}\times {\cal J}$ by the formula
\begin{equation}
\label{ivo} \omega={1\over 2\pi}\int_{D}\delta A\wedge \delta B.
\end{equation}
 $\omega$ vanishes if contracted with a vector field generating a gauge transformation
 of $A$ or $B$, since if one adds an exact form $d\epsilon$ to $\delta A$ or
$\delta B$ and integrates by parts, the result vanishes upon using
the flatness condition (\ref{olpp}).  So $\omega$ descends to a two-form
on ${\cal J}\times {\cal J}$, and this form is nondegenerate by
Poincar\'e duality.  For any $P,P'\subset {\cal J}$, $\omega$
vanishes when restricted to $P\times {\cal J}$ (which we can do by
setting $\delta A=0$) or to ${\cal J}\times P'$ (which we do by
setting $\delta B=0$). Moreover, we have normalized $\omega$ so
that its periods are integer multiples of $2\pi$.

All this being so, there exists a unitary line bundle ${\cal T}$
with connection over ${\cal J}\times {\cal J}$, such that the
curvature of ${\cal T}$ is equal to $-i\omega$.  Moreover, ${\cal
T}$ is uniquely determined up to isomorphism if we ask that it is
trivial (flat and vanishing monodromies) if restricted to $0\times
{\cal J}$ or ${\cal J}\times 0$, where $0\in {\cal J}$ is the
point corresponding to the trivial flat connection.

Such a ${\cal T}$ establishes the self-duality of ${\cal J}$, that
is, ${\cal J}$ is its own dual torus.  In fact, if we pick a
symplectic basis of one-cycles on $D$, to reduce the intersection
form to a sum of $2\times 2$ blocks of the form $\begin{pmatrix}0
& 1\\ -1 & 0\end{pmatrix}$, then this example reduces  to a
product of copies of the elementary example we considered first,
involving a line bundle over  $S\times S'$ with curvature
$-idx\wedge dy/2\pi$.

This also makes it possible to construct ${\cal T}$ explicitly by
elementary formulas (as we did in the toy example of $S\times
S'$). But a more elegant way to construct ${\cal T}$ is to
consider a Chern-Simons gauge theory over the three-manifold
$\Bbb{R}\times D$ in which the gauge group is $U(1)\times U(1)$,
the gauge field consisting of a pair of $U(1)$ gauge fields that
we call $A,B$. We take the action to be
\begin{equation}
\label{ploko} I={1\over 2\pi}\int_{\Bbb{R}\times D}\,A\wedge dB.
\end{equation}
 The classical equations of motion tell us that $dA=dB=0$, and the
classical phase space is the moduli space of solutions up to gauge
transformations, or ${\cal J}\times {\cal J}$. The symplectic
structure on the classical phase space is what we have called
$\omega$, and the first step in quantization is to construct a line
bundle ${\cal T}$ with the properties that we have claimed.  This can
be done directly using the gauge-invariance of the Chern-Simons action
(for example, see \cite{rsw,axelrod}).

As in many constructions that can be carried out in general for a
smooth two-dimensional surface $D$, this one can be given an
important alternative description in case a complex structure is
picked on $D$.  In this case, ${\cal J}$ becomes the Jacobian of
$D$,  which parametrizes holomorphic line bundles over $D$ of
degree zero; we denote it as ${\rm Jac}(D)$. This is a classic
example of a self-dual abelian variety, and very relevant to
Hitchin fibrations, as we will discuss momentarily. Apart from the
above topological construction, the Poincar\'e line bundle of
${\rm Jac}(D)$ can be conveniently constructed using holomorphic
methods; we explain one way to do this in section \ref{ugduality}.

\subsection{The Case Of Unitary Groups} \label{unitaryexample}

Here we will describe explicitly the duality between the fibers
of dual Hitchin fibrations for the dual pair of groups $SU(N)$ and $PSU(N)$, and
similarly the self-duality for  $U(N)$.

In doing this, we meet immediately the fact that $\MH(PSU(N))$ has $N$ components,
labeled by the value of a discrete characteristic class $\xi$ (introduced in section \ref{psun}), while $\MH(SU(N))$ is connected.  
As explained in \cite{HauselThaddeus} and in section 7 of \cite{KW} (see also section \ref{gerbes} below), the dual of $\xi$ is a certain
discrete $B$-field (or gerbe) that can be introduced on $\MH(SU(N))$.  We will postpone this issue  and for now limit
ourselves to the duality between $\MH(SU(N))$ and the component $\MH^{(0)}(PSU(N))$ of $\MH(PSU(N))$ with $\xi=0$.
Similarly, in the case of $U(N)$, we will for now consider Higgs bundles with vanishing first Chern class.  (A formulation
of the duality without these limitations is given in section \ref{dualap}.)

Even  with these restrictions, a further conundrum presents itself.
As we have explained above, if $\cF$ and $\cF'$ are dual complex algebraic tori,
they are always abelian varieties in a canonical way. The fiber
$\FF^{(0)}_{PSU(N)}$  has this property for all $N$, but as we have
seen, $\FF_{SU(N)}$ and $\FF_{U(N)}$ are only abelian varieties in a
natural fashion if $N$ is odd.  For $N$ even, they become abelian
varieties in a natural fashion only once a spin structure
$K^{1/2}$ is picked on $C$.  Therefore, the claim that $\FF_{SU(N)}$ and $\FF^{(0)}_{PSU(N)}$ are
dual (and similarly the claim that $\FF_{U(N)}$ is self-dual)  cannot quite be correct for even $N$.
A corrected statement was formulated in section 10 of \cite{FW}, but will not be described here.
For our purposes, we will just say that for even $N$, the duality statements about $\FF_{SU(N)}$, $\FF^{(0)}_{PSU(N)}$, and $\FF_{U(N)}$ hold
once a choice of $K^{1/2}$ is made.\footnote{\label{brief} In brief, the formulation in \cite{FW} is that, for even $N$, $S$-duality generates
a $B$-field (or gerbe) on $\MH$ that is trivial but not canonically trivial; it can be trivialized by a choice of $K^{1/2}$. (This gerbe is described 
in section \ref{gerbes}.) To state for even $N$ the
duality between $\FF_{SU(N)}$ and $\FF^{(0)}_{PSU(N)}$ or the self-duality of $\FF_{U(N)}$  without making a choice of $K^{1/2}$, one
has to incorporate this $B$-field on one side of the duality.  A direct explanation from four-dimensional gauge theory of why $S$-duality generates
this trivial but not canonically trivial $B$-field has not yet been given.}

We recall that a fiber $\FF_{SU(N)}$ is determined by a spectral cover $\psi:D\to C$, and is parametrized, as we found in  section \ref{whichline},
by the choice of a line bundle $\L\to D$ of degree $N(N-1)(g_C-1)$.
Once a choice of $K^{1/2}$ is made,  $\FF_{SU(N)}$ contains the
canonical point $\psi^*(K^{(N-1)/2})$, and is an abelian variety,
conveniently parametrized by the line bundle
\begin{equation}\label{zumbo}
{\cal L}'={\cal L}\otimes \zeta^*(K^{-(N-1)/2})\end{equation}
which is of degree zero. To describe $\FF_{SU(N)}$ in terms of ${\cal L}'$, one more concept is useful.
Given a covering $\psi:D\to C$ of Riemann surfaces, one defines a
``norm'' map ${\rm Nm}:{\rm Pic}(D)\to{\rm Pic}(C)$. If ${\cal M}\to D$
has divisor $\sum n_i\qq_i$, with integers $n_i$ and $\qq_i\in D$,
then ${\rm Nm}({\cal M})$ has divisor $\sum_in_i\psi(\qq_i)$.  The
norm map differs from the pushforward operation $\cal M\to \det\psi_*(\cal M)$ that we considered
before in that it does {\it not} have a correction for the
ramification. For our $N$-fold covering $\psi:D\to C$, the two are
related by $\det\psi_*({\cal M})={\rm Nm}({\cal M})\otimes
K^{-N(N-1)/2}$, for any ${\cal M}$. (It is enough to verify this
for ${\cal M}={\cal O}$; we have ${\rm Nm}({\cal O})={\cal O}$,
while (\ref{jofo}) implies $\det\psi_*({\cal O})=K^{-N(N-1)/2}$.) So the description
(\ref{yudo}) of the fiber of the Hitchin fibration for $SU(N)$ is
equivalent to saying that $\FF_{SU(N)}$, when parametrized by $\cal L'$, is the subgroup of ${\rm
Jac}(D)$ defined by
\begin{equation}\label{nirk}  \FF_{SU(N)}={\rm Nm}^{-1}({\cal O}).\end{equation}

On the other hand, in terms of ${\cal L}'$, the fiber
$\FF^{(0)}_{PSU(N)}$ of the Hitchin fibration for topologically
trivial $PSU(N)$ bundles is
\begin{equation}\label{lirk}
\FF^{(0)}_{PSU(N)}={\rm Jac}(D)/{\rm Jac}(C).\end{equation}
 That these are dual has
been shown in \cite{HauselThaddeus} in a
fairly elementary way.

A rough explanation is as follows.  If ${\cal J}$ is any self-dual
torus with Poincar\'e line bundle ${\cal P}\to {\cal J}\times
{\cal J}$, and $\cF$ is any subtorus of ${\cal J}$, then the dual of
$\cF$ is ${\cal J}/\cF^\perp$, where $\cF^\perp$ consists of all
$f^\perp\in {\cal J}$ such that ${\cal P}$ restricted to $\cF\times
f^\perp$ is trivial. For ${\cal J}={\rm Jac}(D)$,
$\cF=\FF_{SU(N)}={\rm Nm}^{-1}({\cal O})$, one has
$\cF^\perp=\psi^*({\rm Jac}(C))$, and the dual of $\cF$ is then
$\FF_{PSU(N)}^{(0)}={\cal J}/\cF^\perp={\rm Jac}(D)/{\rm Jac}(C)$. We
explain more in section \ref{topview}, and we describe another
approach to this duality in section \ref{dualap}.

Similarly, when expressed in terms of ${\cal L}'$ instead of
${\cal L}$, (\ref{blonko}) becomes
\begin{equation}\label{pirk}\FF^{(d)}_{PSU(N)}={\rm Pic}_d(D)/{\rm
Jac}(C).\end{equation}
 An equivalent characterization can be found by restating
(\ref{normat}):
\begin{equation}\label{dormat} \FF_{PSU(N)}^{(d)}={\rm
Nm}^{-1}({\cal L}_0)/\Gamma_N,\end{equation} where ${\cal L}_0$ is
any fixed degree $d$ line bundle over $C$. And likewise,
(\ref{palooko}) becomes
\begin{equation}\label{kormat}\hat \FF_{PSU(N)}^{(d)}={\rm
Nm}^{-1}({\cal L}_0).\end{equation}

The analog for $U(N)$ is simply that, when parametrized by the
degree zero line bundle ${\cal L}'$, the fiber of the Hitchin
fibration is a copy of the Jacobian of $D$:
\begin{equation}\label{zormo}\FF_{U(N)}={\rm Jac}(D).\end{equation}
This is certainly a self-dual abelian variety.

\subsection{Topological Viewpoint}
\label{topview}

Here, for $G=SU(N)$, we consider the spectral curve more carefully
from a topological viewpoint.  We will fill in some gaps in
previous arguments and obtain a few useful results.
In addition, given the basic claim that the geometric Langlands
program can be derived from topological field theory, we want to
understand things topologically when possible.

In general, a map $\psi:D\to C$ from one Riemann surface to
another determines an associated map $\psi_*:H_1(D,\Bbb{Z})\to
H_1(C,\Bbb{Z})$.  We simply map the homology class of a loop
$\gamma\subset D$ to the homology class of the corresponding loop
$\psi(\gamma)\subset C$.  We abbreviate the homology groups as
$H_1(D)$ and $H_1(C)$ and denote this map as
\begin{equation}\label{orgok}\psi_*:H_1(D)\to
H_1(C).\end{equation}

The map $\psi_*$ is surjective if every closed loop in $C$ can be
lifted to a closed loop in $D$.  This is not so for every map between Riemann surfaces; for
example, it is not so for an unramified covering $D\to C$.
However, the spectral cover $\psi:D\to C$ has plenty of
ramification, and in this case $\psi_*$ is surjective.  One way to
prove this is to pick a convenient spectral cover, using the fact
that the space of smooth spectral covers is connected, so that any
two smooth spectral covers have the same topology.  We pick a
point $\pp\in C$ and pick a spectral cover for which
$\Tr\,\varphi^n$ vanishes at $\pp$ for $n=2,\dots, N$, with
$\Tr\,\varphi^N$ having only a simple zero at $\pp$.  The equation
of the spectral cover then looks near $\pp$ like $y^N-z=0$, where
$z$ is a local parameter on $C$ that vanishes at $\pp$.  Hence, in
this example, $D$ contains only a single point $\qq$ that lies above
$\pp$. Now, any closed loop in $C$ can be deformed to a path that
begins and ends at $\pp$. Such a path lifts in $D$ to a path that
begins and ends at $\qq$, and so forms a closed loop. So the map
$\psi_*$ is surjective.

We write $\Gamma$ for the kernel of $\psi_*$.  Since $\psi_*$ is
surjective, we get an exact sequence of lattices
\begin{equation}\label{helfgo}0\to \Gamma\to H_1(D)\underarrow{\psi_*}H_1(C)\to
0.\end{equation}  Such a sequence can always be split, though not
canonically:
\begin{equation}\label{norgo}H_1(D)=\Gamma\oplus
H_1(C).\end{equation} Relative to this splitting, the map $\psi_*$
is just
\begin{equation}\label{orgo}\psi_*=0\oplus 1.\end{equation}

We can also define another map $\psi^*:H_1(C)\to H_1(D)$ which
maps the homology class of a closed loop $\gamma\subset C$ to that
of $\psi^{-1}(\gamma)\subset D$.   For any $\gamma\subset C$,
$\psi(\psi^{-1}(\gamma))$ is an $N$-fold cover of $\gamma$, which
means at the level of homology that
\begin{equation}\label{ilbo}\psi_*\circ\psi^*=N.\end{equation}
Relative to the decomposition (\ref{norgo}), we have therefore
\begin{equation}\label{milbo}\psi^*=m\oplus N,\end{equation}
where $m:H_1(C)\to\Gamma$ is some lattice map, and $N:H_1(C)\to
H_1(C)$ is multiplication by $N$.  Because the splitting
(\ref{norgo}) is not canonical, $m$ is only uniquely determined
modulo $N$.

The dual of the exact sequence (\ref{helfgo}) is another exact
sequence
\begin{equation}\label{plygo}0\to H_1(C)^*\underarrow{i} H_1(D)^*\to
\Gamma^*\to 0.\end{equation} Here $\Gamma^*={\rm
Hom}(\Gamma,\Bbb{Z})$ is the dual lattice of a lattice $\Gamma$.
The map $i$ by definition maps $x:H_1(C)\to \Bbb{Z}$ to $x\circ
\psi_*:H_1(D)\to \Bbb{Z}$.  However, $H_1(C)$ and $H_1(D)$ are
self-dual via their intersection pairings, which we denote
$\langle~,~\rangle_C$ and $\langle ~,~\rangle_D$.  If $x:H_1(C)\to
\Bbb{Z}$ is $y\to \langle x',y\rangle_C$ for some $x'$, then
$i(x)$ takes $z\in H_1(D)$ to $\langle x',\psi_*(z)\rangle_C$.
But, for any $x'\in H_1(C)$, $z\in H_1(D)$, we have $\langle
x',\psi_*(z)\rangle_C =\langle \psi^*(x'),z\rangle_D$ (as one can
argue by counting intersections on $C$ and on $D$). This means
that under the self-duality of $H_1(D)$, the element $i(x)\in
H_1(D)^*$ corresponds to  $\psi^*(x')\in H_1(D)$.  So we can write
the exact sequence (\ref{plygo}) in a more useful way:
\begin{equation}\label{lygo}
0\to H_1(C)\underarrow{\psi^*} H_1(D)\to \Gamma^*\to
0.\end{equation} This tells us that $H_1(D)/\psi^* (H_1(C))$ is a
lattice, in fact
\begin{equation}\label{polyfo}
H_1(D)/\psi^*(H_1(C))=\Gamma^*.\end{equation}

The moduli space of flat $U(1)$ bundles on $C$ is ${\cal
J}(C)={\rm Hom}(H_1(C),U(1))$, which we abbreviate as
$H_1(C)^\vee$. Likewise, the moduli space of flat $U(1)$ bundles
on $D$ is ${\cal J}(D)={\rm Hom}(H_1(D),U(1))=H_1(D)^\vee$. There
is a natural map $(\psi_*)^\vee:H_1(C)^\vee\to H_1(D)^\vee$,
taking $\phi:H_1(C)\to U(1)$ to $\phi\circ\psi_*:H_1(D)\to U(1)$.
The fiber of the Hitchin fibration for topologically trivial
$PSU(N)$ bundles is
\begin{equation}\label{doorstop}
\FF^{(0)}_{PSU(N)}={\cal J}(D)/{\cal J}(C)=H_1(D)^\vee/(\psi_*)^\vee
(H_1(C)^\vee).\end{equation} In the context of the splitting in
(\ref{norgo}), $\FF^{(0)}_{PSU(N)}$ parametrizes homomorphisms from
$H_1(D)$ to $U(1)$, except that we do not care what such a
homomorphism does to $H_1(C)$.  So
\begin{equation}\label{hormstop}
\FF^{(0)}_{PSU(N)}={\rm Hom}(\Gamma,U(1))=\Gamma^\vee.\end{equation}

Likewise we define $\psi^*{}^\vee:H_1(D)^\vee\to H_1(C)^\vee$,
taking $\phi:H_1(D)\to U(1)$ to $\phi\circ\psi^*:H_1(C)\to U(1)$.
This is a topological version of the norm map in algebraic
geometry.  The norm map can be defined in algebraic geometry for
line bundles of any degree, but for degree zero it is equivalent
to $\psi^*{}^\vee$, whose definition does not use a complex
structure. The fiber of the Hitchin fibration for $SU(N)$ is
according to (\ref{nirk}) the kernel of the norm map, or
\begin{equation}\label{kerno}\FF_{SU(N)}={\rm
ker}(\psi^*{}^\vee).\end{equation} Concretely, ${\rm
ker}(\psi^*{}^\vee)$ parametrizes homomorphisms $\phi:H_1(D)\to
U(1)$ that are trivial on $\psi^*(H_1(C))$.  These are the same as
homomorphisms to $U(1)$ from $H_1(D)/\psi^*(H_1(C))=\Gamma^*$, so
we have actually
\begin{equation}\label{berno} \FF_{SU(N)}={\rm
Hom}(\Gamma^*,U(1))=(\Gamma^*)^\vee.\end{equation} In particular,
$\FF_{SU(N)}$ is connected, as promised in section \ref{whichline},
and is a torus.

Comparing (\ref{hormstop}) and (\ref{berno}), we see that
$\FF^{(0)}_{PSU(N)} $ and $\FF_{SU(N)}$ are obtained as ${\rm
Hom}(\Gamma,U(1))$ and ${\rm Hom}(\Gamma^*,U(1))$ for two dual
lattices $\Gamma$ and $\Gamma^*$.  Tori obtained in this way are
always canonically dual, as we explained in section \ref{examplo}.

Since $\Gamma$ is a sublattice of $H_1(D)$, the unimodular
intersection pairing on $H_1(D)$ restricts to an integer-valued
bilinear form on $\Gamma$.  This gives a natural injective map
$\Gamma\to \Gamma^*$. If $r_N$ denotes reduction mod $N$, then we
have a natural surjective map $r=r_N\circ \psi_*:H_1(D)\to
\Gamma_N$, where $\Gamma_N=H_1(C,\Bbb{Z}_N)$.  $r_N$ annihilates
$\psi^*(H_1(C))$, in view of (\ref{ilbo}), so it can be regarded
as a map from $\Gamma^*=H_1(D)/\psi^*(H_1(C))$ onto $\Gamma_N$. It
also annihilates $\Gamma$, since $\psi_*$ does, so it really gives
a map from $\Gamma^*/\Gamma$ onto $\Gamma_N$.  Finally, by taking
discriminants, one can show that this map is an isomorphism.  So
we get an exact sequence
\begin{equation}\label{trulfo}0\to
\Gamma\to\Gamma^*\to\Gamma_N\to 0.\end{equation} Taking
homomorphisms to $U(1)$, and using the fact that
$\FF_{PSU(N)}^{(0)}= \Gamma^\vee$ while $\FF_{SU(N)}=(\Gamma^*)^\vee$,
we have
\begin{equation}\label{mikolfo}\FF^{(0)}_{PSU(N)}=\FF_{SU(N)}/\Gamma_N.\end{equation}
This result was explained from the viewpoint of complex geometry
in (\ref{mormot}).

\subsubsection{Characterization of $\FF_{SU(N)}$}
\label{character}

Now, we return to complex geometry and see what we can deduce from
the fact that $\FF_{SU(N)}={\rm Nm}^{-1}({\cal O})$ is a complex
torus.

Let $\pp$ be any point in $C$ and $\qq,\qq'$ two points lying above it
in $D$. Then the line bundle ${\cal O}(\qq)\otimes {\cal
O}(\qq')^{-1}$ over $D$ has norm ${\cal O}$ and so defines a point
in $\FF_{SU(N)}$.

More generally, let $\pp_i$, $i-1,\dots,k$ be any collection of
points in $C$, and for each $i$ let $\qq_i$ and $\qq'_i$ lie above
$\pp_i$ in $D$.  We pick integers $n_i$ and set
${\DD}=\sum_in_i(\qq_i-\qq'_i)$.  Then the line bundle ${\cal L}={\cal
O}({\DD})=\otimes_{i=1}^k\left({\cal O}(\qq_i)\otimes {\cal
O}(\qq'_i)^{-1}\right)^{n_i}$ has trivial norm and lies in
$\FF_{SU(N)}$.

We claim that conversely, if ${\rm Nm}({\cal L})={\cal O}$, then
${\cal L}={\cal O}({\DD})$ for a divisor ${\DD}$  of this form.
The argument depends upon knowing that $\FF_{SU(N)}$ is a complex
torus.  This being so, we can divide by $N$, and find some ${\cal
M}\in \FF_{SU(N)}$ with ${\cal M}^N\cong {\cal L}$. There is some
divisor ${\DD}'=\sum_i n_i\qq_i$ with ${\cal M}={\cal O}({\DD}')$,
and hence ${\cal L}={\cal O}(N{\DD}')$. Now for each $i$, let
$\qq_{i,\alpha}$, $\alpha=1,\dots, N$ be the points in $D$ with
$\psi(\qq_{i,\alpha})=\psi(\qq_i)$ (one of the $q_{i,\alpha}$ is equal to $q_i$; we allow for ramification by
permitting some of the $\qq_{i,\alpha}$ to be equal).  ${\rm
Nm}({\cal M})$ is trivial since $\cal M\in \FF_{SU(N)}$, so $\psi^*({\rm Nm}({\cal M}))$ is also
trivial.  But  $\psi^*({\rm Nm}({\cal M}))={\cal O}({\DD}'')$,
where ${\DD}''=\sum_{i,\alpha}n_i \qq_{i,\alpha}$. So the fact that
$\psi^*({\rm Nm}({\cal M}))={\cal O}$ means that we can
equivalently characterize ${\cal L}$ as ${\cal
O}(N{\DD}'-{\DD}'')$.  But
\begin{equation}\label{helbo}N{\DD}'-{\DD}''=\sum_{i,\alpha}n_i(\qq_i-\qq_{i,\alpha}).\end{equation} This is a
sum of divisors $\qq_i-\qq_i'$ where $\psi(\qq_i)=\psi(\qq'_i)$, so we
have established that every ${\cal L}\in {\rm Nm}^{-1}({\cal O})$
is ${\cal O}({\DD})$ for a divisor ${\DD}$ of that form.

\subsection{The Symplectic Form}
\label{sympform}

The results that we have just obtained are helpful in
understanding another aspect of the geometry of $\MH$ in complex
structure $I$.  At least for unitary groups, we want to describe
in terms of spectral covers the holomorphic symplectic structure
$\Omega_I$, which was introduced via gauge theory in eqn.
(\ref{zolk}). (This detailed description will not be used in the rest of the paper.)
 For more general groups, one would hope to get
similar results based on a more abstract approach to spectral
covers.

 Our approach will be in two steps.  First, we construct
directly a holomorphic symplectic structure on the space of pairs
consisting of a spectral cover $\psi:D\to C$ and a line bundle
${\cal L}\to D$ of the appropriate degree.  Then we will compare
this formula to our expectations from the gauge theory definition.

First we recall the Abel-Jacobi map, which explicitly identifies
the Jacobian of a Riemann surface $D$ as a complex torus.  Let
$g_D$ be the genus of $D$ and pick a basis of $g_D$ holomorphic
one-forms $\omega^a$, $a=1,\dots, g_D$.  Map the first homology
group  $H_1(D,\Bbb{Z})$ to $\Bbb{C}^{g_D}$ by mapping an integral
one-cycle $\bar \gamma$, representing an element of
$H_1(D,\Bbb{Z})$, to the periods
$(\int_{\bar\gamma}\omega^1,\dots,\int_{\bar
\gamma}\omega^{g_D})$.    (By definition, an integral  one-cycle
is a formal linear combination of oriented closed loops with integer
coefficients.) The image of $H_1(D,\Bbb{Z})$ in $\Bbb{C}^{g_D}$ is
a lattice $\Gamma_D$ of rank $2g_D$.   The quotient ${\cal
T}=\Bbb{C}^{g_D}/\Gamma_D$ is a complex torus.

Now suppose that ${\cal L}$ is a line bundle over $D$ of degree
zero. It is isomorphic to ${\cal O}({\DD})$ for some divisor
${\DD}=\sum_in_i\qq_i$, with integers $n_i$ such that $\sum_in_i=0$
and points $\qq_i\in D$.  Since $\sum_in_i=0$, we can find a
one-dimensional ``chain'' $\gamma$ (a formal linear combination of
oriented paths, not necessarily closed, with integral
coefficients) whose boundary $\partial\gamma$ is equal to $\sum_i
n_i\qq_i$. Then we map ${\cal L}$ to the point  $x\in {\cal T}$ with
coordinates $(\int_\gamma \omega^1,\dots,\int_\gamma
\omega^{g_D})$.  This gives a well-defined point in ${\cal T}$,
because $\gamma$ is uniquely determined up to addition of a
one-cycle $\bar\gamma$, whose addition to $\gamma$ will shift the
coordinates of $x$ by an element of the lattice $\Gamma_D$. The
Abel-Jacobi theorem says that the point $x$ depends only on ${\cal
L}$ and not on the choice of a divisor ${\DD}$ representing ${\cal
L}$, and moreover that this map gives an isomorphism of ${\rm
Jac}(D)$ with the complex torus ${\cal T}$.

For spectral curves, this can be implemented in a particularly nice
way.  We begin with the case of $G=U(N)$, and then describe the
minor variations needed for $SU(N)$ and $PSU(N)$.  On the total
space $W$ of the cotangent bundle of $C$, there is a natural
holomophic one-form $\lambda$ which, in terms of a local
coordinate $z$ on $C$, can be written $\lambda=y\, dz$.  Once we
pick a spectral curve $D$, $\lambda$ can be restricted to a
holomorphic differential on $D$.  Of course, $D$ is defined by an
equation $\det(y-\varphi)=0$, which more explicitly takes the form
$\PP(y,z;H_a)=0$, where $\PP$ is a polynomial that is of degree
$N$ in $y$, and in which the commuting Hamiltonians $H_a$ appear
as parameters.

The cohomology class of the restriction of $\lambda$ to $D$
depends on the parameters $H_a$.  One can compute this dependence
by simply differentiating $\lambda$ with respect to $H_a$ at fixed
$z$.  Differentiating the equation $\PP(y,z;H_a)=0$ in this
fashion, we learn that $0=\PP'(\partial y/\partial H_a)+\partial
\PP/\partial H_a$, where $\PP'$ is short for $\partial
\PP/\partial y$.  So
\begin{equation}\label{zonko} {\partial y\over \partial
H_a}=-{1\over \PP'}{\partial \PP\over \partial H_a}\end{equation}
Hence if we set $\omega^a=\partial \lambda/\partial H_a$, we get
\begin{equation}\label{onko} \omega^a={\partial y\over \partial
H_a}dz=-{\partial \PP\over\partial H_a}{dz\over
\PP'}.\end{equation} The objects $\omega^a$ are actually
holomorphic one-forms on $D$. The only point here that is not
completely trivial is that the $\omega^a$ have no pole at
$\PP'=0$. But in fact, for a smooth plane curve $\PP(y,z)=0$, the
differential $dz/(\partial \PP/\partial y)$ is regular at points
with $\PP=\partial \PP/\partial y=0$, as it can also be written
$-dy/(\partial \PP/\partial z)$.

For $G=U(N)$, the number of commuting Hamiltonians is equal to the
genus $g_D$ of $D$.  The $\omega^a$ defined in (\ref{onko}) are
all linearly independent, since the polynomial $\PP(y,z;H_a)$ does
depend nontrivially on all of the $H_a$.  So they give a basis of holomorphic differentials on $D$.

Now we can write down almost by inspection a holomorphic two-form
on $\MH$.  If $D$ is a spectral curve, and ${\cal L}$ a line
bundle over $D$ of degree given in (\ref{ywo}), we write ${\cal
L}=\psi^*(K^{(N-1)/2})\otimes {\cal L}'$, where ${\cal L}'$ has
degree zero.  We represent ${\cal L}'$ by a divisor
${\DD}=\sum_{i=1}^kn_i\qq_i$, and find a one-chain $\gamma$ with
$\partial\gamma={\DD}$.  Then we set
\begin{equation}\label{dunko}
\Omega=\sum_{a=1}^{g_D} dH_a\wedge d\int_\gamma
\omega^a.\end{equation} To see that this does not depend on the
choice of $\gamma$, note that if we add a  one-cycle $\bar\gamma$
to $\gamma$, the change in $\Omega$ is
\begin{align}\label{unko}
\Delta\Omega& =\sum_{a=1}^{g_D} dH_a\wedge d\int_{\bar\gamma}
\omega^a =\sum_{a,b=1}^{g_D}dH_a\wedge dH_b\int_{\bar
\gamma}{\partial \omega^a\over \partial H_b}\\
\nonumber &=\sum_{a,b=1}^{g_D}dH_a\wedge dH_b\int_{\bar
\gamma}{\partial^2 \lambda\over\partial H_a\partial
H_b}=0.\end{align} Similarly,  the choice of  $K^{1/2}$ in the
definition of ${\cal L}'$ does not matter. Changing $K^{1/2}$
has the effect of changing ${\cal L}'$ by a line bundle whose
square is trivial.  So it changes the  vector of periods
$(\int_\gamma\omega^1,\dots,\int_\gamma\omega^{g_D})$ by half of a
lattice vector, or in other words by $1/2$ of $(\int_{\bar
\gamma}\omega^1,\dots,\int_{\bar \gamma}\omega^{g_D})$, for some
one-cycle $\bar\gamma$.  The computation already performed in
(\ref{unko}), but with an extra factor of 1/2, now serves to show
that this operation does not change $\Omega$.

Actually, we can write $\Omega$ in a way that manifestly does not
depend on the choice of one-chain $\gamma$.  In (\ref{dunko}),
when the exterior derivative acts on $\int_\gamma \omega^a$, it
may differentiate either $\gamma$ or $\omega^a$.  However, the
terms in which the one-forms $\omega^a$ are differentiated do not
contribute, again by the same reasoning as in (\ref{unko}).  So we
only have to differentiate the chain $\gamma$, or more precisely,
its endpoints, which are characterized by $\partial\gamma=\sum_i
n_i\qq_i$.  So we have
\begin{equation}\label{wellwrite}\Omega=\sum_{a=1}^{g_D}\sum_{i=1}^kn_i
\,dH_a \wedge (\omega^a,d{\qq_i}).\end{equation}   The meaning of the symbol $d \qq_i$ is as follows.
Since a tangent vector to $\FF_{SU(N)}$ is represented concretely by a first order displacement of the
$\qq_i$, there is, for each $i=1,\dots,N$, a natural map from the tangent space to $\FF_{SU(N)}$ to the tangent
space $TD_{\qq_i}$ to $D$ at $\qq_i$.  Equivalently, for each $i$, there is a natural 1-form on $\FF_{SU(N)}$ with values in $TD_{\qq_i}$.  This has
been denoted $d \qq_i$.
Also,
$(\omega^a,\cdot)$ represents the pairing with $TD$ of the 1-form $\omega^a$ on $D$.  So $(\omega^a,d \qq_i)$ is for each $i$ a 1-form
on $\FF_{SU(N)}$.   And thus $\Omega$ is a 2-form on $\M_H$, as desired.

If we denote the linear coordinates on the fibers of the Hitchin
fibration as $X$, then $\Omega$ is schematically of the form
$dX\wedge dH$.  (The components of $X$ are functions of the $q_i$.) 
This ensures that the functions $H_a$ are
Poisson-commuting and, moreover, by Poisson brackets they generate
linear motion of the $X$'s. Furthermore, under a rescaling of the
Higgs field $\varphi$, the form  $\Omega$ is homogeneous of degree
$1$ (since $\lambda=y\,dz$ has this property and
$dH_a\,\omega^a=dH_a\,\partial\lambda/\partial H_a$ scales in the same way as $\lambda$).
These properties agree with those of the symplectic form
$\Omega_I$ defined in the underlying gauge theory. In section
\ref{compgauge}, we aim to demonstrate  directly that they coincide.

But first we consider the analogous formulas for $G=SU(N)$ and
$PSU(N)$. On the Riemann surface $C$, whose genus is $g_C$, there
are $g_C$ holomorphic differentials $w^\alpha$,
$\alpha=1,\dots,g_C$. Of the $g_D$ holomorphic differentials on
$D$, we can take $g_C$ of them to be pullbacks
$\omega^\alpha=\psi^*(w^\alpha)$ from $C$. There remain $g_D-g_C$
such differentials $\tilde \omega^a$, $a=1,\dots,g_D-g_C$  on $D$
that are not pullbacks from $C$.  We can normalize the
$\tilde\omega^a$ to require that $\psi_*(\tilde\omega^a)=0$ for
all $a$. Here, for a holomorphic differential $\omega=f(y,z)\,dz$
on $D$, $\psi_*(\omega)$ is obtained by pushing $\omega$ forward
to $C$. Concretely, if $y_i$, $i=1,\dots,N$ are the roots of the
equation $\PP(y,z)=0$ (regarded as an equation for $y$ with fixed
$z$), then $\psi_*(\omega)=\sum_i f(y_i,z)\,dz$.

To go from $U(N)$ to $SU(N)$ gauge theory, we remove $N$ commuting
Hamiltonians by setting $\Tr\,\varphi=0$.  The roots $y_i$ of the
characteristic polynomial $\PP(y,z;H_a)$ are the same as the
eigenvalues of $\varphi$, and the fact that $\Tr\,\varphi=0$ means
that $\sum_i y_i=0$.  The holomorphic differential $\lambda=y\,dz$
therefore obeys $\psi_*(\lambda)=\sum_iy_i\,dz=0$.  Since
$\psi_*(\lambda)=0$ for all values of the commuting Hamiltonians
of the $SU(N)$ gauge theory, its derivatives with respect to those
Hamiltonians also vanish.  So
\begin{equation}\label{milbox}0={\partial\over\partial
H_a}\psi_*(\lambda)=\psi_*\left({\partial\lambda\over
\partial H_a} \right),\end{equation}
where here the $H_a$ are the commuting Hamiltonians
of $SU(N)$, not $U(N)$.
Thus,  upon setting $\tilde\omega^a=\partial\lambda/\partial H_a$,
$a=1,\dots, g_D-g_C$, we get precisely the differentials on $D$
that are annihilated by $\psi_*$.

Now let us apply the Abel-Jacobi map to $\FF_{SU(N)}$, which
parametrizes line bundles ${\cal L}'\in{\rm Jac}(D)$ such that
${\rm Nm}({\cal L}')={\cal O}$.  As we have explained  in section
\ref{character}, any such ${\cal L}'$ is isomorphic to ${\cal
O}(\DD)$ where we can take the divisor ${\DD}$ to be
${\DD}=\sum_in_i(\qq_i-\qq_i')$, with $\psi(\qq_i)=\psi(\qq_i')$.  We can
take a chain $\gamma$ with $\partial\gamma=D$ to be a sum of paths
from $\qq_i'$ to $\qq_i$, so that $\psi(\gamma)$ is a closed cycle in
$C$.   As we learned at the end of section \ref{character}, such a closed
cycle in $C$ can be lifted to a closed cycle $\h\gamma\subset D$.  By replacing
$\gamma$ with $\gamma-\h\gamma$, we can assume that $\psi(\gamma)=0$.
The Abel-Jacobi map takes ${\cal L}'$ to the point in
$\Bbb{C}^{g_D}/\Gamma_D$ with coordinates
\begin{equation}\label{gelmo}\left(\int_\gamma \tilde\omega^a,\int_\gamma
\psi^*(w^\alpha)\right).\end{equation} However, if $w$ is a
holomorphic differential on $C$, then $\int_\gamma
\psi^*(w)=\int_{\psi(\gamma)}w=0$, as we have required $\psi(\gamma)=0$. So the image of ${\cal L}'$
under the Abel-Jacobi map is
\begin{equation}\label{elmo}
\left(\int_\gamma\tilde\omega^a,~0\,\right).\end{equation} Having
set the last $g_C$ coordinates to zero, we now are free to shift
$\gamma$ only by an element of the lattice $\Gamma$ which is the
kernel of $\psi_*$ (recall eqn. (\ref{helfgo})).  Shifting
$\gamma$ by an element of $\Gamma$ does not affect the vanishing
of the last $g_C$ entries of the Abel-Jacobi map, since if
$\psi_*(\bar\gamma)=0$ and $w\in H^0(C,K)$, then
$\int_{\bar\gamma}\psi^*(w)=\int_{\psi_*(\bar\gamma)}w=0$.
Shifting $\gamma$ by any other element of $\Gamma$ would destroy
the vanishing of those last $g_C$ entries.

So the Abel-Jacobi map sends $\FF_{SU(N)}$ to
$\Bbb{C}^{g_D-g_C}/\Gamma$. The map is injective because the
Abel-Jacobi map is injective even on the larger space $\FF_{U(N)}$,
and it is surjective on dimensional grounds. So $\FF_{SU(N)}$ is
isomorphic to $\Bbb{C}^{g_D-g_C}/\Gamma$. This is in complete
accord with the assertion in (\ref{berno}) that $\FF_{SU(N)}={\rm
Hom}(\Gamma^*,U(1))$, since ${\rm Hom}(\Gamma^*,U(1))$ is
canonically isomorphic to
$(\Gamma\otimes_{\Bbb{Z}}\Bbb{R})/\Gamma$, which from the point of view
of the complex geometry of  $C$ is the same as
$\Bbb{C}^{g_D-g_C}/\Gamma$.

At this stage, we can imitate either of the two formulas for the
symplectic form that we had in the case of $U(N)$.  For $G=SU(N)$, we define a
holomorphic symplectic form on $\MH$ by
\begin{equation}\label{flunko}
\Omega=\sum_{a=1}^{g_D-g_C} dH_a\wedge d\int_\gamma\tilde
\omega^a=\sum_{a=1}^{g_D-g_C}\sum_{i=1}^kn_i \,dH_a \wedge
(\tilde\omega^a,d\qq_i).\end{equation}

Now what about $G=PSU(N)$?  The component $\MH^{(0)}(PSU(N))$ of
$\MH(PSU(N))$ that parametrizes Higgs bundles of $PSU(N)$ in the
topologically trivial case is just the quotient $\MH(SU(N))/\Gamma_N$, where
$\Gamma_N$ is the group of line bundles of order $N$ on $C$.  The
symplectic form $\Omega$ is invariant under twisting ${\cal L}'$
by a line bundle of order $N$ (as one sees by repeating the
argument in (\ref{unko}) with an extra factor of $1/N$), so the
same formulas give a holomorphic symplectic form on
$\MH^{(0)}(PSU(N))$.

The other components of $\MH(PSU(N))$ similarly are quotients by
$\Gamma_N$ of what we have called $\hat\MH^{(d)}(PSU(N))$ (the
moduli space of solutions of Hitchin's equations with $\xi=d$ up to gauge
transformations that are homotopic to the identity), and again it
suffices to construct the symplectic form over
$\hat\MH^{(d)}(PSU(N))$. We pick a point $\pp\in C$, set ${\cal
L}_0={\cal O}(\pp)^d$, and then, according to (\ref{kormat}), the
fiber of the Hitchin fibration is ${\rm Nm}^{-1}({\cal L}_0)$.  If
${\cal L}'$ is a line bundle with ${\rm Nm}({\cal L}')={\cal
L}_0$, we write ${\cal L}'={\cal O}({\DD}')$ with some divisor
${\DD}'$, and we let ${\DD}=\DD'-d\qq$,  where $\qq\in D$
lies above $\pp\in C$.  Hence ${\rm Nm}(\O(\DD))=0$,   so  the divisor $\DD$ is linearly
equivalent to 
 $\DD''=\sum_in_i(\qq_i-\qq_i')$,  where $\qq_i,\,\qq_i'\in D$ 
 have the same image in $C$. Finding a one-chain $\gamma$ with
$\partial\gamma={\DD''}$, we define the symplectic form by the same
formulas (\ref{flunko}).  

\subsubsection{Comparison To Gauge Theory}
\label{compgauge}

In gauge theory, the holomorphic symplectic form was defined in
eqn. (\ref{zolk}) as
\begin{equation}\label{zinto}\Omega_I={1\over\pi}\int_C|d^2z|\Tr\,\delta\phi_z
\delta A_{\bar z}.\end{equation} Let us remember that the symbol
$\delta$ denotes the exterior derivative on an
infinite-dimensional function space, such as the space of
connections.  When we get down to finite dimensions, we will just
write $d$ for the exterior derivative.

The element $\varphi_z \,dz\in H^0(C,\ad(E)\otimes K)$ is obtained
by pushing down the differential $y\,dz\in H^0(D,K_D)$ from the
spectral cover $D$. In terms of data on $D$, the symplectic form
can be written
\begin{equation}\label{minto}\Omega_I={1\over
\pi}\int_D|d^2z|\delta y\,\delta a_{\bar z},\end{equation} where
the complex structure of the line bundle ${\cal L}'\to D$ is
defined by a $\bar\partial$ operator that can locally be written
(wherever $z$ is a good local parameter, that is, away from
ramification points) as $d\bar z(\partial_{\bar z}+a_{\bar z})$.

The main step in comparing this to our formulas such as
(\ref{wellwrite}) is to find a convenient choice of $a_{\bar z}$.
Of course, $a_{\bar z}$ is only uniquely determined up to gauge
transformation, and we want to find a convenient representative.
If the line bundle ${\cal L}'$ is ${\cal O}(\DD)$, where
$\DD=\sum_in_i\rr_i$,  with $\rr_i\in D$ and $\sum_in_i=0$, then up to an additive
constant, there is a unique solution on $D$ of the equation
$\bar\partial\partial f=i\pi\sum_in_i\,\delta_{\rr_i}$.  Here
$\delta_{\rr_i}$ is a distributional $(1,1)$-form on $D$, supported
at $\rr_i$ and with $\int_D\delta_{\rr_i}=1$.  If for simplicity $\rr_i$
is not a ramification point, so that $z$ is a good local parameter
near $\rr_i$, then near $\rr_i$ we have $f\sim\frac {1}{
2}\ln|z-\rr_i|^2$.  We can define the complex structure of ${\cal
L}'$ by\footnote{This formula means that the unitary connection
$a=dz\,a_z+d{\bar z}\,a_{\bar z}$, with $a_z=-\partial_zf$, has
curvature $f=da=2\pi i\sum_i n_i\delta_{\rr_i}$.  The line bundle
described by a connection with this curvature is indeed ${\cal
O}(\sum_in_i\rr_i)$.} $a_{\bar z}=\partial_{\bar z}f$. This is not a
pure gauge because of the singularity in $f$. Now we can evaluate
$\delta a_{\bar z}$, where $\delta$ refers to the variation in a
change in ${\cal L}'$, that is in the $\rr_i$. We have
\begin{equation}\label{dunkoff}\delta a_{\bar z}d\bar z=
\sum_id
\rr_i{\partial_{r_i}}a_{\bar z}d\bar z= \sum_i d\rr_i\,d\bar
z\,\partial_{\bar z}\partial_{\rr_i} f=-i\sum_i \pi n_i\,
\delta_{\rr_i}\cdot d\rr_i,\end{equation} where the derivative
$\partial_{\rr_i}a_{\bar z}$ was evaluated using the fact that
$a_{\bar z}$ is antiholomorphic away from the $\rr_i$ and has a
known singular behavior at $z=\rr_i$. As in (\ref{wellwrite}), we
think of $d\rr_i$ as a 1-form on $\FF_{SU(N)}$ with values in $TD_{\rr_i}$. In (\ref{dunkoff}), $d\rr_i$ is contracted with the
$(1,1)$-form $\delta_{\rr_i}$ on $D$ to make a 1-form on $\FF_{SU(N)}$ with values in $(0,1)$-forms on $D$.
 The contraction is
denoted $\delta_{\rr_i}\cdot d\rr_i$.  Now when we reduce to $\MH$, we
have   $\delta y \,dz=\sum_a dH_a\,\omega^a$ as in the derivation
of (\ref{dunko}). Using this fact and (\ref{dunkoff}), and also
recalling that $|d^2z|=idz\wedge d\bar z$, we can evaluate
(\ref{minto}) to give
\begin{equation}\label{unkoff} \Omega_I=\sum_{a,i}n_i\, dH_a\wedge
(\omega^a,d\rr_i),\end{equation} in agreement with
(\ref{wellwrite}).

\section{'t Hooft Operators And Hecke Modifications}
\label{thooftheckeop}

\subsection{Eigenbranes}

Our main goal in the rest of this paper is to describe the properties of 't Hooft operators
and Hecke modifications in more depth than was explained in \cite{KW}.   We begin with
 a short introduction, relying on the reader to consult \cite{KW} for more detail.

Starting in four dimensions, one twists $\N=4$ super Yang-Mills theory in two possible ways,
producing two (partial, as in footnote \ref{partial}) topological field theories that we will call the $A$-model and the $B$-model.
Each model has half-BPS line operators -- Wilson operators in the $B$-model and 't Hooft operators in the $A$-model.
The next step is compactification to two dimensions on a Riemann surface $C$.   A four-dimensional line operator, supported
at a point $p\in C$, descends to a line operator in two dimensions.

\begin{figure}[!t]
\begin{center}
\includegraphics[width=3.5in]{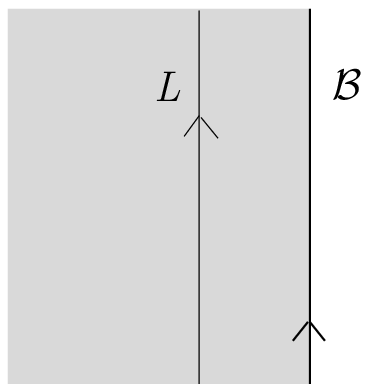}
\end{center}
\caption{ A line operator $L$ approaching a boundary labeled by a brane $\B$.  This gives a new composite boundary condition
$\B'=L\cdot \B$.\label{nuxto}}
\end{figure}

\begin{figure}[!t]
\begin{center}
\includegraphics[width=3.5in]{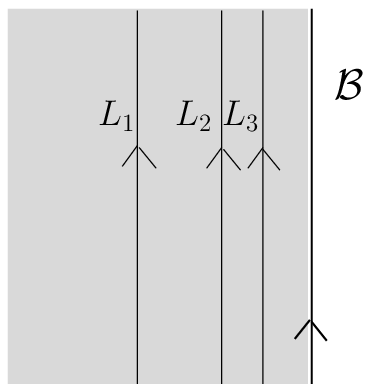}
\end{center}
\caption{Line operators can be brought to the boundary one by one.  This gives an associative action of line operators on branes. \label{buxto}}
\end{figure}

One studies the effective two-dimensional theory that results from
compactification on a two-manifold $\Sigma$ with boundary.  The boundary condition is defined by a brane $\B$. (We will use the same notation for
the brane $\B$ and the corresponding boundary condition.) The key is now to consider the behavior as a line operator 
approaches the boundary (fig. \ref{nuxto}).  Clearly, in a two-dimensional topological field theory, a line operator $L$ approaching a boundary
with boundary condition $\B$ makes a new composite boundary condition $L\B$.  This gives an operation of line operators on branes:
\begin{equation}\label{doofus}L \B =\B'. \end{equation}
One can act on a brane with a succession of line operators that approach the boundary one by one (fig. \ref{buxto}), and clearly the
action of line operators on branes is associative.  Actually, in the present context, there is some commutativity as well.  If $L,L'$ are line operators
supported at distinct points $\pp,\pp'\in C$, then they can be passed through each other in $\Sigma$ without meeting any singularity and therefore
\begin{equation}\label{loggo} LL'\B=L'L\B. \end{equation}  
But actually, in the context of a four-dimensional topological
field theory, a line operator supported at a given point $\pp\in C$ is locally independent of $\pp$ (globally there may be
a nontrivial monodromy if $\pp$ traverses  a noncontractible loop in $C$).  So in trying to more two line operators past each other,
we can always assume that they are inserted at different points in $C$.  Therefore line operators in the 2d theory
that originate from loop operators in four dimensions commute even without the restriction $\pp\not=\pp'$.

Since the line operators of interest do commute, the question arises of whether in some sense they can be simultaneously ``diagonalized.''
To explain the relevant notion requires a few preliminaries.
We think of a brane $\B$ as being represented in the effective two-dimensional description by a brane on $\MH$.  This brane is an $A$-brane or a $B$-brane
depending on which twist we start with in the underlying four-dimensional super Yang-Mills theory.  It is described by a sheaf $\U$ over $\MH$
(whose support may be all of $\MH$ or a submanifold, depending on the brane considered).  There is a natural operation on branes of tensoring
$\U$ with a fixed vector space $V$. Applied to a brane $\B$, this gives a new brane that we call $\B\otimes V$; it is associated to the sheaf $\U\otimes V$
over $\MH$.  If $V$ is of dimension $n$, then roughly speaking $\B\otimes V$ is the sum of $n$ copies of $\B$ (this is a rough description as it does
not take into account the $GL(n,\C)$ group of automorphisms of $V$, which enables one to construct families of branes in which $V$ varies nontrivially).
We say that the brane $\B$ is an eigenbrane of the line operator $L$ if
\begin{equation}\label{zollo}L\B=\B\otimes V \end{equation}
for some vector space $V$.

Unfortunately, it is difficult to find a convenient terminology for the vector space $V$ that appears in this definition.
It plays the role of the eigenvalue $\lambda$ in an ordinary matrix equation $M\uppsi=\lambda \uppsi$, and it is a vector
space.  So one might think of referring to $V$ as an ``eigenvector space,'' but unfortunately
this phrase has another and more elementary
meaning. 

Since the Wilson operators of the $B$-model, or the 't Hooft operators of the $A$-model, commute with each other,
for reasons explained above, it is possible to have a brane $\B$ that is a simultaneous eigenbrane for all Wilson operators
or all 't Hooft operators.  A simultaneous eigenbrane for the Wilson operators is what we call an electric eigenbrane,
and a simultaneous eigenbrane for the 't Hooft operators is what we call a magnetic eigenbrane. 

The geometric Langlands correspondence, as formulated mathematically \cite{BD}, was deduced in \cite{KW} from 
the duality between Wilson operators of $^L\neg G$ and 't Hooft operators of $G$.  In particular, the duality maps
electric eigenbranes of $^L\neg G$ to magnetic eigenbranes of $G$.  But what, concretely, are these branes?

\subsection{The Electric Eigenbranes}\label{eleig}

It is straightforward to find the electric eigenbranes.  In fact, more generally, it is straightforward to describe the action of
Wilson operators on branes. We will give a brief explanation, referring to section 8.1 of \cite{KW} for more detail.
Let $\M$ be the moduli space of $^L\neg G$-bundles on $C$, so that for any $m\in \M$, there is a corresponding
$^L\neg G$-bundle $E_m\to C$.
The ``universal bundle'' over $C$ is a bundle 
$\EE\to C\times \M$ with connection $A$ with the property that for any $m\in \M$, the restriction of $\EE$ to $C\times m$ is isomorphic to $E_m$.
If $^L\neg R$ is an irreducible representation of $^L\neg G$, we write $\EE_{^L\neg R}$ for the associated bundle
in the representation $^L\neg R$.  Actually, if the center of $^L\neg G$ acts nontrivially in the representation $^L\neg R$,
then $\EE_{^L\neg R}\to C\times \M$ is not an ordinary vector bundle but a twisted bundle, twisted by a nontrivial
$B$-field (or gerbe) over $\M$ of finite order.  The physical meaning of this, as explained in \cite{KW} and also in section \ref{gerbes} below, is that
a Wilson operator in the representation $^L\neg R$ carries a discrete electric charge which is measured by the gerbe. This discrete
electric charge  is transformed by $S$-duality
to a discrete magnetic charge carried by the dual 't Hooft operator.

The universal Higgs bundle is an analogous concept.  Let
$\MH$ be the moduli space of $^L\neg G$ Higgs bundles over $C$.  The universal Higgs bundle is an $^L\neg G$-bundle
 $\EE_H\to C\times \MH$, with connection $A$, and endowed with a Higgs field $\phi$ that is a section of $T^*C
\otimes \mathrm{ad}(\EE_H)$ (where here by $T^*C$ we mean really the pullback of $T^*C$ from $C$ to $C\times \MH$) such 
that the restriction of $(A,\phi)$ to $C\times m$ for any $m\in\MH$ is the Higgs bundle $(E,\phi)$ associated to $m$.
(Again $\EE_H$ in general must be understood as a twisted vector bundle.)  By the universal Higgs bundle in representation
$^L\neg R$, we simply mean the bundle $\EE_{H,^L\neg R}\to C\times \MH$ associated to $\EE_H $ in the representation $^L\neg R$.

Given this concept, the action of a Wilson operator on a $B$-brane can be described in general.  Let $\B$ be a $B$-brane
described by a sheaf $\U\to \MH$ and let $W(^L\neg R,\pp)$ be a Wilson operator in the representation $^L\neg R$,
and supported at a point $\pp\in C$.  Then the brane $W(^L\neg R,\pp)\B$ is described by the sheaf $\U\otimes \EE_{H,^L\neg R}|_{\pp\times
\MH}$.  In other words, the action of a Wilson operator  $W(^L\neg R,\pp)$ on the sheaf $\U\to \MH$ that describes a $B$-brane is
\begin{equation}\label{dotto}\U\to \U\otimes  \EE_{H,^L\neg R}|_{\pp\times
\MH}.   \end{equation} 
Here $\EE_{H,^L\neg R}|_{\pp\times \MH}$ is simply the restriction of $\EE_{H,^L\neg R}$ to $\pp\times \MH$; thus
it is a vector bundle over $\MH$ and it makes sense to tensor $\U\to \MH$ with this vector bundle.

The intuition behind the claim in eqn. (\ref{dotto}) is as follows. Let us think of the vertical direction in fig. \ref{nuxto} as
the ``time'' direction.  So a Wilson operator that runs along the boundary is a static, time-independent Wilson operator.
It represents an impurity, in the representation $^L\neg R$, that interacts with the ambient fields.  When the boundary
is in a classical ground state described by a Hitchin pair  $A,\phi$ that corresponds to $m\in\MH$, these are the fields
that the external charge sees.  Eqn. (\ref{dotto}) just says that a Wilson operator supported at $\pp\in C$ interacts
with the fields $A,\phi$ at $\pp$.

Eqn. (\ref{dotto}) shows that a generic $B$-brane is not an electric eigenbrane, because  $\EE_{H,^L\neg R}|_{\pp\times
\MH}$ is a non-trivial vector bundle over $\MH$, not a constant vector space.  An electric eigenbrane will be a brane
$\B$ such that, when restricted to the support of the corresponding sheaf $\U$, $\EE_{H,^L\neg R}|_{\pp\times
\MH}$ becomes (holomorphically) trivial.   This happens if  $\B$ is a zero-brane supported at a point $m\in\MH$.
Then $\U$ is a skyscraper sheaf supported at $m$, and
\begin{equation}\U\otimes  \EE_{H,^L\neg R}|_{\pp\times
\MH}\cong \U\otimes  \EE_{H,^L\neg R}|_{\pp\times m}.   \end{equation} 
On the right hand side, $ \EE_{H,^L\neg R}|_{\pp\times m}$ is the restriction of $\EE_{H,^L\neg R}$ to $\pp\times m\in
C\times \MH$, and in particular is a constant vector space.
Thus a zero-brane $\B$ supported at $m$ is an electric eigenbrane:
\begin{equation}\label{zefor} W(^L\neg R,p)\B=\B\otimes  \EE_{H,^L\neg R}|_{\pp\times m}. \end{equation}

In geometric Langlands, we really want a more precise concept of electric eigenbrane.  First of all, an electric eigenbrane is supposed
to be a brane $\B$ that is a simultaneous eigenbrane for all Wilson operators $W(^L\neg R,\pp)$ for all $^L\neg R$ and all $\pp\in C$.
A convenient way to consider all $\pp$ at once is to restrict $\EE_{H,^L\neg R}$ not to $\pp \times \MH$ but to $C\times \MH$.   Then an electric
eigenbrane should have the property that
\begin{equation}\label{turnof} \U\otimes  \EE_{H,^L\neg R}|_{C\times
\MH}\cong \U\otimes  V^{^L\neg R},   \end{equation} 
where $V^{^L\neg R}$ is the pullback of a vector bundle over $C$.  When restricted to $\pp\in C$ for any $\pp$, this implies the electric eigenbrane
property as stated before.  Furthermore, we want compatibility between these statements for different $^LR$.  The compatibility is that there is a principal
$^LG$ bundle $V\to C$ whose associated bundle in the representation $^L\neg R$ is $V^{^L\neg R}$.

Moreover, we should specify the complex structure on $\MH$ that we are interested in.  Here, crucially, geometric Langlands is a statement about electric
eigenbranes in complex structure $J$, in which $\MH$ parametrizes flat $^LG$ bundles over $C$.  The $B$-model in complex structure $J$ is a reduction
to two dimensions of a four-dimensional topological field theory (very likely a partial one, as explained in footnote \ref{partial}), and this implies that the Wilson
operator $W(^L\neg R,\pp)$ is locally independent of $\pp$ in a natural way.  That translates into the existence of a flat connection on the ``eigenbundle''
$V^{^L\neg R}\to C$ or the corresponding principal bundle $V\to C$.   

A zero-brane supported at a point $m\in \MH$ satisfies these stronger conditions for an electric eigenbrane, with $V$ being the flat bundle that corresponds to
$m$ in complex structure $J$.  These properties characterize a zero-brane in a way that is natural in geometric Langlands.   However, in this paper we will
make use of another property that also characterizes a zero-brane.  This will be described more fully in section \ref{magbranes}, but in brief, a zero-brane
is a brane of type $(B,B,B)$ (a $B$-brane in complex structures $I, J$, and $K$ on $\MH$) that is an electric eigenbrane in each complex structure.

$S$-duality maps branes of type $(B,B,B)$ to branes of type $(B,A,A)$ (a $B$-brane in complex structure $I$ and an $A$-brane in symplectic structure $J$ or $K$)
so in particular, the $S$-dual of a zero-brane will be a brane of type $(B,A,A)$.  These are the magnetic eigenbranes that we will study.
It is not difficult to determine concretely what kind of brane is the $S$-dual of a zero-brane.  Consider a zero-brane $\B$
supported at a point $m\in\MH$ that lies in a particular fiber $^L\neg \FF$ of the Hitchin fibration of $^L\neg G$.
Since $S$-duality acts
as $T$-duality on the fibers of the Hitchin fibration $\MH\to \V$, the $S$-dual of $\B$ will be a rank 1 $A$-brane $\B'$ supported on
the fiber $\FF$ of the Hitchin fibration of $G$ that corresponds\footnote{Fibers of the Hitchin fibration are parametrized by the values of
invariant polynomials in the Higgs field $\varphi$.  As there is a natural correspondence between invariant polynomials on the Lie algebras of $^L\neg G$ and
of $G$, there is a natural correspondence between the fibers of the two Hitchin fibrations.  To be more precise, this correspondence involves a rescaling of the Higgs field that  depends on the gauge coupling constant and will play no
essential role.}  to $^L\neg \FF$.
$\B'$ is described by a flat $\mathrm{Spin}_c$ structure over $\FF$ that encodes the  position of $m$ in $^L\neg \FF$.
We will refer to it as a brane of type $\bf F$.

We will eventually understand more or less explicitly why a rank 1 $A$-brane supported on a fiber of the Hitchin fibration
$\MH\to \V$ is a magnetic eigenbrane.  But we will first develop a better understanding of many properties of 't Hooft operators in general.

\subsection{'t Hooft Operators and Hecke Transformations}\label{thecke}

A half-BPS Wilson operator is represented by the holonomy of the complex connection $\A=A+i\phi.$  In the case of a closed loop $\mathcal S$,
the half-BPS Wilson operator in a representation $^L\neg R$ of $^L\neg G$ is
\begin{equation}\label{wilop}W(^L\neg R,\mathcal S)=\Tr_{^L\neg R}P\exp\left(-\oint_{\mathcal S}(A+i\phi)\right). \end{equation}

Upon $S$-duality to a magnetic description in gauge group $G$, the $A$-dependent part of the holonomy operator is replaced by the Dirac monopole
singularity that defines an 't Hooft operator.  The $\phi$-dependent part remains, and can be interpreted classically as creating a singularity in $\phi$.
The upshot is that a half-BPS 't Hooft operator can be described by a half-BPS solution of the Bogomolny equations.  To be precise, we
consider the four-dimensional spacetime $M=\R^3\times \R$, where $\R$ is the ``time'' direction, parametrized by $s$, and $\R^3$ is parametrized by
a three-vector $\vec x$.  Then an 't Hooft operator at rest at $\vec x=0$ is defined by 
specifying the singularity that the fields should have near $\vec x=0$.
For gauge group $U(1)$ and an 't Hooft operator of charge 1, the solution is characterized by
\begin{align}\label{charec}F & = \frac{i}{2}\star_3 \frac{1}{|\vec x|}\cr  \phi&=\frac{i}{2|\vec x|}d s. \end{align}
In other words, this is a Dirac monopole singularity for $F=dA$, extended to a solution of the Bogomoly equations by including
an analogous point singularity in $\phi$.

In general  \cite{Kapustintwo}, an arbitrary half-BPS 't Hooft operator for gauge group $G$ is defined by picking a homomorphism
\begin{equation}\label{pickhom}\rho:U(1)\to G, \end{equation}
and using this homomorphism to embed the abelian solution (\ref{charec}) in $G$.
In Langlands or GNO duality, the following statement is very fundamental:  
homomorphisms $U(1)\to G$ are classified, up to conjugacy, by dominant highest
weights of the dual group $^L\neg G$, or equivalently by irreducible representations $^L\neg R$ 
of $^L\neg G$.  Thus 't Hooft operators of $G$ are in natural
correspondence with Wilson operators of $^L\neg G$.

$A$-model observables are basically evaluated by solving equations for field configurations with suitable unbroken supersymmetry
and then counting the solutions, or suitably quantizing the space of solutions.  To understand the action of an 't Hooft operator on a brane,
the basic problem to consider is as follows.  The four-dimensional spacetime is $\Sigma\times C$ where $\Sigma$ is a two-manifold with boundary.
The 't Hooft operator of interest runs parallel to the boundary of $\Sigma$, as in fig. \ref{nuxto}.  Near its boundary, we can factorize $\Sigma$
as $I\times \R$ where $\R$ is the ``time'' direction, running along the boundary, and $I$ is a one-manifold with boundary.   The supersymmetric
fields that are relevant to understand the action of the 't Hooft operator on a brane are time-independent, so they are solutions of a gauge
theory equation in three dimensions, namely on $I\times C$.  Though a compact one-manifold $I$ with boundary inevitably has two ends, we focus
attention on just one end, which we denote $\partial I$, as we wish to study an 't Hooft operator near one given boundary of $\Sigma$.  

We will give a brief description of the supersymmetric equations that have to be satisfied, referring to \cite{KW} for more detail.
The relevant equations are most familiar in the case that the  the part of the Higgs field  $\phi$
that is tangent to $C$ vanishes along $\partial I$.  In this case, we can assume that the only nonzero component of $\phi$ is the component
$\phi_s$ in the ``time'' direction, which is forced to be nonzero because it actually has a singular behavior at the position of the 't Hooft operator
(eqn. (\ref{charec})).  One can show a vanishing theorem saying that in a well-behaved solution, the ``time'' component $A_s$ of $A$ vanishes.
The equations then reduce to equations on the three-manifold $I\times C$ for a connection $A$ and adjoint-valued scalar field $\phi_s$.
These equations prove to be the Bogomolny equations:
\begin{equation}\label{zoddo} F=\star D\phi_s. \end{equation}
In the presence of an 't Hooft operator, the relevant solutions of the Bogomolny equations will have Dirac monopole singularities.
(Most of the literature on the Bogomolny equations deals with smooth solutions, but there is also a substantial literature on solutions
with such singularities    
 \cite{kronheimermon,polo,golo,golo2,latwein,bais,hanwit,cherkis}.)

In the more general case that the tangent part of $\phi$ does not vanish, one can still show a vanishing theorem both for $A_s$ and
for the ``normal'' component of $\phi$, that is the component along $I$.  It is convenient to parametrize $I$ by a real variable $y$ and also 
to pick a local coordinate $z$ on $C$. (For $C=\R^2$, $y$ and $z$ can be related to Euclidean coordinates on $\T\times C=\R^3$ by $y=x^1$,
$x=x^2+ix^3$.)   The vanishing theorem says that we can set $A_s=\phi_y=0$.  The supersymmetric equations (at $t=1$ in the notation of
eqn. (3.29) of \cite{KW}; this is a convenient value for studying the $A$-model) in general read 
\begin{equation}\label{irko}0=F-\phi\wedge\phi+\star D_A\phi = D_A\star \phi, \end{equation}
where $D_A=d+[A,\cdot]$ and $\star $ is the Hodge star.  When we specialize to the case $A_s=\phi_y=0$, the equations 
for the remaining fields can be formulated as follows (this description was introduced in \cite{WittenKhov}, section 3.6, and exploited to find some interesting solutions
of the equations).  One defines the three operators 
\begin{align}\label{doff}\D_1 & = \frac{D}{D\bar z}=\partial_{\bar z}  +[A_{\bar z},\,\cdot\, ]\cr
               \D_2& =D_y-i[\phi_s,\,\cdot\,]      \cr 
                  \D_3&=[\phi_z,\,\cdot\,] ,  \end{align}  
and also the ``moment map''
\begin{equation}\label{zolff}\mu=\sum_{i=1}^3[\D_i,\D_i^\dagger].\end{equation} 
The equations for a supersymmetric configuration can then be written
as  a ``complex equation''
\begin{equation}\label{woff} [\D_i,\D_j]=0,~~1\leq i<j\leq 3,\end{equation}
and a ``moment map condition''
\begin{equation}\label{old} \mu=0.\end{equation}  
It will be convenient to refer to the combined system of equations as the extended Bogomolny equations -- extended to include the Higgs field $\varphi=\phi_z\, dz$. 

The complex equation $[\D_i,\D_j]=0$ is invariant under $G_\C$-valued gauge transformations.  Under suitable conditions of semi-stability, which
are satisfied in reasonably simple applications (such as we will consider in this paper),
one expects that a solution of the complex equation modulo complex-valued gauge transformations is 
equivalent to the full system $[\D_i,\D_j]=\mu=0$
modulo $G$-valued gauge transformations.   Thus to determine the moduli space of solutions, one mainly has to understand the complex
equations $[\D_i,\D_j]=0$.

If  $\varphi=0$, then trivially $\D_3=0$.  The remaining
equations $[\D_1,\D_2]=[\D_1,\D_1^\dagger]+[\D_2,\D_2^\dagger]=0$ are the Bogomolny equations (\ref{zoddo}), written in a possibly unfamiliar way.
However, this way of writing the Bogomolny equations is very convenient for application to geometric Langlands.

First we must recall that the $(0,1)$ part of any connection on a $G$-bundle $E\to C$ over a Riemann surface $C$ defines a $\bar\partial$ operator
and turns $E$ (or more precisely its complexification) into a holomorphic $G_\C$-bundle.  Thus in particular, at any given value of $y$,
the operator $\D_1$ is such a $\bar\partial $ operator and gives the bundle $E$ a holomorphic structure.  The equation $[\D_1,\D_2]=0$ tells us
that up to conjugation, the operator $\D_1=D_{\bar z}$ is independent of $y$, since
\begin{equation}\label{zobo}\frac{\partial}{\partial y}D_{\bar z}=-[A_y+i\phi_s,D_{\bar z}].  \end{equation}
Thus as long as this equation is obeyed, the holomorphic structure of the bundle $E$ is independent of $y$.  

Now suppose that there is an 't Hooft operator at some point $y=y_0$, $z=z_0$.  At that point, there is a delta-function source in the Bogomolny
equations, and the condition $[\D_1,\D_2]=0$ is not satisfied.  The holomorphic type of the bundle $E$ actually does jump in crossing $y=y_0$;
it has one type for $y>y_0$ and another for $y<y_0$.  However, the singularity associated with the 
't Hooft operator occurs only at one point $z=z_0$ in $C$.
If we omit this one point from $C$, then the bundle $E $ is unchanged holomorphically even in crossing the past $y=y_0$.  

A modification of a holomorphic $G$-bundle $E\to C$ that is trivial if a single point $\pp$ is removed from $C$ is called in the context of the geometric
Langlands program a Hecke modification of $E$ at $\pp$.  (The terminology is based on an analogy with Hecke operators in number theory.)   
Thus an 't Hooft operator induces a Hecke modification of the holomorphic $G$-bundle $E$ at the point in $C$ at which it is inserted.

An important fact in mathematical approaches to geometric Langlands is that there are different types of possible Hecke modifications for
a $G$-bundle, and that these possible types are classified by the choice of an irreducible representation $^L\neg R$ of the dual group
$^L\neg G$.  What we have described is a physical interpretation of this:  the possible types of Hecke modifications are determined
by the choice of an 't Hooft operator, which determines the precise nature of the singularity in the solution of the Bogomolny equations.
Moreover the 't Hooft operators are indeed classified by the choice of $^L\neg R$.

It is not difficult to include the Higgs field $\varphi$ in this discussion.  At fixed $y$, we have the equation $[\D_1,\D_3]=0$,
which tells us that $[D_{\bar z},\varphi]=0$, in other words the pair $A_{\bar z},\varphi$ is a Hitchin pair and determines a Higgs bundle,
which we regard as a pair consisting of a holomorphic bundle $E$ along with $\varphi\in H^0(C,K\otimes\ade)$.
The equations $[\D_2,\D_1]=[\D_2,\D_3]=0$ or
\begin{equation}\label{dolgo}[D_y-i\phi_s,D_{\bar z}]=[D_y-i\phi_s,\varphi]=0 \end{equation}
tell us that, away from the position of a possible 't Hooft operator, the holomorphic type of the Higgs bundle $(E,\varphi)$ is independent of $y$.
In the presence of an  't Hooft operator at $y=y_0$ and at a point $\pp\in C$, the holomorphic type of the pair $(E,\varphi)$ will jump at $y=y_0$,
but in a way that is trivial if we omit the point $\pp$ from $C$.  We can describe this by saying that in crossing $y=y_0$,
the pair $(E,\varphi)$ undergoes
a Hecke modification at the point $\pp\in C$.  

If we simply forget $\varphi=\D_3$ and the equations it enters, and remember only $\D_1,D_2$ and the condition $[\D_1,\D_2]=0$, we see that a Hecke modification of $(E,\varphi)$
consists, in particular, of an ordinary Hecke modification of $E$.   Now remembering $\varphi$, the equations $[\D_1,\D_1]=[\D_2,\D_3]=0$ imply that the holomorphic
type of $(E,\varphi)$ is independent of $y$ for $y\not=y_0$.  But what happens when we cross $y=y_0$?
The condition $0=[\D_2,\D_3]=[D_y-i\phi_s,\varphi]$ 
 determines what happens to $\varphi$
in crossing $y=y_0$.  If $\varphi$ is chosen generically for, say, $y>y_0$, it will have a pole at $z=z_0$ for $y<y_0$.
 This important fact will be explained in section \ref{inco}.  Given this fact, the possible Hecke modifications of a Hitchin pair $(E,\varphi)$ at a specified point $\pp\in C$ are
a subset of the possible Hecke modifications of $E$ at $\pp$.  We will refer to the Hecke modifications of $(E,\varphi)$ as $\varphi$-invariant
Hecke modifications of $E$.  The rationale for this terminology will become clear.    Hecke modifications of Higgs bundles $(E,\varphi)$ have
been considered mathematically \cite{donagip}.

Before understanding Hecke modifications of the pair $(E,\varphi)$, one should first be familiar with Hecke modifications of $E$ by itself, in the special case $\varphi=0$.
The reader may want to consult sections 9 and 10 of \cite{KW}, where an introduction can be found.  Here we will just recall a few facts which are the minimum
that we will need for our study of magnetic eigenbranes.

\subsection{Basic Examples}\label{basics}

\subsubsection{$G=U(1)$}\label{guone}

The most basic case to consider is the group $G=U(1)$.  In this case, $E$ is a complex
line bundle that we denote as $\L$.  The isomorphism type of $\L$ is going to be constant for $y\not=y_0$,
for some $y_0$.  We denote $\L$ as $\L_-$ for $y<y_0$ and as $\L_+$ for $y>y_0$.

In crossing $y=y_0$, $\L$ will be modified in a way that is trivial except at a point $\pp\in C$. 
We can pick a trivialization of $\L_-$ in a neighborhood of $\pp$ and thus identify it with a trivial line bundle $\O$.
Away from $\pp$, the trivialization of $\L_-$ determines a natural trivialization of $\L_+$ by parallel transport in the $y$ direction, using
the connection $D_y-i\phi_s$.  In other words, if for $y<y_0$ a holomorphic section $s$ of $\L_-$ gives a trivialization of $\L_-$
in a neighborhood $U$ of $\pp$, then parallel transport of $s$ to $y>y_0$ will give a trivialization of $\L_+$ over $U\backslash \pp$ (that is,
over $U$ with the point $\pp$ omitted).  Hence, over $U\backslash p$, $\L_+$ is naturally identified with $\O$.

But the trivialization of $\O$ does not necessarily extend over the point $\pp$.  The general possibility is that, after being parallel transported
to $y>y_0$, the section $s$ may have a zero of order $m$ (or a pole of order $-m$) at $\pp$ for some integer $m$.  
Thus, after identifying $\L_-$ with $\O$,
$\L_+$ may be identified with $\O(m\pp)$.  Here $\O(m\pp)$ is the line bundle whose local sections near $\pp$ are holomorphic
functions that are allowed to have a pole of order $m$ (or required to have a zero of order $-m$) at $\pp$.

It is explained in \cite{KW}, section 4.5, that the Hecke transformation $\O\to \O(m\pp)$ is the result of inserting at the point $\pp\times y_0\in
C\times \R$ an 't Hooft operator $T(m)$ of magnetic charge $m$. (We denote this 't Hooft operator as $T(m;\pp)$ if we wish to specify
the point $\pp\in C$ at which it is inserted.)  The basic idea in showing this is that such an 't Hooft operator creates $m$
units of magnetic flux.  The operation that creates $m$ units of magnetic flux at a point $\pp\in C$ is described in algebraic geometry as
the Hecke transformation $\O\to \O(m\pp)$. 

\subsubsection{$G=U(2)$}\label{gutwo}

For a second example, let us take $G=U(2)$.  The bundle $E$ is now a complex vector bundle of rank 2 and we denote it as $E_-$ or $E_+$ for
$y<y_0$ or $y>y_0$.  Since a holomorphic vector bundle is locally trivial, we can pick an identification of $E_-$ as $\O\oplus \O$ for $y<y_0$.
A basic Hecke operation now transforms $E_-$ to
\begin{equation}\label{zobbox} E_+ =\O(\pp)\oplus \O  \end{equation}
for $y>y_0$.   As explained in \cite{KW}, this is the Hecke transformation implemented by an 't Hooft operator dual to the natural two-dimensional
representation of $^L\neg U(2)\cong U(2)$.  We can think of this as the representation of $U(2)$ with highest weight $^Lw=(1,0)$.  We denote
the corresponding 't Hooft operator as $T(^Lw)$ (or $T(^Lw; \pp)$).  We refer to a Hecke transformation induced by $T(^Lw)$ as a Hecke
transformation of type $^Lw$.

What is exhibited in eqn. (\ref{zobbox}) is a special case of a Hecke transformation induced by $T(^Lw)$ for $^Lw=(1,0)$.  The reason that it is a special
case is that a particular decomposition of $E_-$ as $\O\oplus \O$ was used.  More generally, instead of saying that a section of $E_+$ is
a section of $\O\oplus\O$ that may have a simple pole in the first component, we can allow a simple pole in a specified linear combination
of the two components.  For this, we pick a pair of complex numbers $(u,v)$, not both 0, and we say that a section of $E_+$ is a pair
$(f,g)$ defining a section of $\O\oplus \O$ away from $\pp$, such that $f$ and $g$ are allowed to have a simple pole at $\pp$, but the residue
of this pole must be a multiple of $(u,v)$.  In formulas, if $z$ is a holomorphic function on $C$ with a simple zero at $\pp$,
we require
\begin{equation}\label{zummo} (f,g)=(f_0,g_0)+\frac{\lambda}{z}(u,v), \end{equation}
where $f_0$ and $g_0$ are holomorphic at $\pp$, and $\lambda$ is a possibly nonzero complex number.  The bundle $E_+$ that is defined
by this procedure clearly depends on the pair $(u,v)$ only up to overall scaling, and thus the family of bundles $E_+$ that can be built this
way, starting from $E_-$ and making a Hecke transformation of type $(1,0)$, is  parametrized by a copy of $\CP^1$.  For future
reference, it is convenient to rewrite eqn. (\ref{zummo}) with the sections regarded as column vectors rather than row vectors.  Thus
a holomorphic section $s$ of $E_+$ takes the form 
\begin{equation}\label{onkey}s=s_0+\frac{\lambda}{z}\begin{pmatrix}u \cr v\end{pmatrix}, \end{equation}
where $s_0$ is a holomorphic section of $E_-$ and
the column vector $\begin{pmatrix} u\cr v\end{pmatrix}$ now represents an element of $\CP^1$ with homogeneous coordinates
$u,v$.

We could reach this result more intrinsically without ever picking a local trivialization of $E_-$.  We write $E_{-,p}$ for the fiber of $E_-$ at
$p$, and pick a nonzero vector $\bb\in E_{-,p}$.  Then we characterize $E_+$ by saying that a section of $E_+$ near $p$ has the form
\begin{equation}s=s_0+\frac{\bb}{z}\lambda. \end{equation}
Thus, we allow a simple pole, but its residue must be a multiple of $\bb$.  The bundle obtained this way depends on $\bb$ only up to scaling,
so the family of such bundles is a copy of $\CP^1$ that is obtained by projectiving the two-dimensional vector space $E_{-,p}$.  We denote
this projectivization of $E_{-,p}$ as $\Bbb P(E_{-,p})$.

Thus, there is a natural space of Hecke modification of type (1,0) of given bundle $E_-\to C$ at a specified point $p\in C$.
This space is a compact smooth manifold, which is a copy of $\CP^1$, naturally isomorphic to $\Bbb P(E_{-,p})$. 

Up to a certain point, we can treat arbitrary 't Hooft operators of $U(2)$ in a similar way.  The highest weight of an arbitrary
representation of $^L\neg U(2)\cong U(2)$ is given by a pair of integers $(n,m)$.  The Weyl group acts by exchanging the two weights,
so by a Weyl transformation, we can take $n\geq m$.  

An example of a Hecke modification of type $(n,m)$ is the one that maps $E_-=\O\oplus \O$ to
\begin{equation}\label{doffy}E_+=\O(n\pp)\oplus \O(m\pp). \end{equation}
This is a special case of a Hecke modification at $\pp$ of type $(n,m)$.  The full family of such Hecke modifications has complex
dimension $n-m$.  However, there is an important complication compared to the case $n-m=1$ that was treated above.  The space of
Hecke modifications of type $(n,m)$ for $n-m\geq 2$ is not compact, or better, it has a natural compactification that involves allowing
Hecke modifications of lower weights (of weight $(n-k,m+k)$ for $1\leq k\leq (n-m)/2$).  From the point of view of 't Hooft operators,
this compactification involves ``monopole bubbling,'' in which a smooth BPS monopole, in the field of an 't Hooft operator, shrinks down
near the 't Hooft operator and disappears.\footnote{Such bubbling is relatively familiar for Yang-Mills instantons in four dimensions,
but may be unfamiliar for BPS monopoles as it does not occur in the absence of 't Hooft operators.  More precisely, for gauge group $U(2)$,
it does not occur
except in the presence of 't Hooft operators whose weights obey $n-m\geq 2$.}  The compactified space of Hecke modifications of type $(n,m)$
for $n-m\geq 2$ has singularities associated to monopole bubbling.   In the literature on geometric Langlands, the space of Hecke modifications
of a given type is called a Schubert cell, and its compactification is called a Schubert cycle (in the affine Grassmannian).

An introduction to these matters can be found in section 10 of \cite{KW}.  Here, however, we prefer to avoid the complications associated to
monopole bubbling and the singularities of the compactified space of Hecke modifications.  Accordingly, we will limit ourselves to
't Hooft operators $T(n,m)$ with $n-m\leq 1$.

The condition $n-m\leq 1$ amounts to saying that the representation of $^L\neg U(2)=U(2)$ of highest weight $(n,m)$ is ``minuscule.''
A representation of a compact Lie group $^L\neg G$ is called minuscule if its weights form a single orbit of the Weyl group.  
The representation of $U(2)$ with highest weight $(m,m)$ is a 1-dimensional representation in which $g\in U(2)$ acts by multiplication
by $(\det g)^m$.  The representation with highest weight $(m+1,m)$ is a 2-dimensional representation in which (regarding $g$ as a
$2\times 2$ matrix) $g$ acts by
multiplication by $g(\det g)^m$.  These are minuscule representations, since for example the 2-dimensional representation just mentioned
has precisely 2 weights, which are exchanged by a Weyl transformation.

In general, let $^L\neg R$ be an irreducible representation of any compact group $^L\neg G$, with highest weight $^Lw$.  The space of
Hecke modifications of type $^Lw$ is compact, or equivalently there is no monopole bubbling in the field of an 't Hooft operator $T(^Lw)$,
if and only if the representation $^L\neg R$ is minuscule.  If $^L\neg G$ is semi-simple, minuscule representations are in 1-1 correspondence
with non-trivial characters of the center of $^L\neg G$. Indeed, the smallest representation of $^L\neg G$ that transforms as a given character
of the center of $^L\neg G$ is always minuscule.  For example,  $^L\neg G=SU(2)$ has precisely one non-trivial minuscule representation,
which is the 2-dimensional representation.  (As the Weyl group of $SU(2)$ has only 2 elements, a representation of $SU(2)$ with
dimension greater than 2 cannot possibly be minuscule.)  The group $^L\neg G=SO(3)$ has no non-trivial minuscule representation.

\subsubsection{$G=SO(3)$ or $SU(2)$}\label{morz}

Now let us consider the cases that $G$ is $SO(3)$ or $SU(2)$.

The dual group of $G=SO(3)$ is $^L\neg G=SU(2)$.  Before specializing to $SU(2)$, we make some remarks
about $U(N)$ and $SU(N)$, which will be used in section \ref{zoff}.

The weight lattice of $U(N)$ is spanned by $N$-plets $(m_1,m_2,\dots, m_N)$ of integers $m_1,\dots, m_N$.   The Weyl group
acts by permutations, so up to a Weyl transformation we can impose a dominant weight condition  $m_1\geq m_2\geq \dots\geq m_N$.

The weight lattice of $SU(N)$ is spanned by similar $N$-plets $(m_1,m_2,\dots, m_N)$, but now we take $m_i\in \Z/N$,
but with $m_i-m_j\in \Z$, and we require $\sum_im_i=0$. The Weyl group still acts by permutations, and a dominant weight still obeys $m_1\geq m_2\geq\dots\geq m_N$.

Specializing this description to $^L\neg G=SU(2)$, this means that a dominant weight has the form $^Lw=(n/2,-n/2)$, with an integer $n$.
Now let us consider the Hecke modification associated to an 't Hooft operator $T(^L\neg w)$.  Naively speaking, a typical
such Hecke modification at $p$ maps a bundle $E_-=\O\oplus \O$ to
\begin{equation}\label{lapel}E_+=\O(n\pp/2)\oplus \O(-n\pp/2).  \end{equation}

However, we should ask what this means if $n$ is odd so that $n/2\notin\Z$.  The answer is that for $^L\neg G=SU(2)$, $G=SO(3)$.
The group $SO(3)$ does not have a two-dimensional representation, so we should not try to think of a $G$-bundle as a rank 2 complex
vector bundle.  To make a rigorous statement, instead of a rank 2 complex vector bundle $E$ we should consider the associated
bundle in the adjoint representation of $SO(3)$; this is $V=\ad_0(E)$.  (For a rank $N$ complex vector bundle $E$ with dual $E^*$,
we write $\ad(E)$ for $E\otimes E^*$ and $\ad_0(E)$ for the traceless part of $E\otimes E^*$.)  Note that $V$ is endowed with a holomorphic,
nondegenerate bilinear form, coming from $(v,v')=\Tr\,vv'$, for $v,v'\in \ad_0(E)$.  For $E_-=\O\oplus \O$, we have $V_-=\ad(E_-)
=\O\oplus \O\oplus\O$, and for $E_+$ defined informally as in eqn. (\ref{lapel}), we have $V_+=\ad_0(E_+)=\O\oplus \O(n\pp)\oplus \O(-n\pp)$.
Thus a typical example of the action of $T(^Lw)$ on the $SO(3)$ bundle $V$ is
\begin{equation}\label{apel}\O\oplus\O\oplus \O\to \O\oplus \O(n\pp)\oplus \O(-n\pp). \end{equation}
The quadratic form on the $SO(3)$ bundle $\O\oplus \O(n\pp)\oplus \O(-n\pp)$ pairs $\O$ with itself and $\O(n\pp)$ with $\O(-n\pp)$, so in particular $\O(n\pp)$ and $\O(-n\pp)$
are null subspaces.  

The only nontrivial minuscule representation of $^L\neg G=SU(2)$ corresponds to $n=1$, so that $^Lw=(1/2,-1/2)$ is the highest weight of the
2-dimensional representation of $SU(2)$.   Then eqn. (\ref{apel}) gives the local form of the action of the corresponding 't Hooft operator
$T(^Lw)$ on an $SO(3)$ bundle.  However, a description in terms of the $SO(3)$ bundle $V$, although rigorous, tends to be lengthy,
and it is simpler to relate an $SO(3)$ bundle $V$ to a rank 2 complex vector bundle $E$, possibly of nontrivial determinant, via
$V=\mathrm{ad}_0(E)$.  In doing this, tensoring $E$ with a line bundle $\L$ does not matter (since $\ad_0(E)$ is naturally isomorphic
to $\ad_0(E\otimes \L)$).  Instead of saying that $T(^L\neg w)$ maps $E_-=\O\oplus \O$ to the formal expression written in eqn. (\ref{lapel}),
we could just as well tensor formally with $\L=\O(n\pp/2)$ and say that
\begin{equation}\label{zapel} E_+=\O(n\pp)\oplus \O,\end{equation}
again with the understanding that we are really interested not in $E_+$ but in $V_+=\ad(E_+)$.  

Thus the most convenient way to describe the action of the 't Hooft operator $T(^Lw)$, for $^Lw$ the minuscule weight $(1/2,-1/2)$ of
$SU(2)$, will be to say that locally it maps $E_-=\O\oplus \O$ to
\begin{equation}\label{inrtz}E_+=\O(\pp)\oplus \O. \end{equation}
When we say this, we always bear in mind that this transformation from $E_-$ to $E_+$ is a shorthand way to describe
the Hecke modification from $V_-=\ad_0(E_-)=\O\oplus \O\oplus \O$ to $V_+=\ad_0(E_+)=\O\oplus \O(\pp)\oplus \O(-\pp)$.  
The description by $E_\pm$ is very useful, but not completely canonical, since without changing $V_\pm$, we could replace
$E_\pm$ by $E_\pm \otimes \L$, where $\L$ is a line bundle, for instance $\L=\O(k\pp)$ for some $k\in \Z$.

For $G=SU(2)$, we have $^L\neg G=SO(3)$.  A highest $^Lw$ weight of $SO(3)$ is just a highest weight of $SU(2)$ that is divisible by 2,
so it has the form $(k,-k)$ for some integer $k$.  Thus the generic local action of $T(^Lw;\pp)$ on an $SU(2)$ bundle is
\begin{equation}\label{zobbo}\O\oplus\O\to \O(k\pp)\oplus \O(-k\pp). \end{equation}
This makes sense as a transformation of rank 2 bundles, in keeping with the fact that for $^L\neg G=SO(3)$, we have $G=SU(2)$.
However, the group $SO(3)$ has no nonzero minuscule weights (since its center is trivial), so in studying Hecke modifications for $G=SU(2)$, the complications due to monopole bubbling are inescapable.

\subsubsection{$G=U(N)$, $PSU(N)$, or $SU(N)$}\label{zoff}

The cases that $G$ is $U(N)$, $PSU(N)$, or $SU(N)$ are quite similar to what we have just described for $N=2$.

For $G=U(N)$, the dual group is also $^L\neg G=U(N)$.  As already remarked, a highest weight of $U(N)$ is a sequence of integers
$^L\neg w=(n_1,n_2,\dots, n_N)$ with $n_1\geq n_2\geq\dots
\geq n_N$.  A typical Hecke modification at $\pp$ of type $^L\neg w$ acts by
\begin{equation}\label{omb} \O\oplus \O\oplus \dots \O \to \O(n_1\pp)\oplus \O(n_2\pp)\oplus \dots \oplus \O(n_N\pp). \end{equation}

The representation of weight $^L\neg w$ is minuscule if and only if $n_1-n_N\leq 1$, or equivalently if and only if the integers $n_1,\dots,n_N$
take at most two values.  The case $(n_1,\dots,n_N)=(m,m,\dots,m)$ corresponds to a 1-dimensional representation in which $g\in U(N)$
acts as multiplication by $(\det g)^m$.  More interesting is the case $(n_1,n_2,\dots,n_N)=(m+1,m+1,\dots,m+1,m,m,\dots,m)$, where we will
write $k$ for the number of $m+1$'s.  This corresponds to the $k^{th}$ copy $\wedge ^k W$ of the fundamental $N$-dimensional
representation $W$ of $U(N)$, tensored with a one-dimensional representation,
such that $g\in U(N)$ acts by $(\det g)^m\wedge^kg$.  (Here $\wedge^kg$ is the matrix by which $g$ acts on $\wedge^kW$.)

Let us describe in more detail the Hecke modifications dual to the minuscule representation $\wedge^kV$, which corresponds to the weight
$^Lw=(1,1,\dots,1,0,\dots,0)$, with $k$ 1's.  Specializing eqn. (\ref{omb}) to this case, we see that for some decomposition as
$E_-$ as a direct sum of trivial line bundles, the transformation will be
\begin{equation}\label{gelfix} \O\oplus \O \oplus \dots\oplus \O \to{ \O(\pp)\oplus \O(\pp)\oplus \dots\O(\pp)}\oplus \O\oplus \dots\oplus\O, \end{equation}
with $k$ summands $\O(\pp)$.  However, we can describe this in a more invariant way and thereby describe the space of all Hecke
modifications at $\pp$ that are of type $^Lw$ for this weight.  Let $W$ be an arbitrary $k$-dimensional subspace of $E_{-,\pp}$, the fiber at $\pp$
of $E_-$.  Then we describe $E_+$ by saying that a holomorphic section of $E_+$ near $\pp$ takes the form
\begin{equation}\label{durfy}s= s_0+\frac{\bb}{z} ,\end{equation}
where $s_0$ is a local holomorphic section of $E_-$ near $\pp$, and $\bb$ is a local holomorphic section of $E_-$ such that $\bb(\pp)\in W$.
In other words, $s$ is allowed to have a simple pole at $\pp$, but the residue of the pole must lie in $W$.  Since $W$ is an arbitrary $k$-dimensional
subspace of the $N$-dimensional vector space $E_{-,\pp}$, the family of Hecke modifications of this type is a copy of $\mathbf{Gr}(k,N)$, the
Grassmannian of $k$-planes in $\C^N$.

For $G=PSU(N)$, we have $^L\neg G=SU(N)$.  A highest weight of $^L\neg G$ is an $N$-tuple $^Lw=(n_1,n_2,\dots, n_N)$ now
with $n_i\in\Z/N$ and $n_i-n_j\in\Z$, $\sum_i n_i=0$.  A generic Hecke modification at $\pp$ of a rank $N$ bundle $E_-=\O\oplus\O\dots\O$ of type
$^Lw$
still maps it, formally, to
\begin{equation}\label{zombo}E_+=\O(n_1\pp)\oplus \O(n_2\pp)\oplus\dots\oplus \O(n_N\pp), \end{equation}
but now, because of the fractions in this formula, we shoiuld interpret this as a Hecke modification from $V_-=\ad_0(E_-)$
to $V_+=\ad_0(E_+)$.  The other remarks that we made in the $N=2$ case also have close analogs.  The nonzero minuscule
weights of $SU(N)$ have the form $^Lw=(1-k/N,1-k/N,\dots,1-k/N,-k/N,-k/N,\dots,-k/N)$, with $k$ copies of $1-k/N$.  This
is the highest weight of the $k^{th}$ exterior power of the $N$-dimensional representation of $SU(N)$.  The space of Hecke modifications
dual to this representation is again a copy of $\mathbf{Gr}(k,N)$.

For $G=SU(N)$, and thus $^L\neg G=PSU(N)$, a highest weight $^Lw$ of $^L\neg G$ is an $N$-plet $(n_1,n_2,\dots,n_N)$ with $n_i\in \Z$,
$\sum_i n_i=0$.  The generic local action of $T(^Lw;\pp)$ is
\begin{equation}\label{exc}\O\oplus\O\oplus\dots\oplus\O\to \O(n_1\pp)\oplus \O(n_2\pp)\oplus \dots \oplus \O(n_N\pp). \end{equation}
There are no nonzero minuscule weights.

\subsection{Incorporation Of The Higgs Field}\label{inco}

\subsubsection{Generalities}\label{generalities}

In our application, we are really interesting in Hecke transformations of a Higgs bundle  $(E,\varphi)$, with $\varphi\in H^0(C,\ad(E))$,
and not just of the bundle $E$.  In the presence of an 't Hooft operator at a point $\pp\times y_0\in C\times \R$,
the holomorphic type of the Higgs bundle will jump in crossing $y=y_0$, but in a way that is trivial if we omit the point $\pp$ from $C$.
 We write $(E_-,\varphi_-)$ for the Higgs bundle at $y<y_0$ and $(E_+,\varphi_+) $ for the Higgs bundle at $y>y_0$. 

To understand what happens, we  simply have to consider the implications of eqn. (\ref{dolgo}).
This equation tells us that away from the point $\pp\in C$, parallel transport in the $y$ direction using the connection $D_y-i\phi_s$ gives
a holomorphic isomorphism from $(E_-,\varphi_-)$ to $(E_+,\varphi_+)$.  We already know that this statement does not uniquely determine
$E_+$ in terms of $E_-$:  a Hecke transformation may be made at the point $\pp$, and in general the charge of the 't Hooft operator
at the point $\pp\times y_0$ does not uniquely determine what this Hecke transformation will be.   But away from the point $\pp\in C$,
parallel transport in the $y$ direction gives a distinguished isomorphism $\Theta:E_-\to E_+$.  Moreover, since $\varphi$ is invariant
under that parallel transport, the relation between $\varphi_+$ and $\varphi_-$ is just 
\begin{equation}\label{onk}\varphi_+=\Theta\varphi_-\Theta^{-1}.\end{equation}  In these
statements, $\Theta$ is a gauge transformation away from the point $\pp\in C$, but may have a singularity at $\pp$.

The relation (\ref{onk}) ensures  that $\varphi_+$ is holomorphic away from $\pp$, since $\varphi_-$ is holomorphic
and $\Theta$ and $\Theta^{-1}$ are holomorphic away from $\pp$.
 The extension of $\varphi_+$ over $p$ as a holomorphic section of $K\otimes\ad(E)$
is unique if it exists.  But generically this extension will not exist: $\varphi_+=\Theta\varphi_-\Theta^{-1}$ will have a pole at $\pp$.
Thus the possible Hecke transformations of a pair $(E_-,\varphi_-)$ are simply Hecke transformations of $E_-$ that obey a condition
such that $\varphi_+$ will not have a pole.

Even if $\Theta$ is singular, the conjugacy (\ref{onk}) implies that invariant polynomials in $\varphi_+$ equal the corresponding
invariant polynomials in $\varphi_-$.  For example, for $G$ a unitary group,
\begin{equation}\label{zonk}\Tr\,\varphi_-^s=\Tr\,\varphi_+^s, ~~{\mathrm{for~all}}~s. \end{equation}

The case that $^L\neg G=G=U(1)$ is simple to describe, but too simple to really illustrate some of what we have just explained.
In this case, $\ad(E)$ is a trivial line bundle over $C$
 and $\varphi$ is simply a holomorphic 1-form, acted on trivially by $\Theta$.  So $\varphi_+=\varphi_-$,
and if $\varphi_-$ is holomorphic at $\pp$, then so is $\varphi_+$.  Thus for $U(1)$, a Hecke transformation of the pair $(E,\varphi)$ is 
simply a Hecke transformation of $E$, with no change in $\varphi$.  To illustrate the implications of requiring that $\varphi_+$ has
no pole at $\pp$, we need a nonabelian gauge group such
as $U(2)$.

\subsubsection{Minuscule Representation of $U(2)$}\label{minutwo}

The simplest example that really illustrates the general story is $^L\neg G=U(2)$, where for simplicity -- 
and also because this example is important in our application --
we take $^L\neg w$ to be the minuscule weight $(1,0)$.  The local action of $T(^L\neg w)$, in some basis, is
\begin{equation}\label{omelf} \O\oplus \O\to \O(\pp)\oplus \O, \end{equation}
and this corresponds to  
\begin{equation}\label{zomelf}\Theta=\begin{pmatrix} z & 0 \cr 0 & 1\end{pmatrix}.\end{equation}

Now consider for $y<y_0$ a Higgs field $\varphi \in H^0(C,K\otimes \ad(E_-))$.  With respect to the local trivialization of $E_-$ that is used
in eqn. (\ref{omelf}), we have
\begin{equation}\label{zum}\varphi_-=\begin{pmatrix} a & b  \cr c& d\end{pmatrix},\end{equation}
where $a,b,c$ and $d$ are local holomorphic sections of $K$.  So
\begin{equation}\label{wum}\varphi_+=\Theta\varphi\Theta^{-1}=\begin{pmatrix} a &z b \cr z^{-1} c& d\end{pmatrix}.\end{equation}
Thus $\varphi_+$ has a pole at $\pp$ unless $c(\pp)=0$.

The condition $c(\pp)=0$ has a simple interpretation.   We recall from eqn. (\ref{durfy}) that an arbitrary Hecke modification of $E_-$
of type $^L\neg w=(1,0)$ produces a bundle $E_+$ of which a local holomorphic section is
\begin{equation}\label{zonkey}s=s_0+\frac{\lambda}{z}\begin{pmatrix}u\cr v\end{pmatrix}, \end{equation}
where $\bb=\begin{pmatrix}u\cr v\end{pmatrix}$ represents a point in $\Bbb{CP}^1$.  In making the decomposition (\ref{omelf})
and arriving at the form (\ref{zomelf}) for $\Theta$, we have taken $\bb=
\begin{pmatrix}1\cr 0\end{pmatrix}.$  With this choice of $\bb$, the condition $c(\pp)=0$ is equivalent to
\begin{equation}\label{demino}\varphi_-(\pp)\cdot \bb=0 ~~{\rm{mod}}~
\bb.
\end{equation}  In other words, $\varphi_-(\pp)\cdot \bb$ is a multiple of 
$\bb$.  This criterion for $\varphi_+$ to be holomorphic at $\pp$ holds for any
$\bb$, not necessarily of the form $\begin{pmatrix} 1\cr 0\end{pmatrix}$ with which
we began.

We recall that the space of Hecke modification of $E_-$ at $\pp$ of type $^L\neg w=(1,0)$ is a copy of $\CP^1$ obtained by
projectivizing $E_{-,\pp}$.  The Lie algebra $\ad(E_{-,\pp})$ acts on this $\CP^1$; its generators correspond to holomorphic
vector fields on $\CP^1$.    The condition (\ref{demino}) asserts that the point in $\CP^1$ given by
$\bb=\begin{pmatrix}u\cr v\end{pmatrix}$ is invariant under the symmetry of $\CP^1$ generated\footnote{Since
$\varphi$ is valued not in the Lie algebra $\mathrm{ad}(E)$ but in $\mathrm{ad}(E)\otimes K$, we have to pick a trivialization
of $K$ near $\pp$ to think of $\varphi_-(\pp)$ as generating a symmetry of $\CP^1$.  But the choice of trivialization does
not affect the condition of $\varphi_-(\pp)$ invariance.}  by $\varphi_-(\pp)$.  The action of
$\varphi_-(\pp)$ just rescales the homogeneous coordinates  $u,v$ of this point.    

We can summarize  this as follows.  In order for a Hecke transformation of a Higgs bundle $(E_-,\varphi_-)$ not
to produce a pole in $\varphi_+(\pp)$, the Hecke modification of $E_-$ must be $\varphi_-$-invariant, meaning that it must
be invariant under the symmetry generated by $\varphi_-$.    This is a general condition that holds not just for the particular
group $G=SU(2)$ and weight $^L\neg w=(1,0)$ that we have considered, but for any group and representation.  
The criterion is easiest to understand and implement in the case of a minuscule weight, for then the symmetry
of the space of Hecke modifications at $p$ that is  generated by $\varphi_-$ depends only on $\varphi_-(\pp)$, the value
of $\varphi_-$ at $\pp$.  For a representation that is not minuscule,
 the analysis of the symmetry of the space of Hecke modifications generated by $\varphi$
is more complicated. 

We have carried out this discussion in a way that treats $(E_-,\varphi_-)$ and $(E_+,\varphi_+)$ asymmetrically.
Starting with $(E_-,\varphi_-)$, an 't Hooft operator $T(^L\neg w)$, inserted at some point $\pp\times y_0\in C\times \R$,
induces a Hecke transformation with $(E_+,\varphi_+)$ as output.  Looking at the same picture backwards, one can view
$(E_+,\varphi_+)$ as input and $(E_-,\varphi_-)$ as output.  If $\varphi_-$ and $\varphi_+$ are both free of poles at $\pp$,
then the 't Hooft operator produces a  $\varphi_-$-invariant or $\varphi_+$-invariant Hecke transformation depending
on how one looks at it.  To avoid committing ourselves to one point of view, we sometimes
just say that the Hecke transformation is $\varphi$-invariant. 

Now us consider a Higgs bundle $(E_-,\varphi_-)$ and ask how many $\varphi_-$-invariant Hecke modifications of type
$(1,0)$ are possible at a given point $\pp$.  In answering this question, we will assume that the $2\times 2$ matrix $\varphi_-(\pp)$ has
distinct eigenvalues; an equivalent statement is that  $\Tr\,\varphi(\pp)^2-\frac{1}{2}
\Tr\,\varphi(\pp)^2\not=0$.     
In view of eqn. (\ref{zonk}), this condition is satisfied by  $\varphi_+$ if and only if it is satisfied by $\varphi_-$.

If an $N\times N$ complex matrix has $N$ distinct eigenvalues, we say that it is regular and semisimple.
 Such a matrix can be diagonalized by a complex-valued linear transformation.  So
if $\varphi_-$ is regular and semisimple, then in the right basis, it   can be written near $\pp$ as
\begin{equation}\label{omonk}\varphi_-=\begin{pmatrix}\lambda_1 & 0 \cr 0 & \lambda_2\end{pmatrix},\end{equation}
where $\lambda_1$ and $\lambda_2$ are sections of $K$.  In this basis, a $\varphi_-(\pp)$-invariant Hecke transformation
at $\pp$ simply corresponds to a point in $\Bbb{CP}^1$ with homogeneous coordinates $\begin{pmatrix}1 \cr 0\end{pmatrix}$
or $\begin{pmatrix}0 \cr 1\end{pmatrix}$.    So there are precisely two $\varphi_-(\pp)$-invariant Hecke modifications at $\pp$ if $\varphi_-(\pp)$
has distinct eigenvalues.

Finally, we come to a point that will be crucial in our application.  Under the assumption that we have made, the
 two $\varphi(\pp)$-invariant Hecke modifications of
type $(1,0)$ have a simple interpretation in terms of spectral covers.  We recall that the $U(2)$ Higgs bundle $(E_-,\varphi_-)$ over
$C$
can be derived from a branched double cover $\psi:D\to C$, together with a line bundle $\L\to D$.  According to eqn. (\ref{hytot}).
$E_-$ is reconstructed from this data as $\psi_*(\L)$, and $\varphi_-$ is similarly reconstructed as in eqn. (\ref{vyro}).
Let $\qq',\qq''$ be the two points in $D$ that lie above $\pp\in C$.  They are distinct points since we have assumed that $\varphi(\pp)$ has
distinct eigenvalues, and so the relation $E=\psi_*(\L)$ gives simply
\begin{equation}\label{belf} E_\pp=\L_{\qq'}\oplus \L_{\qq''}. \end{equation}

We can extend this decomposition of $E$ at $\pp$ to a decomposition in a small neighborhood of $\pp$.  In a neighborhood of $\pp\in C$,
there are two sections $s_1:C\to D$ and $s_2:C\to D$, with $s_1(\pp)=\qq'$, $s_2(\pp)=\qq''$.  In a neighborhood of $\pp$, we have
\begin{equation}\label{pelf} E\cong s_1^*(\L)\oplus s_2^*(\L). \end{equation}

To make a Hecke modification of $E$ of type $(1,0)$, we begin with a general decomposition of $E$ as $\O\oplus \O$ and
replace one of the summands with $\O(\pp)$.  However, if we want to get a $\varphi$-invariant Hecke modification, the summand
that we modify must be one of the two $\varphi$-invariant summands in eqn. (\ref{pelf}).

For example, a Hecke modification in which the first summand in eqn. (\ref{pelf}) is modified by allowing a pole
at $\pp$ will replace $s_1^*(\L)$ with $s_1^*(\L)\otimes \O(\pp)=s_1^*(\L(\qq'))$.  Thus it has the same effect as replacing
$\L$ by $\L(\qq')$ on the spectral cover $D$.  An analogous modification of the second summand in eqn. (\ref{pelf}) is equivalent
to replacing $\L$ with $\L(\qq'')$.

Thus the two possible $\varphi$-invariant Hecke modifications of type $^L\neg w=(1,0)$ at a point $\pp$ at which $\varphi$ is regular
and semisimple
have a very simple interpretation 
in terms of the spectral cover $D\to C$.   They correspond to replacing 
$\L\to D$ with either $\L(\qq')$ or $\L(\qq'')$, where $\qq'$ and $\qq''$ are the two
points in $D$ that lie over $\pp$:
\begin{equation}\label{zort}\L \to \begin{cases} \L(\qq') & \mathrm{or }\cr  \L(\qq'').&\end{cases}\end{equation}

\subsubsection{Nonminuscule Weights}\label{nonmin}

The analysis of $\varphi$-invariant Hecke modifications involves much more technicality if $^Lw$ is not a minuscule weight.
In this case, the moduli space of Hecke modifications of type $^Lw$ (the Schubert cell) is not compact, and has a natural compactification (the
Schubert cycle) that
involves monopole bubbling.  One really needs to consider $\varphi$-invariant points on the compactification, and some of 
these points do lie at infinity on the Schubert cycle. 
 So in the non-minuscule case, an analysis of Hecke modifications
of Higgs bundles cannot be made without incorporating monopole bubbling.

In this paper, we will only aim to analyze
the eigenbranes of 't Hooft operators under the simplest conditions in which most of the technicalities do not arise.  To this aim, we will 
consider only the comparatively elementary case that $^Lw$ is a minuscule weight.    We also make the assumption that has already
been introduced above:  we assume that $\varphi_-(\pp)$ is generic, meaning that it has distinct eigenvalues, or equivalently we assume
that the point $\pp\in C$ is not a branch point of the spectral cover.

For $^L\neg G=U(N)$ or $PU(N)=SU(N)/\Z_N$, there are
enough minuscule weights of $^LG$ so that 't Hooft operators dual to minuscule weights
 generate the ring of all 't Hooft operators.  This is far from
true for other groups.

\subsubsection{Analog For $SO(3)$ And For $SU(2)$}\label{anagroups}

For $^L\neg G=SO(3)$, we can take $^Lw$ to be the nonzero minuscule weight
$(1/2,-1/2)$ of $G=SU(2)$.  In this case, again assuming that $\varphi_-(\pp)$ has distinct
eigenvalues (which for $SO(3)=PU(2)$ is equivalent to saying that it is not nilpotent), the analysis
of $\varphi$-invariant Hecke modifications at $\pp$ of type $^Lw$ is the same as for $U(2)$.
There are two of them, and they can be described on the spectral cover as replacing $\L$ by $\L(\qq')$ or $\L(\qq'')$.

For $^L\neg G=SU(2)$, the dual group $G=SO(3)$ has no nonzero minuscule weights, so the technicalities mentioned
in section \ref{nonmin} are inescapable.   Accordingly, we will not analyze magnetic eigenbranes for this group.  

We should note that the technicalities associated with monopole bubbling are still relatively manageable if the representation
$^L\neg R$ has the property that each of its weight spaces has dimension at most 1.  This condition is satisfied for an arbitrary representation
of $SU(2)$ or $SO(3)$, so for these groups one can expect to extend the analysis we give of magnetic eigenbranes to arbitrary 't Hooft
operators without too much
technicality.    The more serious technicalities arise for groups of higher rank.

\subsubsection{Generalization To $U(N)$}

The discussion of section \ref{minutwo} can be readily adapted to $^LG=U(N)$.  Suppose that $\varphi(\pp)$ is regular and semisimple,
meaning that it has $N$ distinct
eigenvalues.  Then near $\pp$, there are $N$ distinct solutions $\uppsi_i$ of
the eigenvector equation
\begin{equation}
\label{oxonik}\varphi\cdot
\uppsi_i=y_i\uppsi_i.
\end{equation}
We can pick these to vary holomorphically in a neighborhood of $\pp$.
 Consider Hecke modifications of weight
$^Lw=(m_1,m_2,\dots,m_N)$, where as usual $m_1\geq
m_2\geq\dots\geq m_N$.   The general frameowrk for such a Hecke modification
was described in section \ref{zoff}, but to find a $\varphi$-invariant Hecke modification of this
type, we have to be more precise.  A $\varphi$-invariant Hecke modification
$E_+$ of the given type can be described by saying that $E_+$ coincides with $E_-$
away from $\pp$, and near $\pp$ a holomorphic section of $E_+$ takes
the form
\begin{equation}
\label{zno}
s=\sum_{i=1}^N z^{-m_i}h_i\uppsi_i,
\end{equation}
 with $z$ a local
parameter at $\pp$, and the functions $h_i$ being holomorphic at
$\pp$.  Choosing the basis functions $\uppsi_i$ to be eigenfunctions of $\varphi$
ensures that  this particular Hecke modification is $\varphi$-invariant.

We can find $N!$ Hecke modifications that are  $\varphi$-invariant
by picking a permutation $\pi$ of the set $\{1,2,\dots,N\}$, and
writing instead
\begin{equation}
\label{znnno}
s=\sum_{i=1}^N z^{-m_i}h_i\uppsi_{\pi(i)}.
\end{equation}
 By arguments similar to those that we
have given already for $U(2)$, one can show that these are the only
$\varphi$-invariant Hecke modifications of weight $^Lw$.  They are
all inequivalent if the $m_i$ are pairwise distinct, but if, say,
$m_i=m_j$, then exchanging $s_i$ and $s_j$ does not change the
Hecke modification. So the number of $\varphi$-invariant Hecke
modifications of $E_-$ is  $N!/\#\Gamma$, where $\#\Gamma$ is the
order of the group $\Gamma$ of permutations of the set
$\{1,2,\dots,N\}$ that leaves the weights $m_i$ invariant.

Alternatively, the number of $\varphi$-invariant Hecke
modifications can be described as follows.  
The Weyl group ${\cal W}$ of $U(N)$ is the group of permutations
of the weights, which has $N!$ elements.  The weights of $^LR$
come in Weyl orbits. The number of weights in the orbit containing
the highest weight is $N!/\#\Gamma$, where $\Gamma$ is the
subgroup of ${\cal W}$ that leaves fixed the highest weight.  This
is the same result found in the last paragraph.

In the case that $\varphi$ has $N$ distinct eigenvalues, the
 $\varphi$-invariant Hecke modifications have the same sort of interpretation on the spectral curve that was described
 for $U(2)$ in section \ref{minutwo}.
This may be seen as follows.

We recall from section \ref{spectral} that the Higgs pair
$(E,\varphi)$ is determined by a line bundle ${\cal L}\to D$,
where $\pi:D\to C$ is the spectral cover  In particular,
$E=\pi_*({\cal L})$.   We recall  also
that a Hecke modification changes $E$ without changing the spectral
curve.  The change in $E$ will come from a change in ${\cal L}$,
which we claim is as follows.  Let $\qq_1,\dots,\qq_N$ be the points
on $D$ that lie above $\pp\in C$.  Then for a suitable ordering of
the $\qq_i$, the effect of the Hecke modification is
\begin{equation}\label{effheck}
{\cal L}\to {\cal L}\otimes \left(\otimes_{i=1}^N{\cal
O}(\qq_i)^{m_i}\right).\end{equation} The different
$\varphi$-invariant Hecke modifications of weight $^Lw$ come from
the different orderings of the $\qq_i$, or equivalently of the
$m_i$.

The idea behind this formula is very simple.  At a point $\pp$ at which the spectral cover $\psi:D\to C$ is unramified, the
inverse image of a small neighborhood $U$ of $p$ in $C$ is a union of small open sets $U_i\subset D$, each containing
one point $\qq_i$ lying over $\pp$.  The whole idea of the spectral cover is that locally, away from ramification points, it reduces
$U(N)$ gauge theory to $U(1)^N$, with one $U(1)$ on each sheet.  A $\varphi$-invariant Hecke modification of $E\to C$
corresponds on the spectral cover to a $U(1)$ Hecke modification on the $i^{th}$ sheet of weight $m_i$ (or more generally
of weight $m_{\pi(i)}$ for some permutation $\pi$).  Thus the Hecke modification acts on the $i^{th}$ sheet by $\L\to \L\otimes \O(m\qq_i)
=\L\otimes \O(\qq_i)^{m_i}$.  This is the claim in eqn. (\ref{effheck}).

The description that we have just given of $\varphi$-invariant Hecke
modifications of type $^L\neg w$ is valid for any $^L\neg w$, but as usual it is less useful if the weight $^L\neg w$ is not minuscule,
for in that case that are additional $\varphi$-invariant points  on the compactification of the Schubert cell associated to monopole bubbling.
Indeed, if the representation $^L\neg R$ is not minuscule,
then weights of this representation that are not on the Weyl orbit of the highest weight are associated to
$\varphi$-invariant Hecke modifications whose description involves monopole bubbling.

\section{Magnetic Eigenbranes}
\label{magbranes} 

\subsection{Preliminaries}\label{preliminaries}

In chapter \ref{eleig}, we analyzed the action of Wilson
lines on branes and identified zero-branes as electric eigenbranes.
Here, we will use the understanding of the action of 't Hooft
operators gained in chapter \ref{thooftheckeop} to make a similar analysis of magnetic
eigenbranes.   We do so only under simplifying assumptions that were stated in section \ref{nonmin}: we consider
only 't Hooft operators that are related to minuscule weights of the dual group, and we insert such an operator at a point
that is not a branch point of the spectral cover.

There is a very important preliminary point about how we will make this analysis.  In the geometric Langlands correspondence,
one is really interested in $B$-branes on $\MH(G,C)$ in complex structure $J$, and their duality with $A$-branes on $\MH(^L\neg G,C)$
in symplectic structure $\omega_K$.   One further wants to compare the  action of Wilson operators on $B$-branes
of $\MH(G,C)$ to the action of 't Hooft operators on $A$-branes of $\MH(^L\neg G,C)$.

However, $\MH(G,C)$  is a hyper-Kahler manifold.  The electric eigenbranes, as described in section \ref{eleig}, are zero-branes
supported at a point $x\in \MH(G,C)$.  A point is a complex submanifold in every complex structure, and in particular a zero-brane is
a brane of type $(B,B,B)$ -- that is, it is a $B$-brane in any of the complex structures on $\MH(G,C)$ that are part of its hyper-Kahler
structure.  Moreover, the natural Wilson operators of $\N=4$ super Yang-Mills theory that act in the $B$-model
of type $J$ are actually half-BPS Wilson operators that  map branes of type $(B,B,B)$ to themselves.
In particular, a zero-brane on $\MH(G,C)$ is not just an electric eigenbrane of type $J$, but is simultaneously an electric
eigenbrane of type $I$, $J$, and $K$ (and more generally it is an electric eigenbrane in the $B$-model of any complex structure on
$\MH(G,C)$ that is a linear combination of $I$, $J$, and $K$).  

The $S$-dual of a half-BPS Wilson operator is a half-BPS 't Hooft operator.  Moreover, $S$-duality maps a brane of type $(B,B,B)$
to a brane of type $(B,A,A)$, that is a brane that is a $B$-brane in complex structure $I$ and simultaneously
 an $A$-brane in symplectic structure $\omega_J$
or $\omega_K$ (or in a linear combination of those two symplectic structures); and the half-BPS 't Hooft operators map
branes of type $(B,A,A)$ to branes of the same type.  For further 
details on these assertions, see \cite{KW}.  We refer to the $A$-model with symplectic structure $\omega_J$ or $\omega_K$ as the
$A$-model of type $J$ or $K$.  In particular, since a zero-brane on $\MH(G,C)$ is of type $(B,B,B)$, its dual
will be a brane on $\MH(^L\neg G,C)$ that is an eigenbrane of type $(B,A,A)$ -- that is, it will be an eigenbrane in each of the
three indicated structures.

Concretely, since $S$-duality of $\N=4$ super Yang-Mills theory acts by $T$-duality of the fibers of the Hitchin fibration,
we can make a simple prediction for what the magnetic eigenbranes must be.  $T$-duality maps a zero-brane supported on a particular
fiber $ ^L\neg \FF$ of the Hitchin fibration of $^L\neg G$ to a rank 1 brane supported on the the corresponding fiber of the Hitchin fibration of $G$, which we  call $\FF$.
This brane has a flat Chan-Paton bundle $\L\to \FF$ of rank 1 .  A brane supported on a fiber of the Hitchin fibration and endowed with a flat Chan-Paton line
bundle is what we will call a brane of type $\bf F$.  Thus, we expect that the magnetic eigenbranes will be branes of type $\bf F$.

In geometric Langlands, one really cares about the magnetic eigenbranes as branes in the $A$-model of type $K$.  But if $\B$
is an $A$-brane of type $K$ that is actually a brane of type $(B,A,A)$, and if $T$ is an 't Hooft operator that maps branes of type $(B,A,A)$
to themselves, then to determine the product $T\cdot \B$, it suffices to identify this product as a brane in the $B$-model of type $I$.
This is much simpler than trying to directly describe $T\cdot \B$ as an $A$-brane of type $K$, and it is the way we will proceed. 

Thus our method of identifying magnetic eigenbranes will use in an essential way the hyper-Kahler structure of $\MH(G,C)$ and $\MH(^L\neg G,C)$.
  We will show that a brane of type $\bf F$  is an eigenbrane in the $B$-model of type $I$, with an ``eigenvalue''
that is of the form required to match expectations from the dual description.  It  automatically follows that such a brane is an eigenbrane
in the $A$-model of type $K$, with the same ``eigenvalue.''  For if some line operator $T$ and brane $\B$ of type $(B,A,A)$ satisfy
$T\cdot \B=\B\otimes V$, then we can read this statement equally either in the $B$-model of type $I$ or in the $A$-model of type $J$ or $K$.

\subsection{Action On A Fiber Of The Hitchin Fibration}
\label{lagcorr}

In eqn. (\ref{zonk}), we showed that the characteristic polynomial of the Higgs
field $\varphi$ is preserved by a Hecke modification -- that is, by the action of an 't Hooft
operator.   The fibers of the Hitchin fibration are labeled by this characteristic polynomial,
and therefore an 't Hooft operator will map a brane supported on a given fiber $\FF$
of the Hitchin fibration to another brane supported on the same fiber. 

We will consider only the case of branes supported on a smooth fiber $\FF$ of the Hitchin
fibration. Such a fiber  is a complex torus.
Moreover, it is a complex Lagrangian submanifold of $\MH(^L\neg G,C)$, and therefore can readily be the support of a brane of type $(B,A,A)$.  We simply endow $\FF$
with a rank 1 Chan-Paton bundle that, from the point of view of complex structure $I$, is a holomorphic
line bundle $\L$ of zero\footnote{Because the normal bundle to $\FF$ in $\MH$ is trivial,
there is no analog of the shift due to the $K$-theory interpretation of branes that was described in section \ref{dolfo} for 
$A$-branes supported on $T^*C$.   To be more exact, the closest such analog is the subtlety involving choice of a spin
structure that is summarized in footnote \ref{brief}.}  first Chern class. $\FF$ endowed with such a line bundle is a $B$-brane $\B$ of type $I$,
but since $\L$ admits a natural flat connnection, the brane $\B$ also has the natural structure of an $A$-brane
of type $J$ or $K$.

Incidentally, the fact that Hecke modifications map a fiber $\FF$ of the Hitchin fibration to itself means that
in a certain sense, Hecke
modifications of Higgs bundles $(E,\varphi)$ are better-behaved
than Hecke modifications of bundles alone.  Starting with any
bundle $E$, possibly stable, repeated Hecke modifications can
produce an arbitrarily unstable bundle -- in fact, they can
produce an arbitrary bundle. However, it can be shown 
that if the characteristic polynomial of $\varphi$ is associated to a generic fiber
of the Hitchin fibration, then every Higgs bundle $(E,\varphi)$ is stable.\footnote{A Higgs
bundle $(E,\varphi)$ is stable if any non-trivial $\varphi$-invariant subbundle of $E$ obeys a condition
that was stated in footnote \ref{semicon}.  This condition is satisfied
if the characteristic polynomial of $\varphi$ is irreducible, for then $E$
has no nontrivial $\varphi$-invariant subbundles at all.}  Thus as long as we start on a generic fiber $\FF$
of the Hitchin fibration, repeated Hecke modifications can be made without leaving the moduli space $\MH$
of stable Higgs bundles.  Indeed, they can be made without leaving the fiber $\FF$.

We can be much more precise than merely saying that a Hecke transformation maps a fiber $\FF$ to itself.  
In eqn. (\ref{zort}) for an 't Hooft operator dual to a minuscule weight $(1,0)$ of $U(2)$, and in eqn. (\ref{effheck}) more
generally for any 't Hooft operator of $U(N)$, we have determined precisely how the Hecke transformation acts on $\FF$.
For instance, for the basic case of the minuscule weight of $U(2)$, the action is by $\L\to \L\otimes \O(\qq')$ or $\L\to \L\otimes \O(\qq'')$,
where $\qq'$ and $\qq''$ are the two points in the spectral cover $\psi:D\to C$ that lie over $\pp\in C$.  Tensoring with a fixed line
bundle $\O(\qq')$ or $\O(\qq'')$ is an automorphism on the fiber $\FF$ of the Hitchin fibration, and this automorphism preserves
the complex symplectic structure of $\FF$.  That is concretely why the action of an 't Hooft operator will map a brane of
type $(B,A,A)$ supported on $\FF$ to another brane of the same type, also supported on $\FF$.  

There is some subtlety in this last statement, because the line bundle $\O(\qq')$ or $\O(\qq'')$ has degree 1, and tensoring
with this line bundle permutes the different components of $\Pic(D)$, the group of holomorphic line bundles over $D$. 
We will take this into account in a more detailed analysis in section \ref{compheck}, but the basic conclusion holds that because tensoring
with a fixed line bundle preserves the complex symplectic structure of $\Pic(D)$, the action of an  't Hooft operator
will map a brane of type $(B,A,A)$ to another brane of the same type.

The basic setup in trying to identify
how the 't Hooft operator acts on the Chan-Paton bundle is that of fig. \ref{nuxto} of section \ref{thooftheckeop}, but now the vertical line $L$
represents an 't Hooft operator parallel to the boundary of the worldsheet $\Sigma$.   The 't Hooft operator does not change the characteristic
polynomial of the Higgs field $\varphi$, so it does not change the spectral cover $D$.  But it changes the line bundle  $\L\to D$ that
determines a given Higgs bundle $(E,\varphi)$.  

Now consider an arbitrary brane $\B$, with Chan-Paton bundle $\U$,
 supported on a chosen fiber $\FF$ of the Hitchin fibration.    We want to act with an 't Hooft operator $T(^L\neg R,\pp)$ and determine the Chan-Paton bundle $\hat \U$ of the resulting
brane $T(^L\neg R,\pp)\cdot \B$.  For simplicity, we explain this first for the case that $^L\neg R$ corresponds to the minuscule weight
$(1,0)$ of $^L\neg G=U(2)$.  

If this line bundle over $D$ that determines a point on the fiber $\FF$
 is $\L$ to the left of the line $L$ in the figure, then it is $\L\otimes\O(\qq')$
or $\L\otimes \O(\qq'')$ near the boundary.  This means that the sheaf $\hat\U$, evaluated in a small neighborhood of a point on $\FF$
corresponding to $\L$, is the direct sum of the sheaf $\U$ evaluated at $\L\otimes \O(\qq')$ or $\L\otimes \O(\qq'')$.
A more succinct way to say that is that
\begin{equation}\label{zelmo} \hat\U=\Phi'{}^*(\U) \oplus \Phi''{}^*(\U).  \end{equation}
Here $\Phi'$ and $\Phi''$  are respectively 
the automorphisms of $\FF$ that correspond to $\L\to \L\otimes\O(\qq')$ and $\L\to \L
\otimes \O(\qq')$.

The generalization of eqn. (\ref{zelmo}) for a minuscule weight of $G=U(N)$ is
\begin{equation}\label{upelmo}\hat \U =\sum_i\Phi_i^*(\U), \end{equation}
where $\Phi_i:\FF\to \FF$ corresponds to the $\varphi$-invariant Hecke transformation $\L\to \L\otimes \O(\sum_i m_i\qq_i)$ of eqn.
(\ref{effheck}).     (As usual, for a non-minuscule weight,
the analogous formula has additional contributions associated to monopole bubbling.) 

For $G=^L\neg G=U(1)$, matters are more simple.  In this case, the spectral cover $D\to C$ is trivial; $D$ simply coincides with $C$.
A charge $n$ 't Hooft operator inserted at $p$ acts on the fiber of the Hitchin fibration by $\L\to \L\otimes \O(np)$.  This gives an automorphism
$\Phi$ of $\Pic(C)$ and the action of the 't Hooft operator is by
\begin{equation}\label{welmo}\U\to \hat \U=\Phi^*(\U).\end{equation}

\subsection{Translation Eigenbundles}
\label{flate}

To analyze the above formulas, we will need to understand the following situation.  $\cF$ is a complex torus equipped  with a flat line bundle $\L\to \cF$.
$\Phi:\cF\to \cF$ is a constant translation.  We want to compare $\L$ to $\Phi^*(\L)$.  

We are not quite in this situation in eqns. (\ref{zelmo}) and (\ref{welmo})
because the operations $\Phi$, $\Phi'$, and $\Phi''$ involve tensoring $\L$ with a line bundle over $D$ of nonzero degree.  We will be in this
situation, however, if we act with a product of 't Hooft operators carrying no net magnetic flux.   (We explain in sections
\ref{dualap} and \ref{compheck} how to think about the case that the 't Hooft operators do carry net magnetic flux.)

First consider simply the case of 
 a torus $\cF$ (with no complex structure assumed) endowed with a flat line bundle ${\cal L}$.
Topologically, such a flat line bundle can be specified by giving
the holonomies around one-cycles in $\cF$.  These holonomies are
obviously invariant under translations on $\cF$, so we conclude that
if $\Phi$ is such a translation, then $\Phi^*({\cal L})$ is
isomorphic to ${\cal L}$ as a flat line bundle.

This remains so if $\cF$ is a complex torus and ${\cal L}$ is a
holomorphic line bundle whose first Chern class vanishes.  Such an
${\cal L}$ admits a flat connection compatible with its
holomorphic structure (compatibility means that the part of the
connection of type $(0,1)$ is the $\bar\partial$ operator
determining the holomorphic structure of ${\cal L}$), and the
reasoning of the last paragraph applies.

We should, however, formulate this carefully.  The group of
translations of $\cF$ is a complex torus $\tilde \cF$.  $\tilde \cF$ is
isomorphic to $\cF$ once a base point $f_0$ in $\cF$ is picked.  Let us denote
a translation of $\cF$ additively as $f\to f+\tilde f$, with $f\in \cF$, $\t f\in \t \cF$.  The statement now that ${\cal L}$ is translation-invariant 
means that for $\tilde f\in \tilde \cF$,
\begin{equation}\label{gaggle}\tilde f^*({\cal L})={\cal L}\otimes
{\cal N}_{\tilde f},\end{equation} where ${\cal N}_{\tilde f}$ is
a one-dimensional vector space that depends holomorphically on
$\tilde f$.  In other words, ${\cal N}_{\tilde f}$ is the fiber at
${\tilde f}$ of a holomorphic line bundle ${\cal N}\to \tilde \cF$.  This holomorphic
bundle is itself non-trivial.

We might describe (\ref{gaggle}) by saying that a flat line bundle
on a complex torus is a ``translation eigenbundle.''  Its pullback under translation by
$\t f$ is isomorphic to itself, but not canonically; the possible isomorphisms
correspond to nonzero vectors in the 1-dimensional vector space ${\cal N}_{\t f}$.

Now let us ask  what holomorphic vector bundles ${\cal Y}\to \cF$
are translation eigenbundles in the same sense:
\begin{equation}\label{aggle} \tilde f^*({\cal Y})={\cal Y}\otimes
{\cal N}_{\tilde f}.\end{equation} Writing ${\cal Y}_f$ for the
fiber of ${\cal Y}$ at $f$, the condition is that there should be
a holomorphically varying isomorphism
\begin{equation}\label{naggle}
{\cal Y}_{f+\tilde f}={\cal Y}_{f}\otimes {\cal N}_{\tilde f}.
\end{equation}
Setting $f=f_0$, $f'=f+\tilde f$, we have
\begin{equation}\label{znaggle}
{\cal Y}_{f'}={\cal Y}_{f_0}\otimes {\cal N}_{f'-f_0},
\end{equation}
where $f'-f_0$ is  the unique element of the translation group
$\tilde \cF$ that maps $f_0$ to $f'$.

Eqn. (\ref{znaggle}) tells us that a general translation
eigenbundle ${\cal Y}$ is the tensor product of a fixed vector
space (namely ${\cal Y}_{f_0}$) with a line bundle (whose fiber at
$f'$ is ${\cal N}_{f'-f_0}$). In more physical terms, it means
that any bundle of rank greater than 1 that is supported on $\cF$ and
is a translation eigenbundle is a direct sum of identical copies of
a fixed rank 1 eigenbundle supported on $\cF$.

These statements also have a partial converse.  A line bundle
${\cal L}\to \cF$ with $c_1({\cal L})\not= 0$ does not admit a
translation-invariant connection. A connection with
translation-invariant curvature and holonomies must be flat (since
curvature forces holonomies around non-contractible loops not to
be translation-invariant).  ${\cal L}$ admits such a connection if
and only if $c_1({\cal L})=0$.

Not only does a translation-invariant line bundle ${\cal L}\to \cF$
have $c_1({\cal L})=0$, but the line bundle ${\cal N}\to \tilde \cF$
that measures its translation ``eigenvalue'' likewise has
$c_1({\cal N})=0$.  Indeed, as we see upon setting ${\cal Y}={\cal
L}$ in eqn. (\ref{znaggle}), ${\cal L}$ and ${\cal N}$ are
essentially isomorphic, up to picking an identification between
$\cF$ and $\tilde \cF$ and tensoring with a fixed one-dimensional
vector space.

\subsection{Branes of type $\bf F$ as   Magnetic Eigenbranes}
\label{brmag}

Now let us consider a brane ${\cal B}$ whose support is a fiber
$\FF$ of the Hitchin fibration, and ask if it can be a magnetic
eigenbrane.

Suppose first that ${\cal U}$ is of rank 1.  Then ${\cal B}$ is
what we have called a brane of type ${\bf F}$, and we expect it to
be a magnetic eigenbrane.  From section \ref{flate}, a flat bundle
of rank 1 over $\FF$ is an eigenbundle for all translations, and in
particular for the $\Phi_i$.  Thus, we have $\Phi_i^*({\cal
U})={\cal U}\otimes {\cal N}_i$, for some one-dimensional complex
vector spaces ${\cal N}_i$.  By virtue of (\ref{upelmo}), it
follows that \begin{equation}\label{cabble}T({}^Lw,\pp){\cal
B}={\cal B}\otimes\left(\oplus_i {\cal N}_i\right),\end{equation}
so that ${\cal B}$ is a magnetic eigenbrane. We will compute the
``eigenvalue'' $\oplus_i{\cal N}_i$ in section \ref{compheck},
after some preliminaries.

Branes of higher rank supported on $\FF$ do not give any essentially
new magnetic eigenbranes.  The 't Hooft operators are plentiful
enough that a joint eigenbrane of the 't Hooft operators is
actually a full translation eigenbrane, and so as in eqn. (\ref{znaggle}),
the Chan-Paton bundle of a general eigenbrane is just the tensor product
of some line bundle over $\FF$ with a fixed vector space.

\section{ Determinant Line Bundles}
\label{backdet}

We will now develop some mathematical techniques that will be
helpful in computing the Hecke ``eigenvalue'' of a brane of type
${\bf F}$.

Associated with a line bundle ${\cal L}$ over a Riemann surface
$D$ are the cohomology groups $H^0(D,{\cal L})$ and $H^1(D,{\cal
L})$. The determinant line of ${\cal L}$ is defined as\footnote{An
opposite convention with $\det\,H^*({\cal L})=\left(H^0(D,{\cal
L})\right)\otimes \det\, H^1(D,{\cal L})^{-1}$ is most often used in algebraic
geometry, for example in \cite{deligne}.  We use here the convention that is most
common in the literature on determinants of $\bar\partial$
operators, for example \cite{quillen}, which is
the most familiar way that the determinant line bundle enters
physics. A  number of our formulas must be reversed, of course, if one
reverses this convention.}
\begin{equation}\label{detline}\det\,H^*({\cal
L})=\det\,H^0(D,{\cal L})^{-1}\otimes \left(\det\,H^1(D,{\cal
L})\right).\end{equation} As ${\cal L}$ varies, its determinant
line varies as the fiber of a  line bundle ${\bf Det}$ over ${\rm
Pic}(D)$ that is known as the determinant line bundle. The
determinant line bundle is most familiar to physicists for its
role in the study of two-dimensional chiral fermions.

\subsection{Pairing Of Line Bundles}\label{linepairing}

For our purposes, the utility of the determinant line bundle is
that it enables one to define a sort of bilinear pairing of line
bundles. If ${\cal L}$ and ${\cal M}$ are two line bundles over
$D$, one defines \cite{deligne} the complex line
\begin{equation}\label{bilpa}
\langle{\cal L},{\cal M}\rangle={\det\,H^*(D,{\cal L})\otimes
\det\,H^*(D,{\cal M})\over \det\,H^*(D,{\cal L}\otimes {\cal
M})\otimes \det\,H^*(D,{\cal O})}.\end{equation} (If $V$ and $W$
are two one-dimensional vector spaces, we write ${V}\otimes
{W}^{-1}$ as a fraction $V/W$.)

There are obvious natural isomorphisms
\begin{align}
\langle {\cal L},{\cal M}\rangle & \cong \langle {\cal M},{\cal
L}\rangle \\ \nonumber \langle {\cal L},{\cal O}\rangle \cong
\langle{\cal O},{\cal L}\rangle& \cong \Bbb{C}.\end{align} Here
and in the rest of this analysis, the symbol $\cong$ refers to an
isomorphism that can be defined in a universal way and which
therefore also gives an isomorphism in families.

The most important property of the symbol $\langle~,~\rangle$ is that
it is bilinear in the sense that for any three line bundles ${\cal
L},\,{\cal M},$ and ${\cal N}$, we have a canonical isomorphism
\begin{equation}\label{duglog} \langle{\cal L},{\cal M}\otimes {\cal 
N}\rangle\cong\langle{\cal L},{\cal M}\rangle \otimes \langle
{\cal L},{\cal N}\rangle.\end{equation} Any line bundle ${\cal N}$
takes the form $\otimes_i{\cal O}(\qq_i)^{n_i}$ for some divisor
${\DD}=\sum_in_i\qq_i$, and (\ref{duglog}) follows by induction in
$\sum_i|n_i|$ if it is true for the special cases ${\cal N}={\cal
O}(\qq)^{\pm 1}$.

For ${\cal N}={\cal O}(\qq)^{-1}$, we have ${\cal L}\otimes {\cal
N}={\cal L}(-\qq)$, the line bundle whose sections are sections of
${\cal L}$ that vanish at $\qq$.  We can compare the cohomology of
${\cal L}$ and ${\cal L}(-\qq)$ using the exact sequence of sheaves
\begin{equation}\label{doglo}
0\to {\cal L}(-\qq)\to {\cal L}\underarrow{r} {\cal L}|_\qq\to 0
\end{equation}
Our notation is a bit informal.  We write ${\cal L}(-\qq)$ and
${\cal L}$ for either the indicated line bundle or the
corresponding sheaf of sections, while ${\cal L}|_\qq$ means the
fiber of ${\cal L}$ at $\qq$ and also the corresponding skyscraper
sheaf at $\qq$. The map $r$ in (\ref{doglo}) is defined by
evaluating a section of ${\cal L}$ at the point $\qq$. From
(\ref{doglo}) we get a long exact sequence of cohomology groups
\begin{align}\label{zonglo}
 0\to & \,H^0(D,{\cal L}(-\qq))\to H^0(D,{\cal L})\to {\cal L}|_\qq \cr
 \to&\,
 H^1(D,{\cal L}(-\qq))\to H^1(D,{\cal L})\to 0.\end{align}
 Now in general, a long exact sequence of vector spaces
\begin{equation}\label{onglo}
0\to A_0\underarrow{d_0}
A_1\underarrow{d_1}A_2\underarrow{d_2}\dots
\underarrow{d_{n-1}}A_n\to 0  \end{equation} determines an
isomorphism
\begin{equation}\label{ponglo}
\otimes_{i=0}^n \left(\det\,
A_i\right)^{(-1)^i}\cong\Bbb{C}.\end{equation} In the present
case, from (\ref{zonglo}), we get an isomorphism
\begin{equation}\label{monglo}
\det\,H^*(D,{\cal L}(-\qq))\cong \det\,H^*(D,{\cal L})\otimes {\cal
L}|_\qq.\end{equation} This is equivalent to
\begin{equation}\label{kongo}
\langle{\cal L},{\cal O}(-\qq)\rangle \cong{\cal L}|_\qq^{-1}\otimes
{\det\,H^*(D,{\cal O}(-\qq))\over \det\,H^*(D,{\cal
O})}.\end{equation} The special case of this with ${\cal L}={\cal
O}$ tells us that
\begin{equation}\label{ubongo}{\det\,H^*(D,{\cal O}(-\qq))\over \det\,H^*(D,{\cal O})}
\cong{\Bbb{C}},
\end{equation}
so actually
\begin{equation}\label{rongo} \langle{\cal L},{\cal O}(-\qq)\rangle\cong {\cal L}|_\qq^{-1}.
\end{equation}

If in (\ref{monglo}) we replace ${\cal L}$ by ${\cal M}$ or by
${\cal L}\otimes {\cal M}$, we learn that
\begin{align}\label{moonglo}
\det\,H^*(D,{\cal M}(-\qq))&\cong \det\,H^*(D,{\cal M})\otimes {\cal
M}|_\qq\\ \nonumber \det\,H^*(D,{\cal L}\otimes {\cal M}(-\qq))&\cong
\det\,H^*(D,{\cal L}\otimes{\cal M})\otimes ({\cal L}\otimes {\cal
M})|_\qq.
\end{align}

Now if one writes out the definition of $\langle {\cal L},{\cal M}\otimes {\cal N}\rangle$
for ${\cal N}={\cal O}(-\qq)$, uses (\ref{moonglo})
to everywhere eliminate  $ {\cal M}(-\qq)$ in
favor of   ${\cal M}$, and evaluates ${\cal L}|_\qq^{-1}$
via (\ref{rongo}), then one arrives at
\begin{equation}\label{onear}\langle{\cal L},{\cal M}(-\qq)\rangle\cong\langle{\cal L},
{\cal M}\rangle\otimes \langle {\cal L},{\cal O}(-\qq)\rangle,\end{equation}
which is (\ref{duglog}) for ${\cal N}={\cal O}(-\qq)$.

If we set ${\cal M}={\cal O}(\qq)$ in (\ref{onear}), we learn that
\begin{equation}\label{gonear}\langle{\cal L},{\cal O}(\qq)\rangle\cong \langle{\cal L},
{\cal O}(-\qq)\rangle^{-1}.
\end{equation}
Replacing ${\cal M}$ by ${\cal M}(\qq)$ in (\ref{onear}), and using
(\ref{gonear}), one arrives at
 (\ref{duglog}) for ${\cal N}={\cal O}(\qq)$.
So by induction, this result holds in general.  We should say, however,
that the explanation we have given is a little naive, as we have not proved that
we get the same isomorphism regardless of how we choose to represent $\N$
as a tensor product of elementary factors $\O(\qq_i)^{\pm 1}$.  A complete treatment
can be found in \cite{deligne}.

Finally, let us note that this result can be written more
symmetrically. For any line bundle ${\cal L}\to D$, let us
abbreviate $\det\,H^*(D,{\cal L})$ as $[{\cal L}]$. As a further
abbreviation, let us omit the symbol for a tensor product.
 Then (\ref{duglog}) amounts to the statement that for any three line bundles
${\cal L}$, ${\cal M}$, ${\cal N}$, one has canonically
\begin{equation}\label{hurglo} {[{\cal L}\,{\cal M}\,{\cal N}]\,[{\cal L}]\,[{\cal M}]\,
[{\cal N}]
\over [{\cal L}\,{\cal M}]\,[{\cal M}\,{\cal N}]\,[{\cal L}\,{\cal N}]\,[{\cal O}]}
\cong\Bbb{C}.
\end{equation}
This has been called the theorem of the cube.  The name reflects the
fact that the eight factors on the left hand side can conveniently
be arranged on the corners of a cube.

\subsection{Interpretation}\label{interpretation}

Now we can evaluate $\langle{\cal L},{\cal M}\rangle$ in general.
Suppose that ${\cal M}=\otimes_{i=1}^k{\cal O}(\qq_i)^{n_i}$, or in
other words ${\cal M} ={\cal O}({\DD})$, where
${\DD}=\sum_in_i\qq_i$ is a divisor of degree $d=\sum_in_i$. An
induction in $\sum_i|n_i|$ based on (\ref{duglog}) and using the
special cases (\ref{rongo}) and (\ref{gonear}) gives
\begin{equation}\label{trifong}
\langle{\cal L},{\cal M}\rangle=\otimes_{i=1}^k {\cal L}|_{\qq_i}^{n_i},\end{equation}
with as usual ${\cal L}|_{\qq_i}$ the fiber of ${\cal L}$ at the point $\qq_i$.

Suppose that ${\cal L}$ is of degree $c$. We want to keep ${\cal
M}$ fixed and let ${\cal L}$ vary.  ${\cal L}$ determines a point
$x\in {\rm Pic}_c(D)$, and to make this explicit, we will write
${\cal L}$ as ${\cal L}_x$. As $x$ varies, we want to interpret
$\langle{\cal L},{\cal M}\rangle$ as the fiber of a line bundle
over ${\rm Pic}_c(D)$. There is a crucial subtlety here, which one often
encounters when one considers moduli spaces of line bundles.
This is that a point $x\in {\rm Pic}_c(D)$ determines a line
bundle ${\cal L}_x\to D$ only up to isomorphism. ${\cal L}_x$
could be replaced by ${\cal L}_x\otimes {\cal R}$, where ${\cal
R}$ is a fixed one-dimensional vector space. This is the general
freedom to change  ${\cal L}_x$, in the following sense.  If
${\cal U}$ and ${\cal U}'$ are isomorphic line bundles on a
Riemann surface $D$, there is no canonical isomorphism between
them (since either one has a $\Bbb{C}^*$ group of automorphisms,
which can be composed with any proposed isomorphism between ${\cal
U}$ and ${\cal U}'$). However, there is a canonical isomorphism
${\cal U}'\cong {\cal U}\otimes {\cal R}$, where ${\cal R}$ is the
one-dimensional vector space $H^0(D,{\cal U}'\otimes {\cal
U}^{-1})$. (The isomorphism is made by observing that a vector in
${\cal R}$ gives, by definition of $H^0(D,{\cal U}'\otimes {\cal
U}^{-1})$, a holomorphic map from ${\cal U}$ to ${\cal U}'$.) So,
once the isomorphism class of ${\cal L}_x$ is given, it is unique
up to ${\cal L}_x\to {\cal L}_x \otimes {\cal R}$.

In view of (\ref{trifong}),
we have
\begin{equation}\label{iftong}
\langle{\cal L}_x\otimes{\cal R},{\cal M}\rangle =\langle{\cal L}_x,{\cal M}\rangle
\otimes{\cal R}^d.
\end{equation}
So $\langle{\cal L}_x,{\cal M}\rangle$ is canonically determined by
the isomorphism class of ${\cal L}_x$
if and only if $d=0$.  A related remark is that an automorphism of ${\cal L}_x$ that
acts as multiplication of a complex number $\lambda\in\Bbb{C}^*$ acts on $\langle{\cal L}_x,
{\cal M}\rangle$ as multiplication by $\lambda^d$.  So automorphisms of ${\cal L}_x$ act
trivially on $\langle{\cal L}_x,{\cal M}\rangle$ if and only if ${\cal M}$ has degree zero.

For $d=0$, therefore, we do get, for each ${\cal M}=\otimes_i \O(\qq_i)^{n_i}$, a line bundle ${\cal N}$
over ${\rm Pic}_c(D)$.
This is the line bundle whose fiber at $x\in {\rm Pic}_c(D)$ is
\begin{equation}\label{zudco}
{\cal N}|_x=\otimes_{i=1}^k {\cal L}_x|_{\qq_i}^{n_i}.\end{equation}
This is an elementary definition that we could have made without first discussing determinant
line bundles.  What we have learned from that discussion is that, up to a natural isomorphism,
the line bundle $\cal N$ defined in this way depends only on the isomorphism class of the line
bundle $\M=\O(\DD)$, and not on the specific divisor $\DD=\sum_i n_i\qq_i$ by which we represent this line bundle. 

What happens if the degree $d$ of ${\cal M}$ is not zero?  To make a similar definition, we have to first pick
a universal bundle line over $D\times \Pic_c(D)$.  We recall this notion from section \ref{eleig}: such a universal  line bundle is
a line bundle $\W\to D\otimes \Pic_c(D)$ whose restriction to $D\times x$, for any $x\in \Pic_c(D)$, is of the isomorphism
type corresponding to $x$.  Such a universal line bundle exists, but is unique only up to
\begin{equation}\label{doffo}\W\to \W\otimes \pi_2^*(\RR),\end{equation}
where $\RR\to \Pic_c(D)$ is some line bundle, and $\pi_2:D\times\Pic_c(D)\to \Pic_c(D)$ is the projection.

Eqn. (\ref{zudco}) can be nicely rewritten in terms of a universal bundle $\W$.  Given $\M=\otimes \O(\qq_i)^{n_i}$, the
corresponding line bundle over $\Pic_c(D)$ is
\begin{equation}\label{offo} \N=\otimes_i \W|^{n_i}_{\qq_i\times \Pic_c(D)}. \end{equation}
To recover eqn. (\ref{zudco}) from this formula, one simply identifies $\W|_{\qq_i\times x}$ with $\L_x|_{\qq_i}$.  This is the best
we can do to generalize eqn. (\ref{zudco}) for the case that $\M$ has a nonzero degree $d$.
However, if $\W$ is transformed to $\W\otimes \RR$, for some line bundle $\RR\to \Pic_c(D)$, then $\N$ is transformed by
\begin{equation}\label{woffo}\N\to \N\otimes \RR^d. \end{equation}
The dependence on the choice of a universal bundle means that $\N$ is not really naturally defined as a line bundle over $\Pic_c(D)$,
but is more naturally understood as a twisted line bundle, twisted by a certain gerbe.  (This is a tautology: one can define a gerbe over
$\Pic_c(D)$ that is naturally trivialized by any choice of a universal line bundle over $D\times \Pic_c(D)$, and then eqn. (\ref{woffo})
means that $\N\to \Pic_c(D)$ is best understood as a twisted line bundle, twisted by the $d^{th}$ power of that gerbe.  See section \ref{gerbes}.)
The motivation for this formulation will hopefully be clear in section \ref{compheck}.

Our discussion so far has been asymmetric.  We have considered the symbol $\langle{\cal L},
{\cal M}\rangle,$ with ${\cal L}$ and ${\cal M}$ being respectively of degree $c$ and degree
$d$.  We have kept ${\cal M}$ fixed and let ${\cal L}$ vary, and found that in this
case, $\langle{\cal L},{\cal M}\rangle$ varies as a twisted line bundle of degree $d$
over ${\rm Pic}_c(D)$.  Obviously, by symmetry, if we keep ${\cal L}$ fixed and let ${\cal M}$
vary, we will get a twisted line bundle of degree $c$ over ${\rm Pic}_d(D)$.
So when we let both ${\cal L}$ and ${\cal M}$ vary, we get a twisted line bundle over
${\rm Pic}_c(D)\times {\rm Pic}_d(D)$ of bidegree $(d,c)$.  After a digression on gerbes, we will  reformulate this
assertion in the language of duality.

\subsection{Gerbes}\label{gerbes}

``Gerbes'' have figured in this paper at several points.  We will give a minimal explanation of this concept, explaining only the simplest
points that might be helpful for understanding this paper (and in particular as background for section \ref{ugduality}).

For our purposes,  a ``gerbe''  is associated to the group $\C^*$ or a subgroup $U(1)$ or $\Z_n$.
A gerbe $\eG$ over a space $X$ is trivial locally but possibly not globally.  (Most gerbes of interest to us are trivial globally but not canonically so.)
Two local trivializations of a gerbe over an open set $U\subset X$ differ by tensoring by a line bundle $\L\to U$.   We will not explain  this statement in an abstract way,
but in examples the meaning will be 
clear.  For a $U(1)$ gerbe, $\L$ has always a hermitian metric (so its structure group reduces to $U(1)$).  For a $\Z_n$ gerbe,
$\L$ always possesses an isomorphism $\L^n\cong \O$.  If a gerbe has a connection, then the line bundle $\L$ associated to a local change
of trivialization will also have a connection.

 Here are the examples that have arisen
in this paper.  If $C$ is a Riemann surface, there is a $\Z_2$ gerbe over $C$, mentioned in footnote \ref{brief},
 whose trivializations are spin structures on $C$.  It is subtle to describe what is a spin structure, but two spin structures differ
 by twisting by a line bundle $\L$ of order 2, so this is a $\Z_2$ gerbe.  Global spin structures exist, so this gerbe is trivial, but not canonically.
 For a second example, consider universal line bundles over $C\times \Pic_c(C)$ for some $c\in \Z$.  Any two such universal
line bundles differ by tensoring with a line bundle $\mathcal R\to \Pic_c(C)$, which tells us that there is  a gerbe over $\Pic_c(C)$ that is trivialized by any choice
of a 
universal line bundle over $C\times \Pic_c(C)$.  Such universal line bundles exist, so the gerbe in question is trivial, but  not canonically.
To give a similar example with a nontrivial gerbe, let $^L\neg G$ be a simple nonabelian group with nontrivial center $\Z_n$.  Then there is a gerbe
over the moduli space  $\M(^L\neg G,C)$ of $^L\neg G$ bundles\footnote{All statements we are about to make apply if $\M(^L\neg G,C)$ is replaced by the corresponding
Higgs bundle moduli space $\MH(^L\neg G,C)$.}
that is trivialized over an open set $U\subset \MH(^L\neg G,C)$  by the choice of a
 universal $^L\neg G$-bundle over $C\times U$.    Such universal bundles exist
locally, and any two choices differ by twisting by a line bundle $\L\to U$ that is of order $n$.  So the problem of finding
a universal bundle in this situation defines a $\Z_n$ gerbe  over $\MH(^L\neg G,C)$  This gerbe is topologically non-trivial,
since a universal bundle does not exist globally.  For more on this example, see section 7 of \cite{KW}.  This is the only nontrivial gerbe that is relevant
in the present paper.

If a gerbe $\eG$ over a space $X$ 
is globally trivial, then a trivialization of it is called a twisted line bundle.  The motivation for this terminology is as follows.  The difference between
two trivializations is a line bundle $\L\to X$,  but any one trivialization is  not such a line bundle, so we call it instead a twisted line bundle,\footnote{If
$\eG$ is a non-trivial  gerbe, then by definition a $\eG$-twisted line bundle does not exist.  However, there is a notion of a $\eG$-twisted
vector bundle, and a non-trivial gerbe may admit a twisted vector bundle.  For instance, given a representation $^L\neg R$ of $^L\neg G$ on which the center of $^L\neg G$ acts non-trivially,
a universal bundle $\EE_{^L\neg R}\to C\times\M(^L\neg G,C)$ it does not exist as a vector bundle, but it does exist as a $\eG$-twisted vector bundle, where $\eG$ is the gerbe that is trivialized locally by a choice of a universal bundle.  This statement is a fancy tautology.
See section 7 of \cite{KW} for an elementary
explanation.} twisted by $\eG$.

Now let us discuss connections   on a $U(1)$ gerbe.   If a $U(1)$ gerbe $\eG$ is trivialized, then a connection can be represented
simply by a two-form $B$.  (This is analogous to the fact that if a line bundle $\L\to X$ is trivialized, then a connection on $\L$ can be represented
by a 1-form $A$.  Connections make sense more generally on non-trivial gerbes, but we will not need this notion.)  A change of trivialization is accomplished by twisting by a line bundle $\L\to X$ that has a connection $A$.  Let $F=dA$ be the
corresponding curvature.  Under the change of trivialization by $\L$, the two-form $B$ is shifted by $B\to B+F$.  
Here $B+F$ is the connection form on $\eG$ relative to the new trivialization.
By a flat $\eG$-twisted line bundle (or equivalently a flat trivialization of $\eG$) we mean a trivialization such that the new connection form $B+F$ is
identically 0.  In order for a flat $\eG$-twisted line bundle to exist,  the three-form curvature $H=dB$ (which is invariant under $B\to B+F$) must
vanish -- in which case we call $\eG$ a flat gerbe -- 
 and the periods of the two-form $B/2\pi$ must be integers.   If flat $\eG$-twisted line bundles exist, there may not be a canonical way
to pick one, but any two differ by twisting by an ordinary flat line bundle.

\subsection{Duality}
\label{ugduality}

In section \ref{dualtori}, we introduced the notion of a duality between complex tori $\cF$ and
$\cF'$.  Such a duality means that $\cF$ parametrizes flat line bundles over $\cF'$, and
vice-versa.  It is most usefully expressed by exhibiting a Poincar\'e line bundle,
which is a line bundle ${\cal T}\to \cF\times \cF'$ whose main property is that its restriction
to $f\times {}\cF'$, for $f\in \cF$, is the line bundle over $\cF'$ labeled by $f$, and similarly
with the roles of $\cF$ and $\cF'$ reversed.  Thus, the restriction to $f\times \cF'$ gives,
as $f$ varies, a universal family of flat line bundles over $\cF'$, and similarly with the
two factors exchanged.

We can extend this to a notion of a twisted duality. 
Suppose  that two tori
$\cF$ and $\cF'$ are endowed with flat $U(1)$ gerbes ${\eG}$ and ${\eG}'$ and suppose that these flat gerbes are trivial (as flat gerbes, not just topologically).
  So ${\eG}$ or ${
\eG}'$ can be trivialized, but not canonically, by choosing a
twisted line bundle, which moreover we can choose to be flat.
What we will call a twisted duality between $\cF$ and $\cF'$ of type
$({\eG},{\eG}')$ identifies $\cF'$ as the moduli space of flat
trivializations of ${\eG}$, and $\cF$ as the moduli space of flat
trivializations of ${\eG}'$.  As in the ordinary case, the most
natural way to describe a twisted duality is to exhibit a twisted
Poincar\'e line bundle, twisted by the gerbe ${\eG}\otimes {}
{\eG}'$ over $\cF\times {}\cF'$, whose restriction to $\cF\times
{}f'$ (or to $f\times {}\cF'$) gives, as $f'$ (or $f$) varies, a
universal family of flat trivializations of ${\eG}$ (or ${\cal
G}'$).

We claim that the symbol $\langle{\cal L},{\cal M}\rangle$ gives a canonical twisted
Poincar\'e line bundle of bidegree $(d,c)$ over ${\rm Pic}_c(D)\times {\rm Pic}_d(D)$.
Since the association ${\cal L},{\cal M}\to \langle{\cal L},{\cal M}\rangle$ is certainly
canonical, and is twisted in the appropriate way, all we have to show
is that it is a duality.
The general case can be mapped, noncanonically, to the case $c=d=0$; we simply map
${\rm Pic}_c(D)$ and ${\rm Pic}_d(D)$ to ${\rm Pic}_0(D)={\rm Jac}(D)$ by picking basepoints
(that is, by picking particular flat line bundles of degree $c$ or $d$ that we map to
the origin in ${\rm Jac}(D)$) and 
trivializations of the gerbes.  So
to show that $\langle{\cal L},{\cal M}\rangle$ defines a canonical twisted duality of
degree $(d,c)$, we just have to show that it does define a duality for $c=d=0$.

As this is a significant fact, we will explain it in two ways.
First, in differential geometry, to describe a family
of holomorphic line bundles over a Riemann surface $D$, one can consider a fixed smooth
complex line bundle ${\cal U}$ with a family of unitary connections.  If we write the
connection on ${\cal U}$ as $U$, then 
\cite{quillen} the determinant line
bundle over a family of complex line bundles obtained by letting $U$ vary
 has a natural hermitian metric and a natural unitary
 connection whose curvature is
\begin{equation}\label{pliko}
{i\over 4\pi}\int_D \delta U\wedge \delta U.\end{equation} Now we
recall the definition of $\langle{\cal L},{\cal M}\rangle$:
\begin{equation}\label{zunko}
\langle{\cal L},{\cal M}\rangle={\det\,H^*(D,{\cal L})\otimes
\det\,H^*(D,{\cal M})\over \det\,H^*(D,{\cal L}\otimes {\cal
M})\otimes \det\,H^*(D,{\cal O})}.\end{equation} We think of
${\cal L}$ and ${\cal M}$ as fixed smooth line bundles with
connections $A$ and $B$, respectively.  Then ${\cal L}\otimes
{\cal M}$ is a fixed smooth line bundle with connection $A+B$. The
determinant line bundles of ${\cal L}$, ${\cal M}$, and ${\cal L}
\otimes{\cal M}$ have natural connections whose curvatures are
obtained, respectively, by substituting $A$, $B$, or $A+B$ for $U$
in (\ref{pliko}).  Then by taking the appropriate linear
combination of these curvatures in view of the definition of
$\langle{\cal L},{\cal M} \rangle$, we learn that the line bundle
given by $\langle{\cal L},{\cal M}\rangle$ has a natural
connection of curvature
\begin{equation}\label{likko}
-{i\over 2\pi}\int_D\delta A\wedge \delta B.\end{equation}

So far we have a line bundle over $\AA\times \AA'$, where the two
factors are, respectively,  the spaces of all connections on ${\cal L}$ and on ${\cal
M}$. We want to descend to a line bundle over ${\rm Jac}(D)\times
{\rm Jac}(D)$. The group of gauge transformations of ${\cal L}$
and ${\cal M}$ acts on $\AA\times \AA'$, and this action lifts to
an action on the line bundle $\langle{\cal L},{\cal M}\rangle$. We
want to divide by the gauge group. For this, we have to take
${\cal L}$ and ${\cal M}$ to have degree zero. Otherwise, the
constant gauge transformations, which act trivially on $A$ and
$B$, will act nontrivially on $\langle{\cal L},{\cal M}\rangle$,
as was explained following (\ref{iftong}), and there will be no
reasonable quotient. Once we restrict ${\cal L}$ and ${\cal M}$ to
have degree zero, we can restrict the connections $A$ and $B$ to
be flat, and then the curvature form (\ref{likko}) descends to the
quotient by the group of gauge transformations, as we explained
following eqn. (\ref{ivo}). At this point, we can take the
quotient by the gauge transformations acting on the line bundle
$\langle{\cal L},{\cal M}\rangle$ and its connection, to get a
line bundle over ${\rm Jac}(D)\times {\rm Jac}(D)$ whose curvature
can be represented by the same formula (\ref{likko}). As we noted
in discussing eqn. (\ref{ivo}), a unitary line bundle over ${\rm
Jac}(D)\times {\rm Jac}(D)$ with this curvature is a Poincar\'e
line bundle.

For an alternative approach closer to algebraic geometry, pick a
basepoint $Q\in D$ and consider the embedding $\theta:D\to{\rm
Jac}(D)$ that maps $R\in D$ to the point in ${\rm Jac}(D)$
corresponding to the degree zero line bundle ${\cal O}(R)\otimes
{\cal O}(Q)^{-1}$.  Since this map gives an isomorphism
$H_1(D,\Bbb{Z}) \cong H_1({\rm Jac}(D),\Bbb{Z})$, and flat line
bundles are classified by homomorphisms of $H_1$ into $U(1)$,
there is  a one-to-one correspondence $W\to \theta^*(W)$ between flat line bundles
$W\to{\rm Jac}(D)$ and flat line bundles ${\cal M}\to D$, or
equivalently holomorphic line bundles ${\cal M}\to D$ of degree
zero.

To establish the duality,
we think of the association ${\cal L}\to \langle{\cal L},
{\cal M}\rangle$, with ${\cal M}$ held fixed and ${\cal L}$ allowed to vary, as defining
a line bundle $W_{\cal M}$ over ${\rm Jac}(D)$.  We want to show that, up to isomorphism,
each flat line bundle  $W\to {\rm Jac}(D)$ is isomorphic to precisely one of the
$W_{\cal M}$.  But in view of the remark in the last paragraph, it suffices
to show that $\theta^*(W)=\theta^*(W_{\cal M})$ for some unique ${\cal M}$.
We claim that, for any degree zero line bundle ${\cal M}\to D$, we have
\begin{equation}\label{yormy}\theta^*(W_{\cal M})\cong {\cal M}.\end{equation}
So we will have $\theta^*(W)\cong \theta^*(W_{\cal M})$
if and only if we take ${\cal M}=\theta^*(W)$.

To establish the claim (\ref{yormy}),
we must understand $\langle{\cal L},{\cal M}\rangle$
for ${\cal L}={\cal O}(R)\otimes {\cal O}(Q)^{-1}$.
For this, we simply use (\ref{trifong}),
 but now with the roles of ${\cal L}$ and ${\cal M}$
reversed, to learn that $\langle{\cal O}(R)\otimes {\cal
O}(Q^{-1}),{\cal M}\rangle \cong {\cal M}|_R\otimes {\cal
M}|_Q^{-1}$. We want to keep $Q$ fixed (so ${\cal M}|_Q^{-1}$ is
an inessential fixed one-dimensional vector space ${\cal R}$) and
let $R$ vary. The line bundle whose fiber at $R$ is ${\cal
M}|_R\otimes {\cal M}|_Q^{-1}$ is simply ${\cal M}\otimes {\cal
R}$. In other words, it is isomorphic to ${\cal M}$, as we aimed
to show.

\subsection{More On Duality Of Hitchin Fibers}
\label{dualap}

In the context of the spectral cover $\psi:D\to C$, self-duality of the
Jacobian of  $D$ means that the fiber of the Hitchin fibration is
self-dual for the self-dual group $U(N)$.\footnote{As explained in
section \ref{unitaryexample} (see footnote \ref{brief}), if $N$ is even, identifying the fiber of the Hitchin fibration
with the Jacobian depends upon a choice of square root of the
canonical bundle $K$.  We likewise use below a description of the
fiber for $SU(N)$ that for even $N$ depends on a choice of spin
structure.} Now let us reconsider
duality of the Hitchin fibers for the groups $SU(N)$ and $PSU(N)$,
using the spectral cover $\psi:D\to C$.

We have already explained that the symbol $\langle{\cal L},{\cal
M}\rangle$, for ${\cal L}\in {\rm Pic}_c(D)$, ${\cal M}\in{\rm
Pic}_d(D)$, gives a twisted duality of degrees $(d,c)$ between
${\rm Pic}_c(D)$ and ${\rm Pic}_d(D)$.  This is the self-duality
of the Hitchin fibers for $G=U(N)$, generalized to allow electric
and magnetic fluxes.  (This interpretation is explained in section
\ref{abcase}.)  Now we claim that if we restrict the first
variable to lie in the Hitchin fiber for $SU(N)$, the second will
naturally project to $PSU(N)$, and we will get the desired twisted
duality between $SU(N)$ and $PSU(N)$. So we require ${\cal L}$ to
take values in $\FF_{SU(N)}$; in other words, we take ${\cal L}\in
{\rm Jac}(D)$ and require ${\rm Nm}({\cal L})$ to be trivial.  As
we will show shortly, if ${\cal M}_0$ is a line bundle over $C$ of
degree zero, then $\langle{\cal L},\psi^*({\cal M}_0)\rangle$
(where we allow ${\cal L}$ to vary while keeping fixed ${\rm
Nm}({\cal L})$ and ${\cal M}_0$) is trivial as a line bundle over
$\FF_{SU(N)}$. Given this, the bilinearity of the pairing
$\langle~,~\rangle$ implies that $\langle{\cal L},{\cal
M}\rangle\cong\langle{\cal L},{\cal M}\otimes \psi^*({\cal
M}_0)\rangle$ for ${\cal M}_0\in {\rm Jac}(C)$.
 Hence
with ${\cal L}$ so restricted, we can consider ${\cal M}$ to take
values in $\FF^{(d)}_{PSU(N)}={\rm Pic}_d(D)/{\rm Jac}(C)$.
  Then the symbol
$\langle{\cal L},{\cal M}\rangle$ defines a twisted line bundle of
degree $(d,0)$ over $\FF_{SU(N)}\times \FF^{(d)}_{PSU(N)}$. Thus,
$\FF_{SU(N)}$ parametrizes a family of ordinary line bundles over
$\FF^{(d)}_{PSU(N)}$, and $\FF^{(d)}_{PSU(N)}$ parametrizes a family
of twisted line bundles of degree $d$ over $\FF_{SU(N)}$. This is
the expected duality.  The fact that it is a duality, and not just
a pairing, depends upon the fact that the symbol $\langle {\cal
L},{\cal M}\rangle$  has no additional symmetry in the second
variable except what we have already accounted for; we discuss
this briefly at the end of this section.

We can generalize this to let ${\cal L}$ have degree $c$.  Now we
pick a fixed line bundle ${\cal L}_0\to C$ of degree $c$ and
consider ${\cal L}\in {\rm Pic}_{c}(D)$ with ${\rm Nm}({\cal
L})={\cal L}_0$. We claim that $\langle{\cal L},{\cal M}\rangle$
is still invariant to twisting ${\cal M}$ by the pullback of a
degree zero line bundle ${\cal M}_0\to C$. This being so, the
symbol $\langle{\cal L},{\cal M}\rangle$ gives the expected
duality of degree $(d,c)$ between $\hat \FF^{(c)}_{PSU(N)}$ and
$\FF^{(d)}_{PSU(N)}$. 

In these statements, the integers $c$ and $d$ correspond\footnote{The reader may wish to return to this paragraph
after reading a more detailed explanation of the same question for the case of $G=U(1)$ in section \ref{compheck}.}  to the discrete electric
and magnetic charges that can arise in gauge theory of $PSU(N)$ or $SU(N)$.
We recall that for these groups, discrete $\Z_N$-valued electric and magnetic charges are possible.
$\Pic_c(D)$ parametrizes Higgs bundles with a value $c$ of the discrete magnetic charge, and a twisted
line bundle over $\Pic_c(D)$ that is twisted by the $d^{th}$ power of the universal bundle can be the Chan-Paton
bundle of a brane whose discrete electric charge is $d$.  In the duality between $PSU(N)$ and $SU(N)$, the roles of
$c$ and $d$ are exchanged, as one would expect.  

It remains to show that if ${\rm Nm}({\cal L})$ is fixed, then
$\langle{\cal L},\psi^*({\cal M}_0)\rangle$ is constant as ${\cal
L}$ varies. For this,
 we need to know a few  facts. First, for
$\psi:D\to C$ a map between Riemann surfaces, we have simply
$H^*(D,{\cal L})=H^*(C,\psi_*{\cal L})$, so
\begin{equation}\label{luto} \det\,H^*(D,{\cal
L})=\det\,H^*(C,\psi_*{\cal L}).\end{equation} The right hand
side, of course, is $\det\, H^*(C,E)$, where $E=\psi_*({\cal L})$.
 Furthermore, for any vector bundle
$E$ over a Riemann surface $C$, we have
\begin{equation}\label{unofto}\det\,H^*(C,E)=\det\,H^*(C,\det\,E).\end{equation}
This can be proved by induction in the rank of $E$.  Any bundle
$E\to C$ has a holomorphic line sub-bundle ${\cal M}$, and so
appears in an exact sequence \begin{equation}\label{gonno} 0\to
{\cal M}\to E\to U\to 0.\end{equation} From this, we get (as in
the discussion of (\ref{onglo})) an isomorphism $\det\, E\cong
\det {\cal M}\otimes {\rm det}\,U$.  In addition, we can derive
from (\ref{gonno}) a long exact sequence of cohomology groups,
which, as in the derivation of (\ref{monglo}), gives an
isomorphism $\det\,H^*(C,E)\cong \det\,H^*(C,{\cal M})\otimes
\det\,H^*(C,U)$. Combining these, we obtain (\ref{unofto}) by
induction in the rank of $E$.

Now for line bundles ${\cal L}\to D$, ${\cal M}_0\to C$, we have
$\det\,H^*(D,{\cal L}\otimes \psi^*({\cal
M}_0))=\det\,H^*(C,\psi_*({\cal L}\otimes \psi^*({\cal M}_0)))
=\det\,H^*(C,E\otimes {\cal M}_0)=\det\,H^*(C,\det (E\otimes {\cal
M}_0))$. Similarly, $\det\,H^*(D,{\cal L})=\det\,H^*(C,\det E)$.
  If we vary ${\cal L}$ keeping
${\rm Nm}({\cal L})$ fixed, then $\det E$ is also fixed, so
finally $\det\,H^*(D,{\cal L}\otimes {\cal M})\otimes
\det\,H^*(D,{\cal L})^{-1}$ is also fixed.  But in the definition
of $\langle{\cal L},\psi^*({\cal M}_0)\rangle$, this is the factor
that depends on ${\cal L}$.  So  $\langle{\cal L},\psi^*({\cal
M}_0)\rangle$  remains fixed, as claimed, when ${\cal L}$ varies
keeping fixed ${\rm Nm}({\cal L})$ and ${\cal M}_0$.

Finally, let us show that if, for ${\cal M}$ of degree zero,
$\langle{\cal L},{\cal M}\rangle$ is trivial for ${\cal L}\in
\FF_{SU(N)}$, then ${\cal M}$ is a pullback from $C$. We showed in section \ref{character} than if ${\rm
Nm}({\cal L})={\cal O}$, then ${\cal L}$ is  a tensor product of
line bundles ${\cal O}(\qq)\otimes {\cal O}(\qq')^{-1}$, where
$\psi(\qq)=\psi(\qq')$.  By the bilinearity of the pairing,
$\langle{\cal L},{\cal M}\rangle$ is trivial for ${\cal L}$ such a
tensor product if and only if it is trivial for a single factor.
So we set ${\cal L}={\cal O}(\qq)\otimes {\cal O}(\qq')^{-1}$, and get
$\langle{\cal L},{\cal M}\rangle={\cal M}(\qq)\otimes {\cal
M}(\qq')^{-1}$.  So a trivialization of $\langle{\cal L},{\cal
M}\rangle$ as ${\cal L}$ varies gives an identification of ${\cal
M}(\qq)$ with ${\cal M}(\qq')$ whenever $\qq$ and $\qq'$ lie over the same
point in $C$. Existence of such an identification implies that
${\cal M}$ is a pullback from $C$.

We have implicitly used here the fact that the variety
parametrizing pairs $\qq,\qq'\in D$ with $\psi(\qq)=\psi(\qq')$ is
irreducible.  One can show this by finding a local model of the
spectral curve with sufficiently large monodromy, for example
$y^N+\epsilon yz+z=0$, with $z$ a local parameter on $C$ and
$\epsilon$ a complex constant.

\section{Computing The Hecke Eigenvalue}
\label{compheck}

Finally we will use what we have  learned to compute the
``eigenvalue'' with which an 't Hooft operator acts on a magnetic
eigenbrane. We begin with the abelian case in section
\ref{abcase}. This is the obvious place to start, and certainly
leads to the most straightforward calculations, though we will run
into some subtle questions of interpretation because of the
elementary fact that -- unlike a simple non-abelian group -- the
group $U(1)$ has a center and a fundamental group that are both of
infinite order. 

Geometric Langlands duality in abelian gauge theory is most
commonly described by a somewhat different argument, attributed to
Deligne, that can be found for example in
\cite{frenkel}.  We follow a route that will give a useful starting point for the
nonabelian generalization, to which we turn in section
\ref{actsun}.

\subsection{The Abelian Case}
\label{abcase}
\subsubsection{Calculation}

For $G=U(1)$, a Higgs bundle over a Riemann surface $C$ is just a
pair $({\cal L},\varphi)$, where ${\cal L}$ is a complex line
bundle, and $\varphi\in H^0(C,K)$ is a holomorphic differential.
$\varphi$ will actually play little role in the analysis, the
reason being that as $U(1)$ is abelian, Hitchin's equations are
linear, the gauge field and Higgs field are decoupled, $\varphi$
is invariant under Hecke transformations, and the interesting
action of $S$-duality is just on the gauge field.

One important point, however, also discussed in
section \ref{unitary}, is that Hitchin's moduli
space $\MH(U(1),C)$ parametrizes Higgs pairs $(\L,\varphi)$ for which
the line bundle $\L$ has degree zero.  This condition is not preserved, in general,
by the action of 't Hooft operators, so some discussion is required.

We consider
$U(1)$ gauge theory in a Hamiltonian framework on a three-manifold $W=I\times C$,
where $I$ is an interval.  Boundary conditions on the right hand end of $W$ are defined
by a brane  that we choose to be an electric or magnetic eigenbrane.
We act on an electric eigenbrane
 ${\cal B}$ with Wilson operators $W^{(d_i)}(y_i,\pp_i)$ of charges $d_i$,
inserted at points $y_i\times \pp_i\in I\times C$.
Or dually we act on a magnetic eigenbrane $\hat{\cal B}$ with 't Hooft operators
$T^{(d_i)}(y_i,\pp_i)$.  The two-dimensional effective picture was sketched in fig.
\ref{nuxto} in section \ref{thooftheckeop}.

Concretely, we pick an electric eigenbrane ${\cal B}$ that is a zero-brane supported
at some point $x\in \MH$ that corresponds to a Higgs bundle $({\cal L},\varphi)$. And
we consider a product of Wilson operators acting on ${\cal B}$:
\begin{equation}\label{pilkob}{\prod_{i=1}^n} \,W^{(d_i)}(y_i,\pp_i)\,{\cal B}.\end{equation}
As we explained in section \ref{eleig}, 
the zero-brane ${\cal B}$ is an electric
eigenbrane, with
\begin{equation}\label{nilkob}{\prod_{i=1}^n} \,W^{(d_i)}(y_i,\pp_i)\,{\cal B}=
{\cal B}\otimes\left(\otimes_{i=1}^n{\cal \L}|_{\pp_i}^{d_i}\right).\end{equation}
(Momentarily we will rewrite this in a way that is closer to eqn. (\ref{zefor}) of
section \ref{eleig}.)

An important subtlety here springs from the familiar fact that
a point $x\in \MH$ only determines a corresponding line bundle
${\cal L}$ up
to isomorphism.  We are free to tensor ${\cal L}$ with a fixed one-dimensional
vector space ${\cal R}$.  This causes no trouble if $d=\sum_id_i$ is equal to zero,
but in general the ``eigenvalue'' with which the given product of Wilson operators
acts on ${\cal B}$ is tensored with ${\cal R}^d$. 
Of course, this ambiguity is
independent of the choice of points $\pp_i$, and we encountered it in a different guise in section \ref{interpretation}.
As in that discussion, the most illuminating way to proceed is to pick a universal bundle $\W\to C\times {\mathrm{Jac}}(C)$
and express eqn. (\ref{nilkob}) in terms of this universal bundle:
\begin{equation}\label{wilkob} {\prod_{i=1}^n} \,W^{(d_i)}(y_i,\pp_i)\,{\cal B}=
{\cal B}\otimes\left(\otimes_{i=1}^n{\W}|_{\pp_i\times x}^{d_i}\right).\end{equation}

For $d\not=0$, this result does depend on the choice of the universal bundle $\W\to C\times {\mathrm{Jac}}(C)$.
If we transform $\W$ to $\W\otimes \RR$, where $\RR$ is the pullback to $C\times \mathrm{Jac}(C)$ of a line bundle over the
second factor, then the right hand side of eqn. (\ref{wilkob}) is tensored with $\RR^d$.

The physical meaning is as follows.  The integer $d$ is the total electric charge  carried by the product of Wilson operators 
${\prod_{i=1}^n} \,W^{(d_i)}(y_i,\pp_i)$ with which we are acting.  If the initial brane $\B$ is electrically neutral, then the brane
that results from acting with this product of Wilson operators has an electric charge\footnote{In the effective
$1+1$-dimensional desription, a brane is a boundary condition at the end of a string.  Saying that a brane $\B$ has charge $d$
simply means that with this boundary condition, the electric charge operator receives a contribution $d$ at the end of the string.} $d$. 
The $\sigma$-model with target $\MH(U(1),C)$ only describes neutral degrees of freedom, and to describe a brane that carries
charge, we cannot just use this $\sigma$-model; to describe a charged brane, we need to include the unbroken $U(1)$ gauge multiplet in the effective two-dimensional description. We say more on this in section \ref{elmag}.

 In general, a neutral brane over $\MH(U(1),C)$
has a Chan-Paton bundle $\U$ that is an ordinary vector bundle or sheaf.  A brane of electric charge
 $d$ has instead a Chan-Paton bundle $\U$
that is a twisted vector bundle or sheaf.   The twisting is most easily described by saying that $\U$ can be described as an ordinary vector
bundle or sheaf once a universal bundle $\W=C\times \mathrm{Jac}(C)$ is picked, but $\U$ depends on the choice of $\W$:
transforming $\W$ to $\W\otimes \RR$ will transform $\U$ to $\U\otimes \RR^d$.  Clearly with this statement of what sort of object is
the Chan-Paton bundle of a brane that carries a net electric charge, eqn. (\ref{wilkob}) is consistent with the way that we expect
acting with a Wilson operator to transform the electric charge of a brane.

The dual of an electric eigenbrane ${\cal B}$ is a brane $\hat{\cal B}$
of type ${\bf F}$ that we expect to be a magnetic eigenbrane. It is supported
on a fiber $\FF$
of the Hitchin fibration for which the Higgs field 
 $\tilde \varphi $ is a multiple of the original Higgs field $\varphi$ on the electric side.\footnote{The multiple is $\mathrm{Im}\,\tau$,
 where $\tau=\theta/2\pi + 4\pi i/e^2$ is the  usual gauge coupling parameter.  It is usually just scaled out and plays no essential role
 in what follows.}
 $\FF$ is a copy of the Jacobian ${\rm Jac}(C)$.  The brane $\hat \B$ is 
characterized by  a flat  Chan-Paton line bundle that we will discuss shortly.  We want to calculate the action of a product of 't
Hooft operators on $\hat{\cal B}$:
\begin{equation}\label{doppy}  {\prod_{i=1}^n} \,T^{(d_i)}(y_i,\pp_i)\,\hat{\cal B}.
\end{equation}
Again, there is a subtlety; the product of the 't Hooft operators
shifts the degree of the line bundle by $d=\sum_id_i$, so if $d$
is nonzero, we are mapped out of $\MH$. As in the electric case,
this means that a brane can carry magnetic flux, which is not incorporated in the two-dimensional $\sigma$-model with target $\MH$. 

For simple or semi-simple
nonabelian $G$, the analogous notion of magnetic flux is a discrete topological invariant $\xi$ (introduced in section \ref{psun})
of a Higgs bundle.  Because it is a discrete invariant, it can be carried
by a flat bundle, or by a solution of the Hitchin equations for $G$.  Thus $\MH(G,C)$ for simple or semi-simple
$G$ has components that are labeled by $\xi$.  But for $G=U(1)$, the magnetic flux is not conveniently
described in a $\sigma$-model with target $\MH(G,C)$; to describe it, one has to retain the unbroken $U(1)$ gauge symmetry in the low
energy description.
 We will
explain in section \ref{elmag} what sort of brane carries magnetic
flux for $G=U(1)$.

Leaving aside questions of interpretation, to calculate the action
of the 't Hooft operators on $\hat{\cal B}$ is actually a simple
exercise.   According to our reassessment of duality in section \ref{dualap},
a zero-brane supported at a point in $\Jac(C)$ corresponding to a line bundle $\L\to C$ should have for its dual a brane $\hat \B$ of type $\bf F$
whose Chan-Paton bundle is $\langle\L,\M\rangle$.  Here this expression is viewed as a line bundle over a copy of $\Jac(C)$ parametrized
by $\M$, with $\L$ kept fixed.   
According to eqn. (\ref{welmo}), the Chan-Paton bundle of the 
brane $\prod_{i=1}^n
\,T^{(d_i)}(y_i,\pp_i)\,\hat{\cal B}$ obtained by acting on
$\hat{\cal B}$ with the indicated product of 't Hooft operators is $\tilde \U=\Phi^*(\U)$, where $\Phi$ is the automorphism
$\M\to \M\otimes \O(\pp_i)^{d_i}$ of $\Jac(C)$.

We have $\Phi^*(\langle\L,\M\rangle)=\langle\L,\M\otimes_i \O(\pp_i)^{d_i}\rangle$.  Using the bilinearity of the $\langle~,~\rangle$ symbol,
this is the same as $\langle\L,\M\rangle\otimes \langle\L,\otimes_i\O(\pp_i)^{d_i}\rangle$.  In other words,
\begin{equation}\label{menz}\hat\U=\U\otimes\langle\L,\otimes _i\O(\pp_i)^{d_i}\rangle. \end{equation}
This confirms that the brane $\hat \B$ is a magnetic eigenbrane with ``eigenvalue'' the one-dimensional vector space
$\langle\L,\otimes _i\O(\pp_i)^{d_i}\rangle=\otimes_i \L_{\pp_i}^{d_i}$, where we have made use of eqn. (\ref{trifong}) to evaluate the $\langle~,~\rangle$
symbol.  Thus the brane  $\hat {\cal B}$ is a magnetic eigenbrane with the ``eigenvalue'' that one would expect from the electric formula (\ref{nilkob}):
\begin{equation}\label{rgo}
{\prod_{i=1}^n} \,T^{(d_i)}(y_i,\pp_i)\,\hat{\cal B}=\hat{\cal B}\otimes
\left(\otimes_{i=1}^n{\cal L}|_{\pp_i}^{d_i}\right).
\end{equation}

This establishes the  duality.  The 't Hooft operators act on $\hat{\cal B}$ in
the same way that the Wilson operators act on ${\cal B}$.  It remains only to explain what
the calculations mean in case $d=\sum_id_i$ is nonzero.

\subsubsection{Electric And Magnetic Fields}
\label{elmag}

Higgs pairs $({\cal L},\varphi)$ with $c_1({\cal L})\not= 0$
inevitably appear when we act
with 't Hooft operators.  But they do not correspond to points in
$\MH$.  How then do they enter the gauge theory?  They correspond
to new branes of a sort that we have not yet considered.

The basic subtlety was actually pointed out in section \ref{prelims}.  The argument  that 
a four-dimensional gauge theory compactified on a Riemann surface $C$ reduces at low energies to a two-dimensional $\sigma$-model with target $\MH(G,C)$
assumes that the gauge symmetry is completely broken in  situations of interest.  For a
simple nonabelian gauge group $G$, this is so (apart from a finite group, the center of $G$) as long as we avoid
singularities of $\MH$ -- at which new degrees of freedom become
relevant. However, $G=U(1)$ is different since constant gauge
transformations act trivially and  every adjoint-valued field has a continuous group of gauge symmetries.   In any compactification of $\N=4$ super Yang-Mills
theory with $U(1)$ gauge group from four to two dimensions, the $U(1)$ gauge group remains unbroken and it should be considered in the low energy description.

As a result, the reduction to a low energy $\sigma$-model  is not quite valid for $G=U(1)$.  The low energy theory is a product
of a $\sigma$-model of target $\MH$ and a supersymmetric gauge theory with gauge group $U(1)$.
It is not possible to describe branes entirely in the $\sigma$-model.  One also has to take the
gauge theory into account.  

Let us consider the extended Bogomolny equations\footnote{Equivalently, we consider the supersymmetric equations
(3.29) of \cite{KW} at $t=1$, a convenient value for studying the $A$-model of type $K$.} (\ref{woff}) and (\ref{old})  on the three-manifold
 $W=I\times C$.   We view these as equations that describe time-independent supersymmetric
 configurations relevant to quantization on $W$.
  We endow $W$ with a
metric $dy^2 +d\Omega^2$, where $y=x^1$ is a Euclidean coordinate on
$I$ and $d\Omega^2$ is a $y$-independent Kahler metric on
$C$.  For gauge group $U(1)$, these equations have solutions of a kind quite
different than what we have considered so far. Let $\omega$ be a
multiple of the Kahler form of $C$, normalized so that
$\int_C\omega = 1$.  We can solve the extended Bogomolny equations
by picking a connection $A$ on a degree $d$ line bundle ${\cal
L}\to C$ whose curvature is\footnote{To avoid unnatural-looking
factors of $i$, we here will take $F$ and $\phi$ to be
real-valued.  So the curvature of a unitary connection $D$ on a
line bundle is $F=-iD^2$.} $F=2\pi  d\,\omega$. Then we pull back
${\cal L}$ and $A$ to $W$ and take\footnote{Here $\phi_0$ is just the ``time'' component of the Higgs field $\phi=\sum_{i=0}^3\phi_i dx^i$.} $\phi_0=2\pi y\cdot  d$. We take
all other fields to vanish. In this fashion, for $G=U(1)$, we get
a supersymmetric configuration with a line bundle on $C$ of any
degree.

Returning to the question of what happens in eqn. (\ref{doppy}) if
$d=\sum_id_i$ is nonzero, the answer is that in that case, to the
left of all of the 't Hooft operators, the line bundle ${\cal L}$
has nonzero degree and a supersymmetric configuration will be
created based on the solution that we have just described.  Hence,
in acting on $\hat {\cal B}$, a product of 't Hooft operators with
nonzero total $d$ creates a brane different from those we have so
far considered. This is a supersymmetric brane for which the
boundary conditions require $c_1({\cal L})=d$ and a normal
derivative of $\phi_0$ equal to $2\pi  d$. We might call this a
brane with magnetic charge $d$.

This construction does not have a close analog for semi-simple
nonabelian $G$, as long as we keep away from singularities of
$\MH$, because the extended Bogomolny equations would force
$\phi_0$ to commute with all the other fields and to generate a
symmetry of the solution. To get an analog for simple non-abelian
$G$ of what we have found for $U(1)$, we can take $C$ to have
genus zero or one, in which case a generic stable Higgs bundle $(E,\varphi)$
can leave a continuous group of unbroken gauge symmetries.  For genus 0,
the semistable Higgs bundle with $E$ trivial and $\varphi=0$ is generic (it has no deformations
as a semistable Higgs bundle) and leaves the gauge symmetry completely unbroken.
Thus the description of branes in genus\footnote{Here and in the following discussion in genus 1, we consider -- as throughout this paper -- the unramified
case of geometric Langlands.  Ramification points will reduce the gauge symmetry, as in \cite{GW}.} 0 will be based on a low energy theory that is a gauge
theory -- with gauge group $G$ -- with no accompanying $\sigma$-model.  In genus 1, the generic
stable Higgs bundle for the case that the bundle $E$ is topologically trivial leaves unbroken a subgroup
of $G$ that is isomorphic to its maximal torus.  So in genus 1, the situation for a simple nonabelian $G$ is somewhat
like what happens for $G=U(1)$:  the generic description of branes will use an abelian gauge theory combined with a $\sigma$-model.
In genus $>1$, a generic brane can be described in the $\sigma$-model language, but this description will break down for branes supported
at a singularity of $\MH$, at which there is enhanced gauge symmetry.  (The case that the enhanced gauge symmetry is a finite group
was studied in \cite{FW}.)

Going back to $G=U(1)$, the $S$-dual of a brane with magnetic flux
$d$ can be described classically as another novel kind of brane, which we will call a brane
with electric flux $d$. For this sort of brane, $\phi_0$ is still
a linear function of $x^1$, but now there is a constant electric
field instead of a  constant magnetic field.  For supersymmetry,
such fields must obey the appropriate
supersymmetric equations (eqn. (3.29) of \cite{KW}, which should be taken at $t=i$, which is the 
appropriate value for studying Wilson lines).  In abelian gauge
theory, those equations reduce (in Euclidean signature) to
$F+i\,d\phi=0$.  On a four-manifold $M=\Bbb{R}^2\times C$, with
Euclidean coordinates $x^0,$ $x^1$ on $\Bbb{R}^2$, these equations
can be solved with $\phi_0$ a linear function of $x^1$, $F_{01}$
constant, and everything else vanishing. In Euclidean signature,
$F_{01}$ has to be imaginary, but in Lorentz signature, which of
course is the real home of physics, the factor of $i$ disappears.

The field $\phi_0$ that was important in this analysis is part of the supersymmetric
multiplet -- known as a vector multiplet --
that contains the $U(1)$ gauge field.  The branes that carry electric and
magnetic charge can be described in the combined low energy theory consisting of the 
$\sigma$-model coupled to the $U(1)$ vector multiplet.

\subsection{Nonabelian Generalization}
\label{actsun}
\subsubsection{Calculation}

Now we move on to the nonabelian generalization.  To consider the
basic idea, we take $^LG=G=U(N)$, and we take a magnetic weight
$^Lw=(1,0,0,\dots,0)$ that is the highest weight of the
$N$-dimensional representation ${\cal V}$.

We let ${\cal B}$ be a zero-brane supported at a point $x\in \MH$
that corresponds to a Higgs bundle ${\cal E}=(E,\varphi)$.  The
Wilson operator $W({}^Lw;y,\pp)$, inserted at a point $y\times \pp\in
I\times C$, acts by
\begin{equation}\label{howacts} W({}^Lw;y,\pp){\cal B}={\cal
B}\otimes \EE_{H,\cal V}|_{p\times x}\end{equation} Here $\EE_{H,\cal V}\to
C\times \MH$ is the universal Higgs bundle  in the representation ${\cal
V}$ (the notion was introduced in section \ref{eleig}), and
$\EE_{H,\cal V}|_{p\times x}$ is its restriction to $\pp\times x\in C\times \MH$.  So $\EE_{H,\cal
V}|_{\pp\times x}$ is a fixed vector space, and (\ref{howacts}) expresses the
fact that ${\cal B}$ is an electric eigenbrane.

We pick $\pp$ and ${\cal E}$ generically so that $\varphi(\pp)$ is
regular and semisimple.  There consequently are $N$ distinct
solutions to the eigenvalue equation near $\pp$,
\begin{equation}\label{regsol}
\varphi\uppsi_i=y_i\uppsi_i,~i=1,\dots,N.\end{equation}
Correspondingly, $\EE_{H,\cal V}|_{\pp\times x}$ decomposes as a sum of
the eigenspaces of $\varphi$, which we denote as $\EE^{(i)}_{H,\cal V}|_{\pp\times x}.$
These eigenspaces have a
familiar interpretation in terms of the spectral cover $\psi:D\to C$.  The Higgs
bundle ${\cal E}=(E,\varphi)$ is determined by a degree zero line
bundle ${\cal M}\to D$.  The points $\qq_i\in D$ that lie above
$\pp\in C$ are in one-to-one correspondence with the eigenvectors of
$\varphi(\pp)$, and the definition of ${\cal M}$ is such that
$\EE^{(i)}_{H,\cal V}|_{\pp\times x}$ is the same as ${\cal M}|_{\qq_i}$, the fiber
of ${\cal M}$ at $\qq_i$.  And so the eigenbrane equation
(\ref{howacts}) is more explicitly
\begin{equation}\label{owacts} W({}^Lw;y,\pp){\cal B}={\cal
B}\otimes \left(\oplus_{i=1}^N{\cal M}|_{\qq_i}\right).\end{equation}

The magnetic eigenbrane $\hat {\cal B}$ that is dual to ${\cal B}$
is supported on the fiber $\FF$ of the Hitchin fibration that
contains the point $x$.  Its Chan-Paton bundle ${\cal U}$ is determined
by the analysis of duality in section \ref{backdet}.  The fiber of
$\U$ at a point in $\FF$ corresponding to a degree zero line bundle
${\cal L}\to D$ is ${\cal U}|_{\cal L}=\langle{\cal L},{\cal M}\rangle$.

Now we can determine the action of the 't Hooft operator
$T({}^Lw;y,\pp)$.  The operator $T({}^Lw;y,\pp)$ can act by any of $N$ possible
$\varphi$-invariant Hecke modifications of ${\cal E}=(E,\varphi)$.
 They correspond to holomorphic maps $\Phi_i:\FF\to \FF$, and
are in natural one-to-one correspondence with the points $\qq_i\in
D$ that lie over $\pp$.  $\Phi_i$ acts by ${\cal L}\to {\cal
L}(\qq_i)$, so it maps ${\cal W}$ to a new Chan-Paton bundle whose fiber at ${\cal L}$
is  $\langle{\cal L}(\qq_i),{\cal M}\rangle,$
which, by virtue of the bilinearity of the pairing
$\langle~,~\rangle$, is the same as $\langle{\cal L},{\cal
M}\rangle\otimes {\cal M}|_{\qq_i}$. So $\Phi_i$ maps $\hat \B$ to
$\hat \B\otimes {\cal M}|_{\qq_i}$, and after summing over all
choices of $\Phi_i$, we get
\begin{equation}\label{gladdo}T({}^Lw;y,\pp)\hat{\cal B}=\hat{\cal
B}\otimes \left(\oplus_{i=1}^N{\cal M}|_{\qq_i}\right).\end{equation} This is in good parallel
with (\ref{howacts}), so we have verified the expected duality in
this example.

Other minuscule weights of $^LG=U(N)$ can be considered in a
similar fashion. As in the discussion of eqn. (\ref{gelfix}), let
$^Lw(k)=(1,1,\dots,1,0,\dots,0)$ (with the number of 1's being $k$) be the highest
weight of the the representation $\wedge^k{\cal V}$ of $U(N)$.
First let us consider the action of the Wilson operator
$W({}^Lw(k);y,\pp)$. It acts on the zero-brane ${\cal B}$ by
\begin{equation}\label{owactos} W({}^Lw(k);y,\pp){\cal B}={\cal
B}\otimes \EE_{H,\wedge^k\cal V}|_{\pp\times x}\end{equation}
$\EE_{H,\wedge^k\cal V}$ is the universal Higgs bundle in the representation $\wedge^k\cal V$, and
$\EE_{H,\wedge^k\cal V}|_{\pp\times x}$ is its restriction to $\pp\times x$.
Once ${\cal V}$ is
decomposed as the direct sum of the one-dimensional eigenspaces ${\cal M}|_{\qq_i}$,
$\wedge^k{\cal V}$ has an analogous decomposition as
$\oplus_{\alpha} {\cal M}|_\alpha$, where $\alpha$ is a subset of
the set $\{1,2,\dots,N\}$ of cardinality $k$, and ${\cal
M}|_\alpha=\otimes_{i\in\alpha}{\cal M}|_{\qq_i}$. So we can write
the eigenvalue equation as
\begin{equation}\label{turme}W({}^Lw(k);y,\pp){\cal B}={\cal
B}\otimes \left(\oplus_{\alpha}{\cal M}|_\alpha\right).\end{equation}

On the magnetic side, the 't Hooft operator $T({}^Lw(k);y,\pp)$ acts
by a $\varphi$-invariant Hecke transformation.  For each
$\alpha$, there is
a $\varphi$-invariant Hecke transformation $\Phi_\alpha$ that
acts by  ${\cal L}\to {\cal L}_\alpha={\cal
L}\otimes\left(\otimes_{i\in \alpha}{\cal O}(\qq_i)\right)$. The
Chan-Paton bundle $\langle{\cal L},{\cal M}\rangle$ is mapped by
$\Phi_\alpha$ to $\langle{\cal L}_\alpha,{\cal M}\rangle$, which,
again using the bilinearity of the pairing, is the same as
$\langle{\cal L},{\cal M}\rangle\otimes {\cal M}|_\alpha$.  Thus,
once we sum over $\alpha$, the 't Hooft operator acts by
\begin{equation}\label{urme}T({}^Lw(k);y,\pp)\hat{\cal B}=\hat{\cal
B}\otimes\left(\oplus_\alpha {\cal M}|_\alpha\right).\end{equation}  Again, in
comparing (\ref{turme}) and (\ref{urme}), we see the expected
duality.

The most general minuscule weight for $G=U(N)$ is obtained from a
slight generalization of this.  We introduce an integer $r$ and
let $^Lw(k;r)=(r+1,r+1,\dots,r+1,r,\dots,r)$, with $k$ weights
equal to $r+1$ and the rest equal to $r$.  For example, if $r=-1$
and $k=N-1$, we get $^Lw=(0,0,\dots,0,-1)$, which is the highest
weight of the representation ${\cal V}^*$ that is dual to ${\cal
V}$.  In general, $^Lw(k;r)$ is the highest weight of the
representation $\wedge^k{\cal V}\otimes (\det\,{\cal V})^r$. On the
electric side, the effect of including $r$ is that the right hand
side of (\ref{turme}) must be tensored with $(\det\,\EE_{H,{\cal
V}}|_{\pp\times x})^r$, which is the same as $(\otimes_{i=1}^N{\cal M}|_{\qq_i})^r$:
\begin{equation}\label{zurme}W({}^Lw(k;r);y,\pp){\cal B}={\cal
B}\otimes \left(\oplus_{\alpha}{\cal
M}|_\alpha\right)\otimes\left(\otimes_{i=1}^N{\cal
M}|_{\qq_i}\right)^r.\end{equation} On the magnetic side, the
$\varphi$-invariant Hecke modifications now act by ${\cal L}\to
{\cal L}_\alpha\otimes(\otimes_{i=1}^N{\cal O}(\qq_i))^r$.  Repeating the
derivation of (\ref{urme}), we now get on the magnetic side
\begin{equation}\label{ozurme}T({}^Lw(k;r))\hat{\cal B}=\hat{\cal
B}\otimes\left(\oplus_{\alpha}{\cal
M}|_\alpha\right)\otimes\left(\otimes_{i=1}^N{\cal
M}|_{\qq_i}\right)^r.\end{equation}
This again exhibits the duality.

\subsubsection{Interpretation}\label{interp}

The sum of the weights of  $^Lw(k;r)$ is $k+rN$; this can be either
positive or negative.  Consider acting with a product of 't Hooft operators
determined by the minuscule weights $^Lw(k_\sigma;r_\sigma)$, $\sigma=1,\dots,s$.
This product of
't Hooft operators changes $c_1(E)$ by
$\Delta=\sum_\sigma\left(k_\sigma+Nr_\sigma\right)$.
Now we recall that although a Higgs bundle ${\cal E}=(E,\varphi)$ for $G=U(N)$
may have any value of $c_1(E)$, the hyper-Kahler manifold $\MH$ which is the target of the
$\sigma$ model parametrizes precisely the Higgs bundles of $c_1(E)=0$.

If $\Delta=0$, the given product of 't Hooft operators keeps us within this class.
Otherwise, we run into the same phenomenon that we discussed for $U(1)$ gauge theory in
section \ref{elmag}.  For $G=U(N)$, a generic Higgs bundle has $U(1)$ symmetry, simply
because the center of $U(N)$ is $U(1)$.   So the low energy theory is the product of a
$\sigma$-model with target $\MH$ and a supersymmetric $U(1)$ gauge theory.  As
in section \ref{elmag}, when the $U(1)$ theory
is included, there are branes with nonvanishing electric
or magnetic charge.  In the context of $G=U(N)$,
these are branes in which $\Tr\,\phi_0$ is a linear function of $x^1$.

A product of 't Hooft operators with $\Delta\not=0$ shifts the magnetic charge of a brane and maps
 a brane that can be described
purely in the $\sigma$-model to a brane whose description really requires us to include the
$U(1)$ vector multiplet.  Similarly, a product of Wilson operators with $\Delta\not=0$ shifts
the electric charge of a brane.
As always, a brane carrying electric charge has a Chan-Paton bundle that is a twisted
line bundle (or twisted vector bundle), not an ordinary one, over $\MH$.  This is reflected
in the
above analysis in the fact that
the ``eigenvalue'' by which a Wilson operator acts on a zero-brane ${\cal B}$
depends on the line bundle ${\cal M}$, not just its isomorphism class.
This dependence cancels out if we act with a product of Wilson operators of
$\Delta=0$.   As usual, the branes whose Chan-Paton
wavefunctions depend on a choice of universal bundle (which in the present context means
a choice of ${\cal M}$) are the ones that carry electric charge.
But for the groups $U(N)$ or $U(1)$ whose center has positive dimension, a proper description
of the branes carrying electric or magnetic charge requires including the $U(1)$ vector
multiplet in the description.

These issues arise entirely because $U(N)$ has a center $U(1)$ that is not of finite
order.  We can avoid these issues, accordingly, while maintaining the spirit of the above
demonstration of geometric Langlands duality, if we take $^LG=SU(N)$ rather than $U(N)$.
This corresponds, of course, to $G=SU(N)/\Z_N$.  Geometric Langlands duality for minuscule
representations of $SU(N)$ can be analyzed in precisely the way that we have done for $U(N)$.

\vskip.5cm
Research supported in part by NSF Grant PHY-1606531. 
\bibliographystyle{unsrt}

\end{document}